\documentclass[useAMS,usenatbib,aas_macros]{mn2e}
\usepackage{aas_macros}
\usepackage{amsmath}
\usepackage{graphicx}
\usepackage{color}
\newcommand{\adb}[1]{\textcolor{black}{#1}}   

\newcommand{\equ}[1]{eq.~(\ref{eq:#1})}
\newcommand{\equs}[1]{eqs.~(\ref{eq:#1})}

\newcommand{\se}[1]{\S\ref{sec:#1}}
\newcommand{\fig}[1]{Fig.~\ref{fig:#1}}
\newcommand{\figs}[1]{Figs.~\ref{fig:#1}}
\newcommand{\Fig}[1]{Figure~\ref{fig:#1}}

\newcommand{\tab}[1]{Table~\ref{tab:#1}}
\newcommand{\be}{\begin{equation}}
\newcommand{\ee}{\end{equation}}
\newcommand{\ba}{\begin{align}}
\newcommand{\ea}{\end{align}}
\newcommand{\bad}{\begin{equation} \begin{aligned}}
\newcommand{\ead}{\end{aligned} \end{equation}}
\newcommand{\bea}{\begin{eqnarray}}
\newcommand{\eea}{\end{eqnarray}}

\def\la{\langle}

\def\ssim{\!\sim\!}
\def\seq{\!=\!}
\def\sgt{\!>\!}
\def\slt{\!<\!}
\def\sdash{\!-\!}
\def\stimes{\!\times\!}

\newcommand{\msun}{M_\odot}
\newcommand{\Msun}{M_\odot}

\newcommand{\ifm}[1]{\relax\ifmmode#1\else$\mathsurround=0pt #1$\fi}
\newcommand{\kms}{\ifmmode\,{\rm km}\,{\rm s}^{-1}\else km$\,$s$^{-1}$\fi}

\newcommand{\Mpc}{\,{\rm Mpc}}
\newcommand{\kpc}{\,{\rm kpc}}
\newcommand{\pc}{\,{\rm pc}}
\newcommand{\cm}{\,{\rm cm}}
\newcommand{\Gyr}{\,{\rm Gyr}}

\newcommand{\Myr}{\,{\rm Myr}}

\newcommand{\yr}{\,{\rm yr}}
\newcommand{\erg}{\,{\rm erg}}
\newcommand{\ergs}{\,{\rm erg}\,{\rm s}^{-1}}

\newcommand{\cmc}{\,{\rm cm}^{-3}}
\newcommand{\cms}{\,{\rm cm}^{-2}}

\newcommand{\ltsima}{$\; \buildrel < \over \sim \;$}
\newcommand{\lsim}{\lower.5ex\hbox{\ltsima}}
\newcommand{\gtsima}{$\; \buildrel > \over \sim \;$}
\newcommand{\gsim}{\lower.5ex\hbox{\gtsima}}
\newcommand{\prop}{\propto}
\newcommand{\dd}{\rm d}
\newcommand{\pa}{\partial}

\def\omm{\Omega_{\rm m}}
\def\oml{\Omega_{\Lambda}}
\def\omb{\Omega_{\rm b}}

\def\Rv{R_{\rm v}}
\def\Vv{V_{\rm v}}

\def\Mg{M_{\rm g}}
\def\Ms{M_{\rm s}}
\def\Re{R_{\rm e}}

\def\Sig1{\Sigma_1}

\def\Rd{R_{\rm d}}

\def\Vrot{V_{\rm rot}}

\def\tinf{t_{\rm inf}}
\def\tacc{t_{\rm acc}}
\def\tsfr{t_{\rm sfr}}

\def\brho{\bar\rho}

\def\Rd{R_{\rm d}}
\def\Hd{H_{\rm d}}

\def\drhos{\rho_{\rm sfr}}

\def\rhog{\rho_{\rm g}}

\def\epsf{\epsilon_{\rm ff}}
\def\eps2{\epsilon_{-2}}

\def\tff{t_{\rm ff}}

\def\torb{t_{\rm orb}}

\def\Mv{M_{\rm v}}

\def\M11{M_{\rm v,11}}

\def\Md{M_{\rm d}}

\def\f16b{f_{\rm b,0.16}}
\def\sv25{\phi_{11,-2.5}}

\def\epsf{\epsilon_{\rm ff}}

\title[Rings around massive centres]
{Origin of Star-Forming Rings around Massive Centres
in Massive Galaxies at $z\!<\!4$}

\author[Dekel et al.]
{\parbox[t]{\textwidth}
{Avishai Dekel$^{1,2}$\thanks{E-mail: dekel@huji.ac.il},
Sharon Lapiner$^1$,
Omri Ginzburg$^1$,
Jonathan Freundlich$^1$,
Fangzhou Jiang$^1$,
\adb{Bar Finish$^1$,}
\adb{Michael Kretschmer$^{1,3}$,}
Doug Lin$^4$,
Daniel Ceverino$^5$,
Joel Primack$^6$,
Mauro Giavalisco$^7$,
Zhiyuan Ji$^7$
}
\\ \\
$^1$Racah Institute of Physics, The Hebrew University, Jerusalem 91904 Israel\\
$^2$SCIPP, University of California, Santa Cruz, CA 95064, USA\\ 
$^3$Institute for Computational Science, University of Zurich, 8057 Zurich,
Switzerland\\
$^4$Department of Astronomy and Astrophysics, University of California, 
Santa Cruz, CA 95064, USA\\
$^5$Departamento de Fisica Teorica, Facultad de Ciencias, Universidad Autonoma
de Madrid, Cantoblanco, 28049 Madrid, Spain\\ 
$^6$Physics Department, University of California, Santa Cruz,  
Santa Cruz, CA 95064, USA\\
$^7$Department of Astronomy, University of Massachusetts, Amherst, MA 01002,
USA
}

\begin{document}

\large

\pagerange{\pageref{firstpage}--\pageref{lastpage}} \pubyear{2002}

\maketitle

\label{firstpage}

\begin{abstract}
Using analytic modeling and simulations, we address the origin of an abundance 
of star-forming, clumpy, extended gas rings about massive central bodies 
in massive galaxies at $z\slt 4$. Rings form by high-angular-momentum streams 
and survive in galaxies of $M_{\rm star}\sgt 10^{9.5-10}\msun$ where 
merger-driven spin flips and supernova feedback are ineffective.  The rings 
survive after events of compaction to central nuggets.  Ring longevity was 
unexpected based on inward mass transport driven by torques from violent disc 
instability.  However, evaluating the torques from a tightly wound spiral 
structure, we find that the timescale for transport per orbital time is long 
and $\prop\! \delta_{\rm d}^{-3}$, with $\delta_{\rm d}$ the cold-to-total mass
ratio interior to the ring.  A long-lived ring forms when the ring 
{\it transport} is slower than its replenishment by {\it accretion} and the 
interior {\it depletion} by SFR, both valid for $\delta_{\rm d}\slt 0.3$.  
The central mass that lowers $\delta_{\rm d}$ is a compaction-driven bulge 
and/or dark matter, aided by the lower gas fraction at $z\slt 4$, 
\adb{provided that it is not too low.}  
The ring is Toomre unstable for clump and star formation.  
\adb{The high-$z$ dynamic rings are not likely to arise form secular resonances
or collisions. AGN feedback is not expected to affect the rings.} 
Mock images of simulated rings through dust indicate qualitative consistency 
with observed rings about bulges in massive $z\!\sim\!0.5\!-\!3$ galaxies, 
in $H_{\alpha}$ and deep HST imaging.  ALMA mock images indicate that 
$z\!\sim\!0.5\!-\!1$ rings should be detectable.  We quote expected observable 
properties of rings and their central nuggets.
\end{abstract}


\section{Introduction}
\label{sec:intro}

\begin{figure*} 
\centering
\includegraphics[width=0.44\textwidth]
{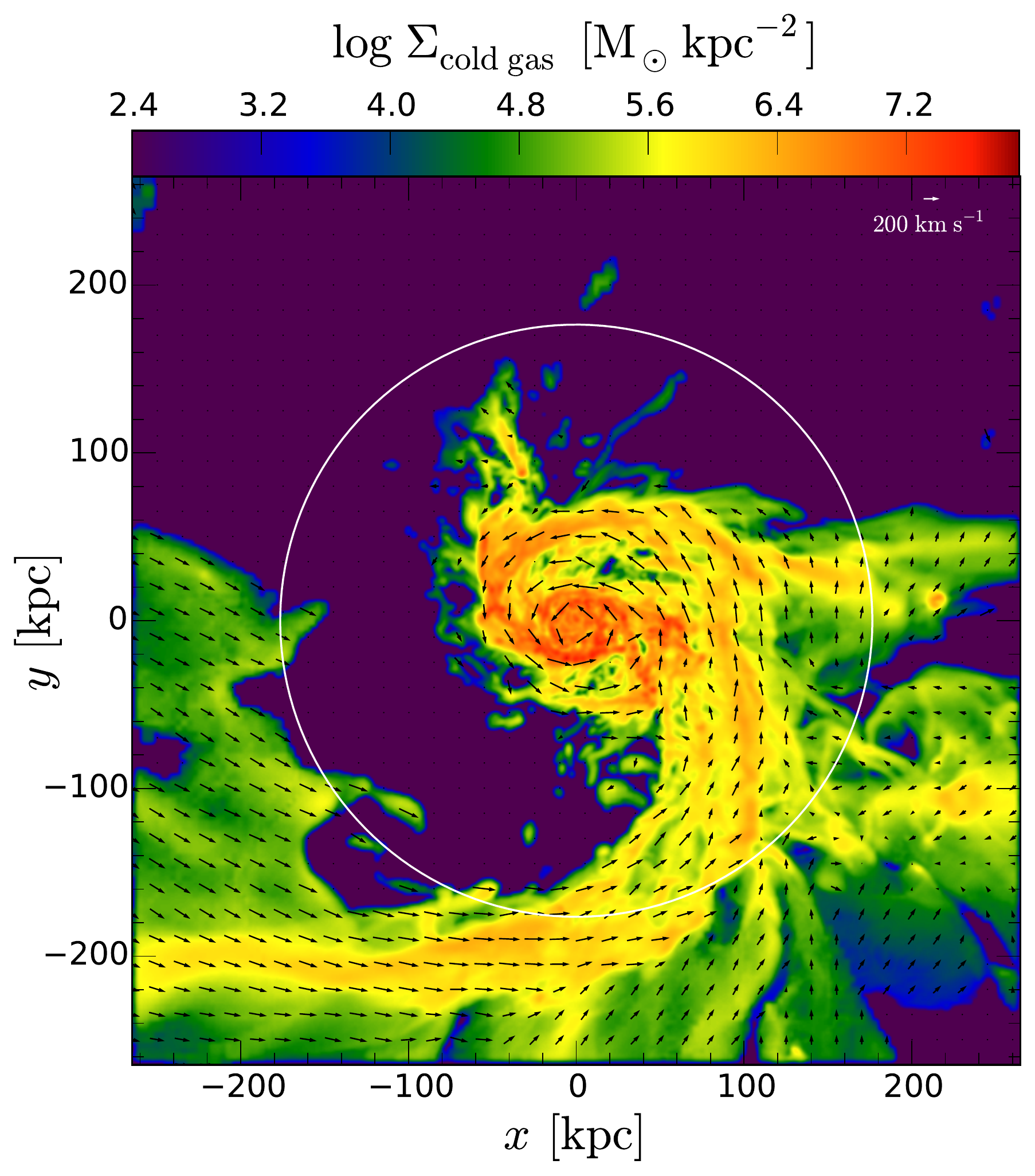}
\quad
\includegraphics[width=0.45\textwidth]
{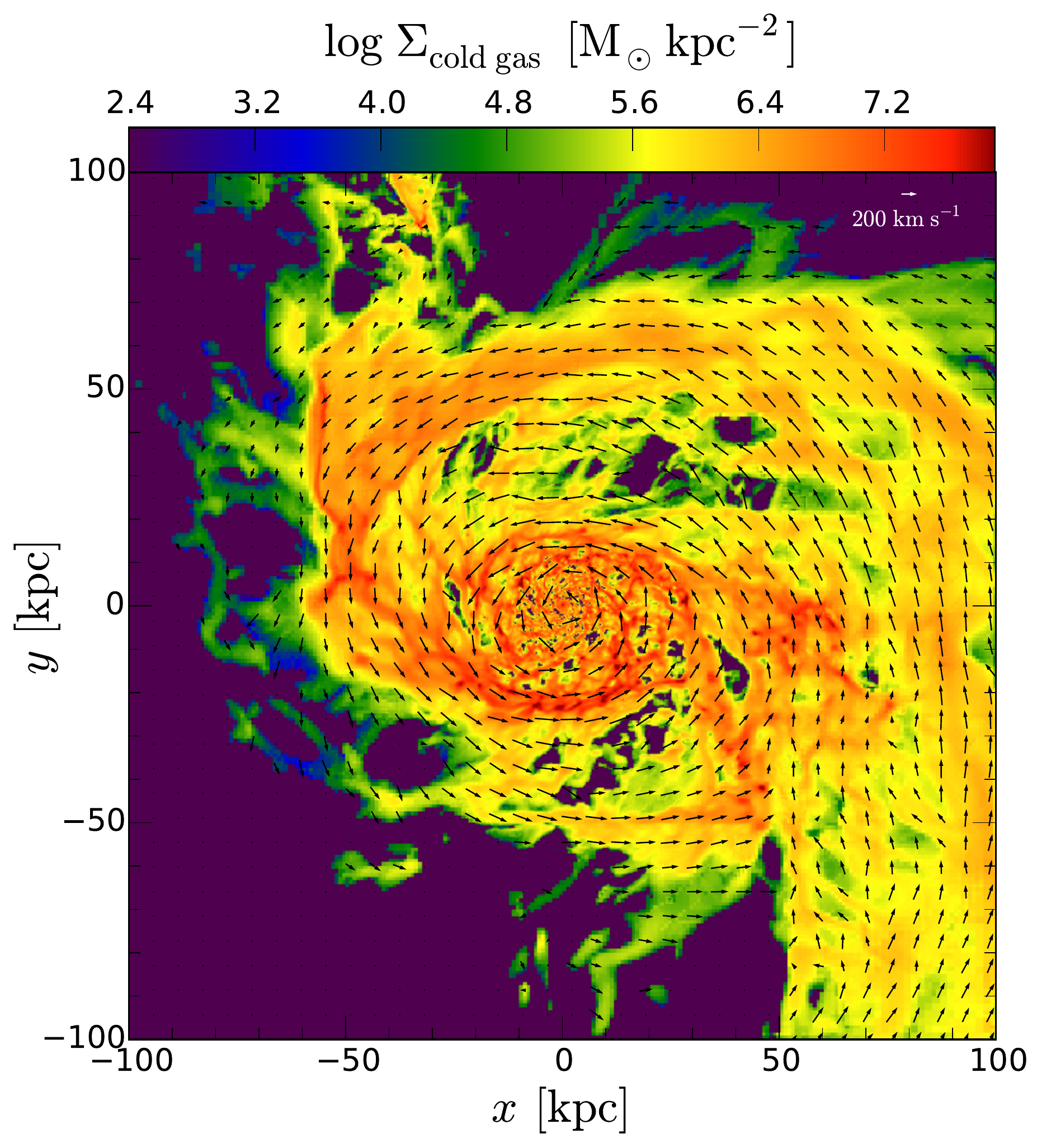}
\caption{
Ring buildup by a high-angular-momentum stream from the cosmic web.
Shown is the face-on projected gas density (color) in a VELA simulated galaxy
(V07 at $z\!=\!1.08$) and the 2D velocity field (arrows).
The virial radius is marked by a white circle.
{\bf Left:} Zoom-out on the stream extending out to $>\!300\kpc$ with an
initial impact parameter comparable to the virial radius.
{\bf Right:} Zoom-in on the spiraling-in to a ring at $\sim\! 10\!-\!20\kpc$.
As illustrated in a cartoon in Fig.~20 of \citet{danovich15},
the AM gained outside the halo by tidal torques from the cosmic web
is gradually lost in the inner halo by torques from friction, other streams
and the tilted disc, causing the buildup of a ring.
The question is what prevents the ring from contracting further.
}
\label{fig:stream_inspiral}
\end{figure*}

High-redshift galaxies are predicted to be fed by cold gas streams from the 
cosmic web \citep{bd03,keres05,db06,keres09,dekel09}. 
According to cosmological simulations,
these streams enter the dark-matter (DM) halo with high angular momentum 
(AM), which they lose in the inner halo and spiral in into
an extended gas ring \citep{danovich15}, as seen in \fig{stream_inspiral}. 

\smallskip 
This ring, like the inner disc, is at a constant risk of being disrupted by
a major merger of galaxies at nodes of the cosmic web, 
which typically involves a change in the pattern of AM-feeding streams. 
In \citet{dekel20_flip}, we showed that in haloes below a 
critical virial mass of $\Mv \!\sim\! 10^{11}\msun$, the merger-driven spin
flips are indeed disruptive as they tend to be gas rich and more frequent than 
the disc/ring orbital frequency.
In more massive haloes, the mergers are less frequent, thus possibly allowing 
the rings/discs to survive for many orbital times. 
The additional disruptive effects of supernova feedback, which could be strong 
below a similar critical mass where the potential well is shallow compared 
to the energy deposited by supernovae in the ISM \citep{ds86}, 
are also expected to be weak above this threshold mass, 
where the gas binding energy is higher.

\smallskip 
Another process that could in principle disrupt discs even above the mass 
threshold is the inward mass transport associated with violent
disc instability (VDI). When the gas fraction is high, and the bulge is not
massive,
this process has been estimated to be efficient, such that the disc or ring 
were expected to be evacuated inwards in a few orbital times 
\citep{noguchi99,immeli04_b,bournaud07c,genzel08,dsc09}. 
In contrast, as we will see below, the simulations 
\citep[indicated already in][]{cdb10,genel12},
and observations 
(below and in \se{obs}), 
show many long-lived rings in massive galaxies, thus
posing a theoretical puzzle that is our main concern here.

\smallskip 
Related to this is the phenomenon of wet compaction into a blue nugget (BN), 
which
tends to occur in most galaxies near a characteristic mass of a similar value
\citep{zolotov15,tacchella16_prof,tacchella16_ms,tomassetti16,dekel19_gold}.
This process is sometimes driven by mergers and in other times by other 
mechanisms such as counter-rotating streams.
The blue-nuggets are observed as compact star-forming galaxies 
at $z\!\sim\!1\!-\!3$ \citep[e.g.][]{barro17},
with a preferred stellar mass of $\Ms\!\sim\!10^{10}\msun$ \citep{huertas18}.
We learn from the simulations that the rings tend to form and survive especially
after a compaction event, namely above the threshold mass.
{\it The way the compaction could give rise to a ring and stabilize it against 
inward mass transport is the main issue addressed here.}
We study below the ring survival by analyzing the torques exerted by a spiral
structure. We find that the post-compaction massive bulge could be the main 
reason for slowing down the mass transport while keeping the ring Toomre
unstable for giant clumps and star formation.
By comparing the inward transport rate of the ring to its replenishment 
by external accretion and the interior depletion to star formation,
we work out the conditions for ring formation and longevity.


\smallskip 
In parallel to our analytic modeling,
we utilize a suite of VELA zoom-in hydro-cosmological simulations, which are
described in appendix \se{app_vela}, \tab{sample} and in earlier papers
\citep[e.g.][]{ceverino14,zolotov15,dekel19_ks,dekel20_flip}.
\adb{Here, we bring only a brief summary of the relevant features of these
simulations.}
The simulations are based on the Adaptive Refinement
Tree (ART) code \citep{krav97,ceverino09}. The suite consists of 34
galaxies that were evolved to $z\!\sim\! 1$, with a unique maximum spatial
resolution ranging from $17.5$ to $35 \pc$ at all times.
The dark-matter halo masses at $z\!=\!2$ range from $10^{11}$ to $10^{12}\msun$,
thus avoiding dwarf galaxies at $z\!<\!4$.
The galaxies were selected at $z\!=\!1$ such that their dark-matter haloes
did not suffer a major merger near that epoch,
which turned out to eliminate less than ten percent of the haloes.

\smallskip 
Besides gravity and hydrodynamics, the code incorporates physical process
relevant for galaxy formation such as gas cooling by atomic hydrogen and
helium, metal and molecular hydrogen cooling, photoionization heating by the
UV background with partial self-shielding, star formation, stellar mass loss,
metal enrichment of the ISM and stellar feedback. Supernovae and stellar winds
are implemented by local injection of thermal energy,
and radiation-pressure stellar feedback is implemented at a moderate level.
In general, the feedback as implemented in this suite is on the weak side
of the range of feedback strengths in common cosmological simulations, and no
AGN feedback is incorporated.

\smallskip 
In the analysis of the simulations,
the disc plane and dimensions are determined iteratively, as detailed in
\citet{mandelker14}, yielding a disc radius $\Rd$ and half-height $\Hd$
(listed in \tab{sample} at $z\!=\!2$)
that contain 85\% of the cold ($T\slt 1.5\stimes10^4$K)
gas mass out to $0.15\Rv$, where $\Rv$ is the halo virial radius. 
The level of ``disciness" is measured by the kinematic ratio of rotation 
velocity to velocity dispersion $\Vrot/\sigma$, or similarly by $\Rd/\Hd$.
Rings are identified and their properties are quantified as described in
\se{ring_prop} and in appendix \se{app_ring_detection}.
Mock images of simulated galaxies as observed through dust
are generated \citep[based on][]{snyder15} for a preliminary comparison
to galaxies observed in deep fields of the HST-CANDELS survey.
Corresponding mock ALMA images are also generated and mock 
$H_{\alpha}$ properties are computed. 

\smallskip 
There are robust observational detections in $H_{\alpha}$ for abundant 
star-forming rings about massive central bodies at $z\ssim 1\sdash 2$ 
\citep[][Genzel et al., in prep.]{genzel14_rings,genzel17,genzel20}, 
which seem to be qualitatively matched by the simulated rings described here.
Furthermore,
contrary to earlier impressions from HST-CANDELS images, 
similar star-forming rings about massive bulges are being detected 
in non-negligible abundances in massive galaxies at $z\ssim 0.5 \sdash 3$ 
when properly focusing on the deepest fields (Ji et al., in prep.).
Possibly related is the abundance of rings about massive bulges in low-redshift 
S0 galaxies entering the ``Green Valley" \citep{salim12}, also beautifully
seen in IR images of nearby galaxies with massive bulges such as M31 and 
the Sombrero galaxy.
Towards the end of this paper, we
attempt very preliminary comparisons between the theoretical and observed
high-z rings, and quote certain predicted observable ring properties for 
more rigorous comparisons with observations, to be performed beyond the scope
of this theory paper.

\smallskip 
The paper is organized as follows.
In a second introductory section, \se{ring_formation},
we elaborate on the formation of rings from the cosmic-web streams, 
the threshold mass for long-lived discs/rings due to merger-driven spin flips, 
and the expected disc disruption by inward mass transport driven by VDI.
In \se{compaction_ring} we demonstrate using the simulations the effect of 
compaction to a blue nugget on the generation of long-lived 
extended discs and then rings above the threshold mass.
In particular, in \se{ring_prop}, we quantify the ring properties
and demonstrate their correlation with the compaction events. 
In \se{ring_toy}, {\it the heart of this paper},
we attempt to understand the stabilization of an extended ring by a massive 
central body via an analytic derivation of the torques exerted on the ring by
a perturbed disc with a tightly wound spiral-arm structure. 
The condition for a long-lived ring is evaluated by comparing the inward 
transport rate to the rates of accretion and star formation in \se{survival}, 
and the model is tested against the simulations in \se{model_sims}. 
In \se{comp_obs} we make first steps of comparing the simulated rings to
observations, where we show example mock images and the corresponding profiles 
from the simulations, both in the HST bands and for ALMA, and show a sneak
preview of rings plus bulges detected in deep CANDELS fields.
In \se{conc} we summarize our conclusions.
Certain more technical matters are deferred to appendices.
In \se{app_vela} we describe the VELA simulations.
In \se{app_ring_properties} we elaborate on how we measure ring 
properties and present the distributions of certain properties.
In \se{app_prolate} we evaluate the possible torques from a prolate central body.
In \se{app_more_figures} we bring complementary images of rings in the
simulations and observations.

\section{Ring formation}
\label{sec:ring_formation}

This more detailed introductory section elaborates on the background and
motivation for the analysis of compaction-driven rings and their longevity. 

\subsection{Ring formation from cosmic-web streams}
\label{sec:danovich} 

The buildup of an extended ring is a natural result of the feeding of
high-redshift galaxies by streams of cold gas that ride the dark-matter
filaments of the cosmic web into its nodes. 
At sufficiently high redshifts, even in massive haloes the streams can
penetrate cold through the halo virial radius without being heated by a stable 
virial shock because their higher density induces efficient post-shock cooling 
that does not allow pressure support for the shock against gravitational
collapse 
\citep{bd03,keres05,db06,cattaneo06,ocvirk08,keres09,dekel09,danovich12}.
The evolution of cold-gas AM leading to the buildup of a ring 
is described in four stages in \citet{danovich15}, 
as summarized in a cartoon in their Fig.~20. 
\Fig{stream_inspiral} demonstrates the buildup of a ring by a high-AM stream,
focusing on one dominant stream in an example galaxy from the VELA suite of
zoom-in cosmological simulations.
The streams acquire excessive AM by tidal torques from the cosmic web while 
outside the halo \citep{white84}, expressed in terms of a velocity
comparable to the virial velocity and an impact parameter that could be on 
the order of the virial radius. 
As the stream penetrates into the halo it spirals in and settles 
into an extended ring at $\sim\!0.15\Rv$. The significant AM loss is by 
torques due to friction against the circum-galactic medium (CGM) and disc gas 
as well as by torques from the central galaxy and other streams. 
As the virial radius is growing with time, the initial impact parameter 
and the resultant ring radius become more extended in time.
Results of related nature were obtained from other simulations
\citep{pichon11,codis12,stewart13}.

\begin{figure*} 
\centering
\includegraphics[width=0.50\textwidth]
{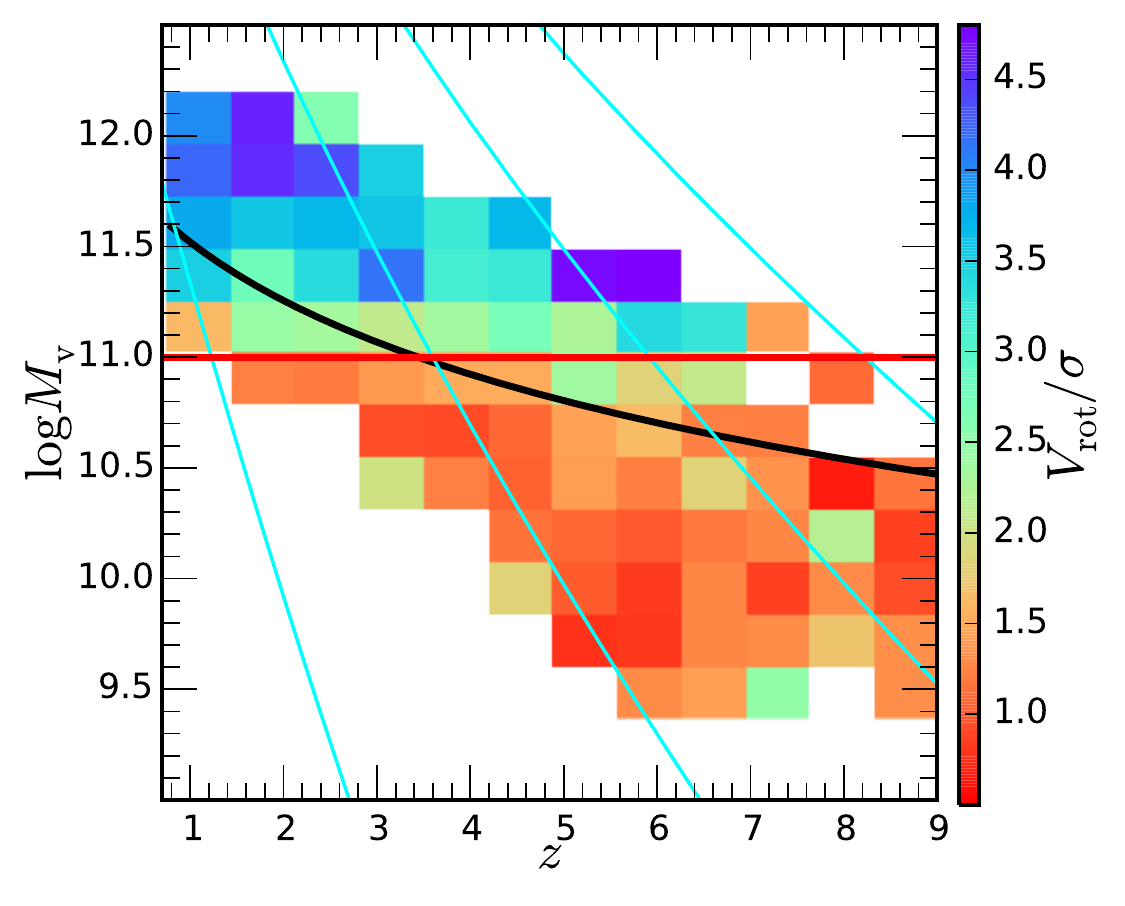}
\includegraphics[width=0.403\textwidth,trim={0 {0.025\textwidth} 0 0}]
{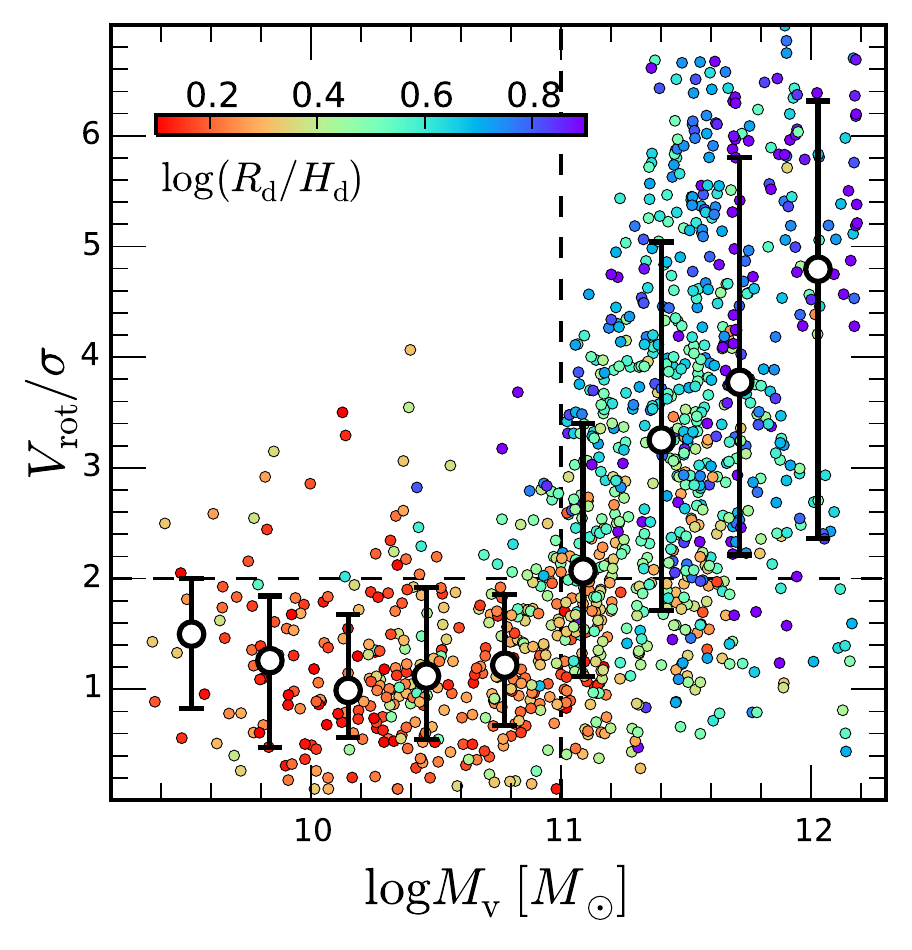}
\caption{
Disc disruption below a characteristic mass.
Shown is the degree of disciness in terms of $\Vrot/\sigma$ in the VELA 
simulations.
{\bf Left:}
$\Vrot/\sigma$ (color) is averaged within bins in the mass-redshift plane.
The black curve refers to the upper limit for effective supernova feedback at
a virial velocity $\Vv\!=\!120\kms$.
The cyan curves refer to the Press-Schechter $\nu\sigma$ peaks,
for $\nu=1,2,3,4$ from left to right, respectively.
{\bf Right:}
$\Vrot/\sigma$ as a function of halo mass $\Mv$ for all the snapshots of all
the 34 evolving galaxies.
Each point refers to a snapshot, with the median and $1\sigma$ scatter
(16\% and 84\% percentiles) marked in bins of $\Mv$.
The color refers to the alternative measure of disciness by shape using
$\Rd/\Hd$,
showing consistency between the two alternative measures of disciness.
In both figures
we see a marked transition from non-discs to discs or rings at a threshold
mass of $\Mv \simeq 10^{11}\msun$, with negligible redshift dependence,
as predicted analytically in \citet{dekel20_flip}.
}
\label{fig:disc_Mz_bins}
\end{figure*}

\smallskip
Preliminary observational kinematic studies of cold gas, 
via Lyman-alpha absorption along the line of sight to a background quasar,
or Lyman-alpha emission that is typically stimulated by a nearby quasar, 
indeed indicate detections of cold inflowing gas with high AM, consistent 
with the simulation predictions \citep[e.g.][]{martin19}. 
A recent observed system at $z\seq 2.9$, that is not illuminated by a nearby 
quasar, also indicates three cold inflowing streams 
(Daddi et al., in prep.)\footnote{In two of these streams 
the line-of-sight velocity is decreasing from large projected radii toward the 
center and even reversing its sign along the way.
Rather than interpreting this as a deceleration of the stream as it falls in,
this behavior is consistent with viewing an in-spiraling stream similar to the 
one shown in \fig{stream_inspiral} from the left or the right, where the
velocity becomes perpendicular to the line of sight as the stream enters the
inner halo, and it may even reverse its line-of-sight velocity on the other 
side of the central galaxy.}
These observations provide preliminary confirmation for the natural buildup 
of an extended ring, as seen in the simulations.

\subsection{Mass threshold for discs by merger-driven spin flips}
\label{sec:flip} 

The extended rings that form are expected to be fragile.
In \citet{dekel20_flip} 
we used the simulations and analytic estimates to explore 
how discs and rings populate the $\Mv\!-\!z$ plane. 
The disc disruption below a characteristic mass is shown in
\fig{disc_Mz_bins}, which displays the distribution of a kinematic 
measure of gas disciness, $\Vrot/\sigma$, in the VELA simulations. 
On the left, this ratio (color) is averaged over 
the simulation snapshots in bins of $\Mv$ and $z$.
On the right, this ratio is shown as a function of halo mass for every
snapshot.
We see a systematic gradient of disciness with mass,
and a division between the zones of non-disc and disc 
dominance at a critical mass of $\Mv\!\simeq\!(1\!-\!2)\times 10^{11}\msun$,
where $\Vrot/\sigma\!\simeq\!2$.
No significant redshift dependence is seen.
A measure of disciness by shape reveals similar results, with a transition
at $\Rd/\Hd\!\simeq\!2.5$

\smallskip 
In particular, major mergers are expected to disrupt rotation-supported systems
if the orbital AM and the spin of the merging galaxy are not aligned with the
spin of the galaxy. This is expected to be the case in mergers
of high-sigma nodes of the cosmic web, when the pattern of feeding streams
drastically changes. 
Figures 4 and 5 of \citet{dekel20_flip} demonstrate that the disruption below
the critical mass is largely due to merger-driven spin flips in less than an
orbital time.  
The mass threshold is derived by a simple analytic model, 
contrary to the naive expectation of a redshift threshold based
on halo merger rates, where the time between mergers with respect to the
halo orbital time is $t_{\rm mer}/\torb \!\prop\! (1+z)^{-1}$
\citep{neistein08_m,dekel13}.
This turns into a mass threshold when taking into account the increase of the
ratio of baryonic galaxy mass to its halo mass with mass and redshift.
While the external inflow (and merger) rate of mass and AM  
that could damage a disc is primarily determined by the total halo mass,
the AM of the existing central galaxy
that it is affecting is increasing with its baryonic mass,
so the damaging relative change in AM 
is expected to be larger for lower-mass galaxies and at
lower redshifts. 
This introduces a strong mass dependence in the disc survivability 
and weaken its redshift dependence, as seen in the simulations.
 
\smallskip
We thus learn in \citet{dekel20_flip} 
that above a threshold mass the discs and rings are expected not to suffer 
disruptions by merger-driven spin flips on timescales comparable to their
orbital times.

\smallskip 
Given this threshold mass,
the expected abundance of gas discs/rings in a given redshift
can be estimated by the number density of haloes above the threshold mass.
For the LCDM cosmology
the Press-Schechter formalism yields a comoving number density of 
$n\!>\!10^{-2}\Mpc^{-3}$ in the redshift range $z\!=\!0\!-\!2$,
and $n\!>\! 2.8\times 10^{-3},\, 5.2\times 10^{-4},\, 
3.2\times 10^{-5}\Mpc^{-3}$ at $z\!\simeq\!4,6,10$ respectively.
%

\subsection{Rapid inward mass transport driven by violent disc instability}
\label{sec:vdi}


In addition to mergers,
another risk for the long-term survival of an extended disc or ring is
the inward mass transport associated with VDI.
In a gravitationally unstable gas disc, the non-cylindrically symmetric
density perturbations exert torques on the rest of the disc, which typically
cause transport of AM outward.
Then, AM conservation, or the angular Euler equation
\citep[eq. 6.34 of][hereafter BT]{bt08}, implies an associated mass transport
inward, in terms of clump migration and gas inflow through the disc
\citep[e.g.][]{noguchi99,gammie01,bournaud07c,elmegreen08b,dsc09,
cdb10,bournaud11,forbes12,ceverino12,goldbaum15,goldbaum16}.

\smallskip
Considering mutual encounters between the giant clumps in a VDI disc,
\citet[][eq. 21]{dsc09} evaluated the evacuation time of the disc to be
\be
\tinf \simeq 1.7\, \alpha_{0.2}^{-1}\, Q^2\, \delta_{\rm d}^{-2}\,
\torb
\, ,
\label{eq:vdi1}
\ee
where $\alpha\!=\!0.2\alpha_{0.2}$ is the instantaneous fraction of the 
cold disc mass in clumps and $Q\! \sim\! 1$ is the Toomre parameter.
The most meaningful variable here is $\delta_{\rm d}$, the mass ratio of cold 
disc
to total mass within the sphere of radius $r$ where the timescale is evaluated,
\be
\delta_{\rm d} \equiv \frac{M_{\rm d}}{M_{\rm tot}} \, .
\label{eq:deltad}
\ee
The disc mass $M_{\rm d}$ refers to the ``cold" mass that participates in the
gravitational instability. In principle it includes the cold gas and the
young stars, but it can be approximated to within a factor of two
by the cold gas, as the young stars typically contribute about half of the gas
mass.
The total mass includes also the ``hot" stars in the disc and bulge and the
dark-matter mass within $r$.
This quantity, $\delta_{\rm d}$, will turn out to also play a major role in our 
analytical modeling of rings in \se{ring_toy} below.

\smallskip
An alternative estimate of the inflow time has been obtained in
\citet[][eq. 24]{dsc09}
based on the shear-driven mass-inflow rate of \citet{shakura73} and the
maximum dimensionless AM flux density $\tilde{\alpha}$ obtained from
simulations by \citet{gammie01},
yielding a lower-limit of
\be
\tinf \sim 1.2\, Q^{-2}\, \delta_{\rm d}^{-2}\, \torb \, .
\label{eq:vdi2}
\ee
With $Q$ between unity and $0.68$, appropriate for marginal instability of a
thick disc \citep{goldreich65_thick}, this is comparable to the estimate
in \equ{vdi1} despite the opposite dependence on $Q$.

\smallskip
In a VDI disc with $\delta_{\rm d} \! \sim\! 0.3\!-\!0.5$,
we thus expect an inward mass transport within a few orbital times.
With such a rapid inflow rate, one would not expect the extended
discs to survive for a long time the way they do for massive, post-compaction
galaxies in the simulations (see below). This is
given that the expected average timescale for accretion into the ring is much
longer, $\tacc\! \sim\! 20\, (1+z)^{-1} \torb$, as estimated in \equ{tacc} 
below.
This puzzling low inflow rate of the post-compaction rings in the
simulated galaxies has been a long-standing theoretical challenge,
which we seek to solve here.
It turns out that the same quantity, $\delta_{\rm d}$, will play a major role 
also in our analytical modeling of rings in \se{ring_toy}.

\section{Post-Compaction Discs and Rings in Simulations}
\label{sec:compaction_ring}

The mass threshold for survival of discs and rings, 
which we interpreted in \citet{dekel20_flip} as largely due to merger-driven 
spin flips on an orbital timescale, is apparently associated with another 
physical process that tends to occur near a similar characteristic mass, 
that of a major compaction event.
We describe here how long-lived discs and rings tend to appear in the
simulations after such a compaction event, once a massive bulge has formed.   

\subsection{Compaction to a blue-nugget}
\label{sec:compaction}

\begin{figure*} 
\centering
\includegraphics[width=0.5\textwidth,height=0.432\textwidth,trim={0 -0.73cm 0 0}]
{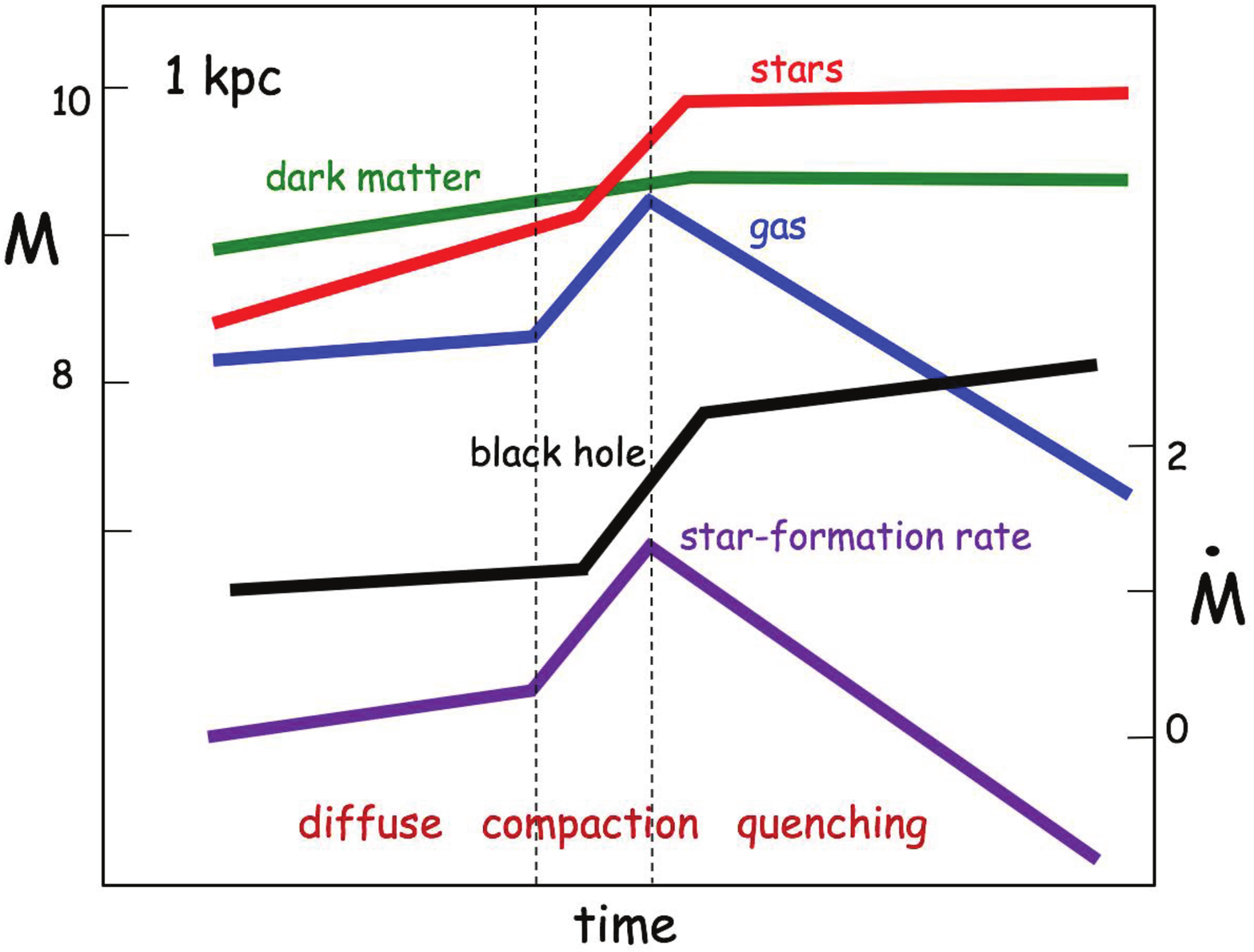}
\includegraphics[width=0.46\textwidth,trim={0 0.3cm 0 0}]
{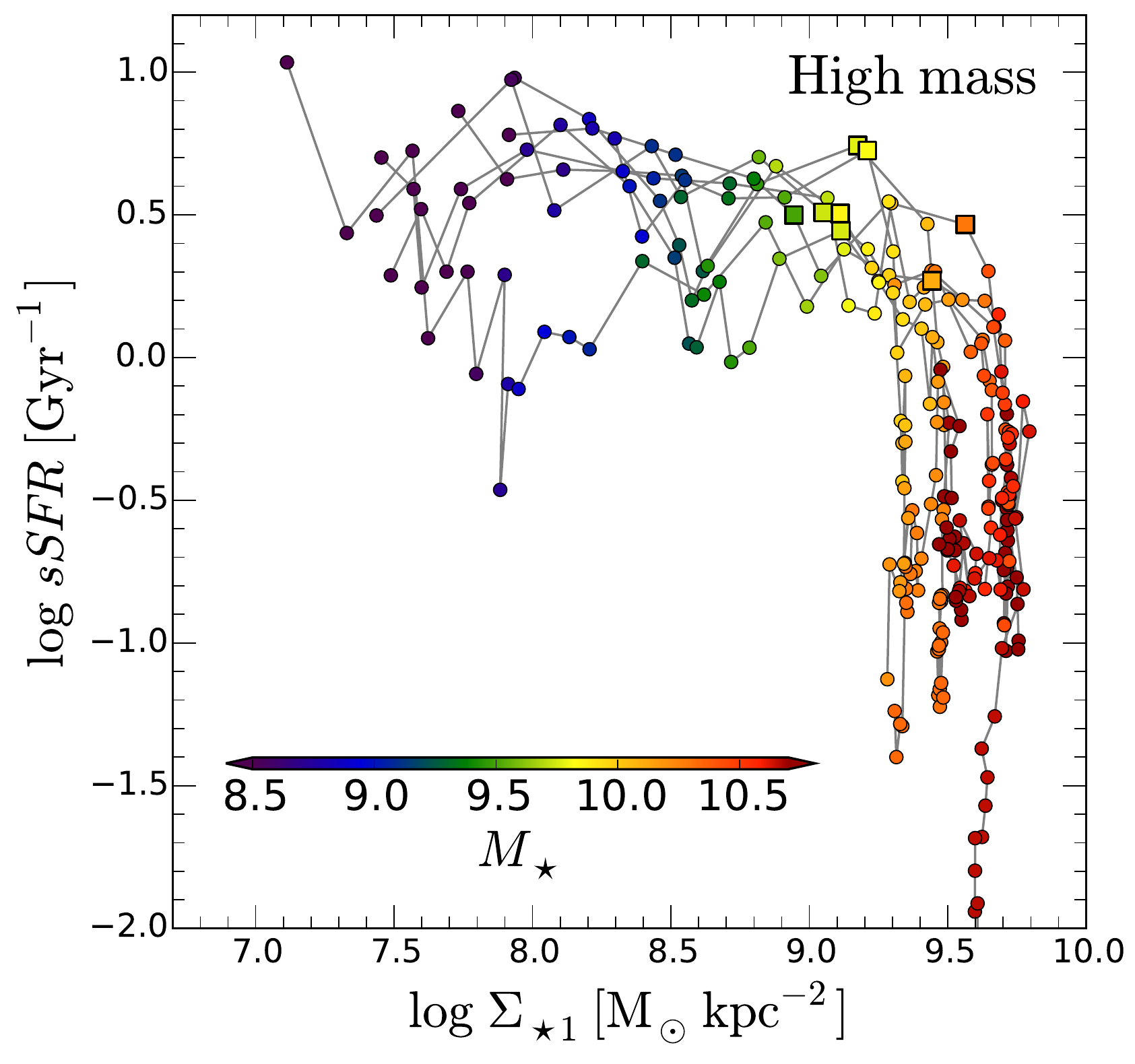}
\caption{ 
Compaction to a blue nugget and quenching in cosmological simulations.
{\bf Left:}
A cartoon describing a typical wet-compaction event (see images in
\fig{mosaic_v07}), 
showing the evolution of masses (in $\log (M/\msun)$)
within the inner $1\kpc$ \citep[following][]{zolotov15}.
The compaction is the steep rise of gas mass (blue), by an order of magnitude
during $\sim 0.3\, t_{\rm Hubble}$, reaching a peak as a blue
nugget, and soon after declining as the central gas is depleted by star
formation and outflows with no replenishment. 
The SFR (magenta, in $\log (\msun \yr^{-1})$) follows closely,
demonstrating post-BN compaction-triggered central quenching.
The central stellar mass (red) is rising accordingly during the compaction,
and it flattens off post-BN.
The inner $1\kpc$ is dominated by dark matter (green) pre compaction
and by baryons (stars, red) post compaction.
The ``disc" kinematics is dispersion-dominated pre-BN and rotation-dominated
post-BN (\fig{vela_V}).
The time of the major BN event is typically when the galaxy is near the golden
mass, $\Ms \sim 10^{10}\msun$, separating between the pre-compaction supernova
phase and the post-compaction hot-CGM phase.
The black-hole growth (black), which is suppressed by supernova
feedback pre compaction, is growing during and after the compaction in
the hot-CGM phase above the golden mass. The onset of rapid black-hole
growth is driven by the compaction event \citep{dekel19_gold}.
{\bf Right:}
The universal L-shape evolution track of eight VELA simulated galaxies in the
plane of sSFR and stellar surface density within $1\kpc$,
$\Sigma_1$, which serves as a measure of compactness
(following Lapiner, Dekel et al., in preparation).
The compactness is growing at a roughly constant sSFR (horizontally)
before and during the compaction event, turning over at the blue-nugget phase
(the ``knee", marked by a square symbol) to quenching at a constant
$\Sigma_1$ (vertically).
A similar behavior is seen observationally \citep[][Fig.~7]{barro17},
with the value of $\Sigma_1$ at the BN phase weakly increasing with redshift.
Note that this phenomenon is not caused by AGN feedback.
}
\label{fig:compaction}
\end{figure*}

\begin{figure*} 
\centering
\includegraphics[width=0.9\textwidth]
{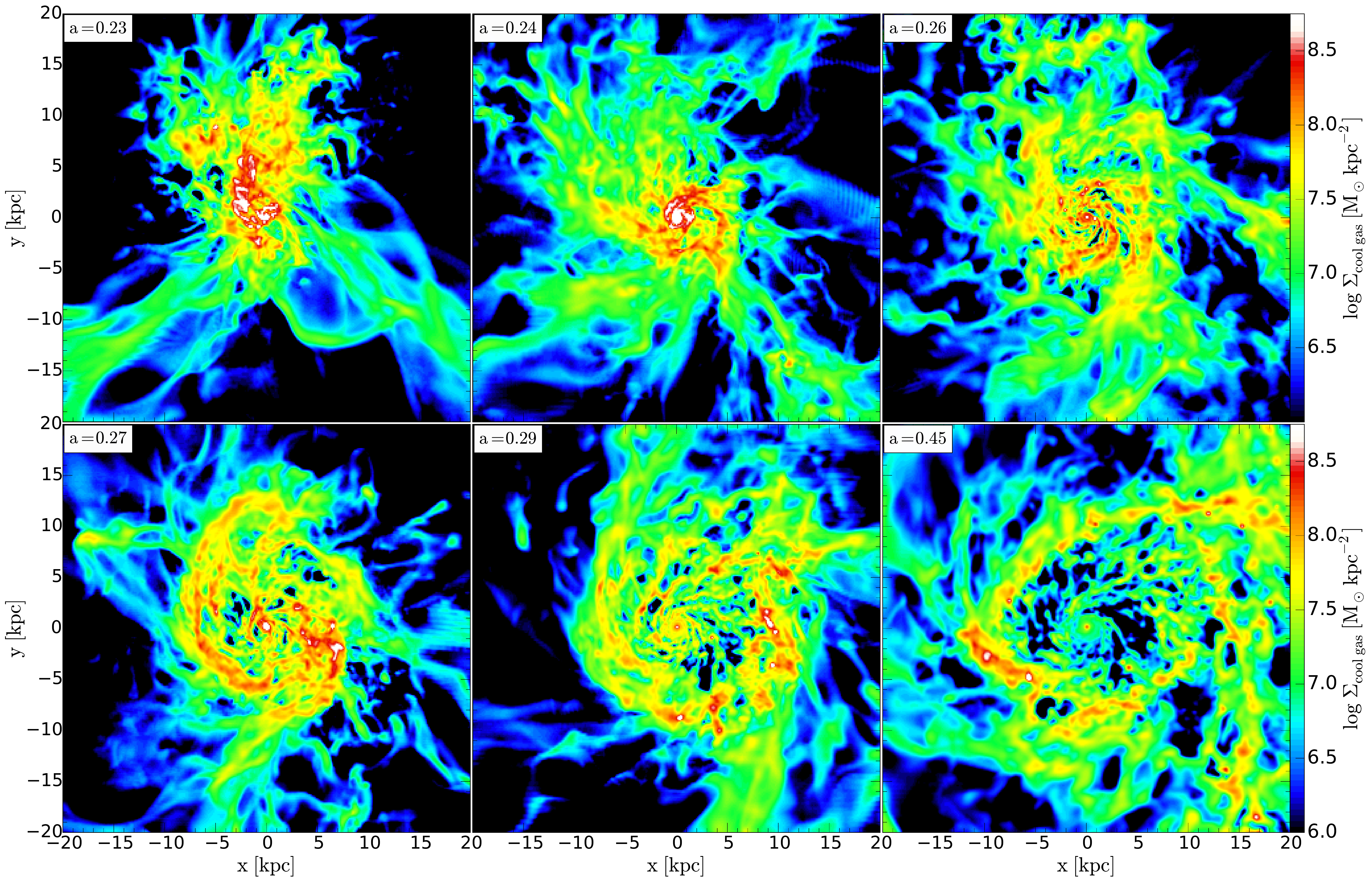}
\includegraphics[width=0.9\textwidth]
{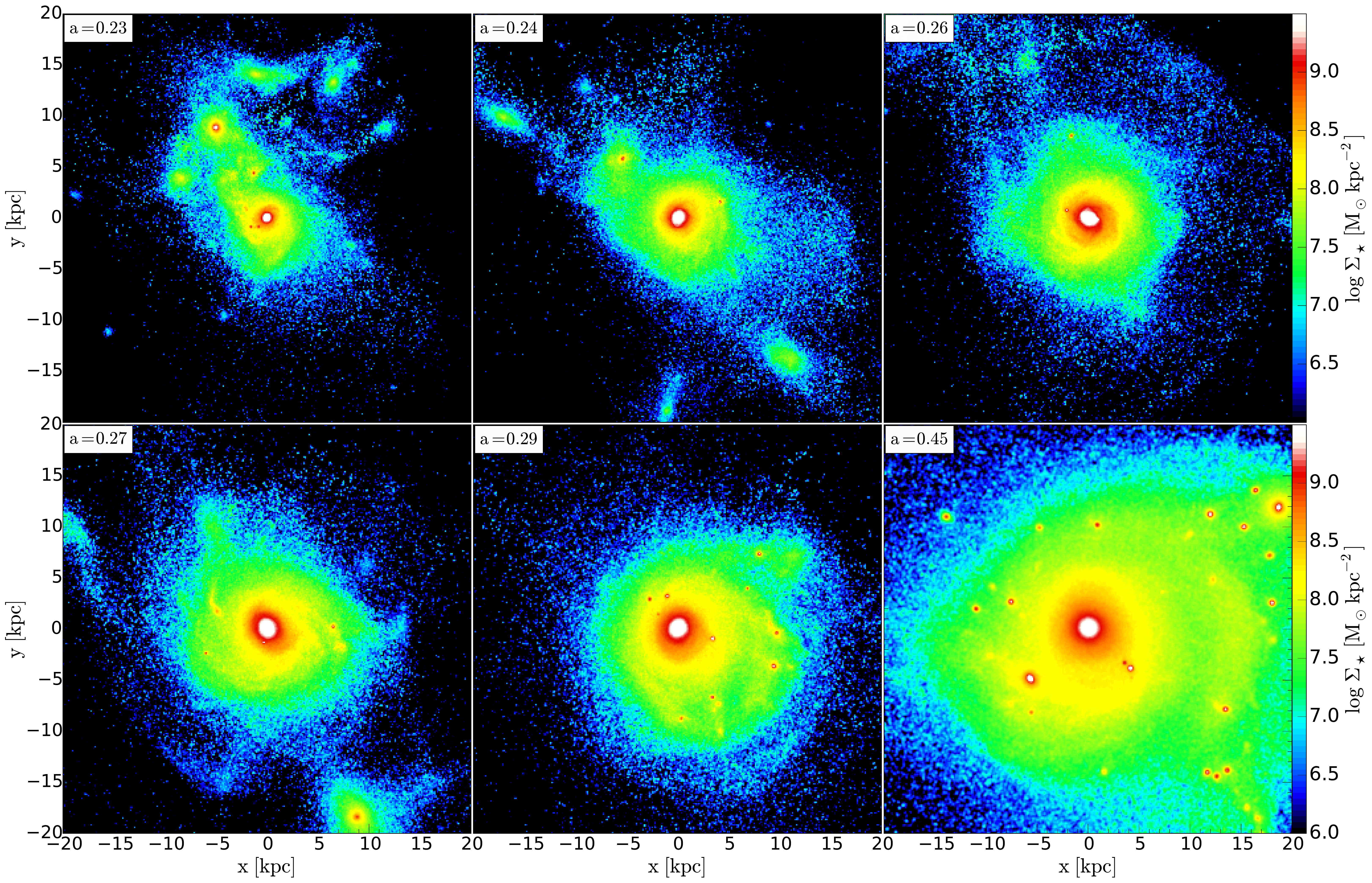}
\caption{
Compaction to a blue nugget and post-compaction gas disc and ring surrounding a
massive compact bulge.
Shown are the projected densities of gas (top) and stars (bottom)
in different phases [expansion factor $a=(1+z)^{-1}$ is marked]
during the evolution of one of the VELA simulated galaxies (V07).
The projections are face on with respect to the AM.
From top to bottom, left to right.
First: during the compaction process
($\log \Ms \seq 10.0$, $\log \Mv\seq 11.6$).
Second: at the blue-nugget phase (10.3, 11.7).
Third and fourth: post-compaction VDI disc (10.5, 11.8).
Fifth and sixth: post-compaction, clumpy, long-lived ring,
fed by incoming streams (10.8, 12.1).
The stellar compact red nugget forms during and soon after the
compaction and the resulting bulge remains compact and grows massive
thereafter.
}
\label{fig:mosaic_v07}
\end{figure*}

Cosmological simulations show that most galaxies evolve through a dramatic
wet-compaction event, which tends to occur at its maximum strength when the
galaxy mass is near or above the golden value, $\Mv\!\sim\! 10^{11.5}\msun$ and
$\Ms\!\sim\! 10^{9.5}\msun$,
especially at $z\!=\!1\!-\!5$ when the gas fraction is high
\citep{zolotov15,tacchella16_ms,tomassetti16,dekel19_gold}.
The wet compaction is a significant gaseous contraction into a compact central
star-forming core within the inner $1\kpc$, termed ``a blue nugget" (BN).
The gas consumption by star formation and the associated gas ejection by
stellar and supernova feedback trigger central gas depletion and
inside-out quenching of star-formation rate (SFR) 
\citep{tacchella16_prof}.
The cartoon in \fig{compaction} illustrates the main features of this
sequence of events as seen in the simulations 
via the evolution of gas mass, stellar mass and SFR within the inner 
kiloparsec.
\adb{The compaction is identified in each simulated galaxy primarily by the 
peak in gas mass within $1\kpc$ that is followed by a significant depletion. 
Secondary tracers of compaction are the end of the steep rise in stellar mass 
within $1\kpc$ where it turns into a plateau, and the transition from central 
dark-matter dominance to baryon dominance.}
In order to more directly compare to observations, the right panel of
\fig{compaction} shows simulated evolution tracks of 
galaxies in the plane of specific SFR (sSFR) versus compactness as measured 
by the stellar surface density within $1\kpc$, termed $\Sigma_1$.
A compaction at a roughly constant sSFR turns into quenching at a constant
$\Sigma_1$, generating an L-shape evolution track with the ``knee" marking the
blue-nugget phase.
This characteristic L-shape evolution track has been confirmed observationally
\citep[e.g.][Fig.~7]{barro17}.
\Fig{mosaic_v07} illustrates through images of gas and stellar surface
density the evolution through the compaction, blue-nugget and post-blue-nugget 
phases in an example VELA galaxy (V07), to be discussed below.
\Fig{app_mosaic12_v07} in appendix \se{app_more_figures} 
brings a more detailed sequence of the evolution.

\smallskip 
Observationally,
it became evident that the massive, passive galaxies, which are
already abundant at $z\!\sim\! 2\!-\!3$, are typically compact,
encompassing $\sim\! 10^{10}\msun$ of stars within $1\kpc$,
termed ``red nuggets"
\citep{dokkum08,damjanov09,newman10,
dokkum10,damjanov11,whitaker12,bruce12,dokkum14,dokkum15}.
Their effective radii are typically one percent of their halo virial radii,
which is smaller than one would expect had the gas started in the halo with a
standard spin of $\lambda \!\sim\! 0.035$ and conserved AM during the infall.
This indicates dissipative inflow associated with AM loss, namely a wet 
compaction \citep{db14}, and it implies the presence of gaseous blue nuggets as 
the immediate progenitors of the red nuggets.
Indeed, star-forming blue nuggets have been convincingly observed, with
masses, structure, kinematics and abundance consistent with being the
progenitors of the red nuggets
\citep{barro13,barro14_bn_rn,barro14_kin,williams14,barro15_kin,dokkum15,
williams15,barro16_alma,barro16_kin,barro17_alma,barro17}.
In particular,
a machine-learning study, after being trained on mock dusty images of the blue 
nuggets as identified in the simulations, recognized with high confidence 
similar blue nuggets in the CANDELS-HST multi-color imaging survey of 
$z\!=\!1\!-\!3$ galaxies \citep{huertas18}.
\adb{The comoving number density of blue nuggets at $z \ssim 2$ is
estimated to be $n \ssim 2 \times 10^{-4} \Mpc^{3}$ 
both in the simulations and in the observations \citep{barro17}. 
}

\smallskip 
The AM loss leading to compaction is found in the simulations 
to be caused either by wet mergers ($\sim\! 40\%$ by major plus minor mergers),
by colliding counter-rotating streams, by recycling fountains
or by other processes (in prep.),
and to be possibly associated with VDI \citep{db14}.
These processes preferentially occur at high redshifts,
where the overall accretion is at a higher rate and more gaseous,
leading to deeper compaction events.

\begin{figure*} 
\centering
\includegraphics[width=0.80\textwidth]
{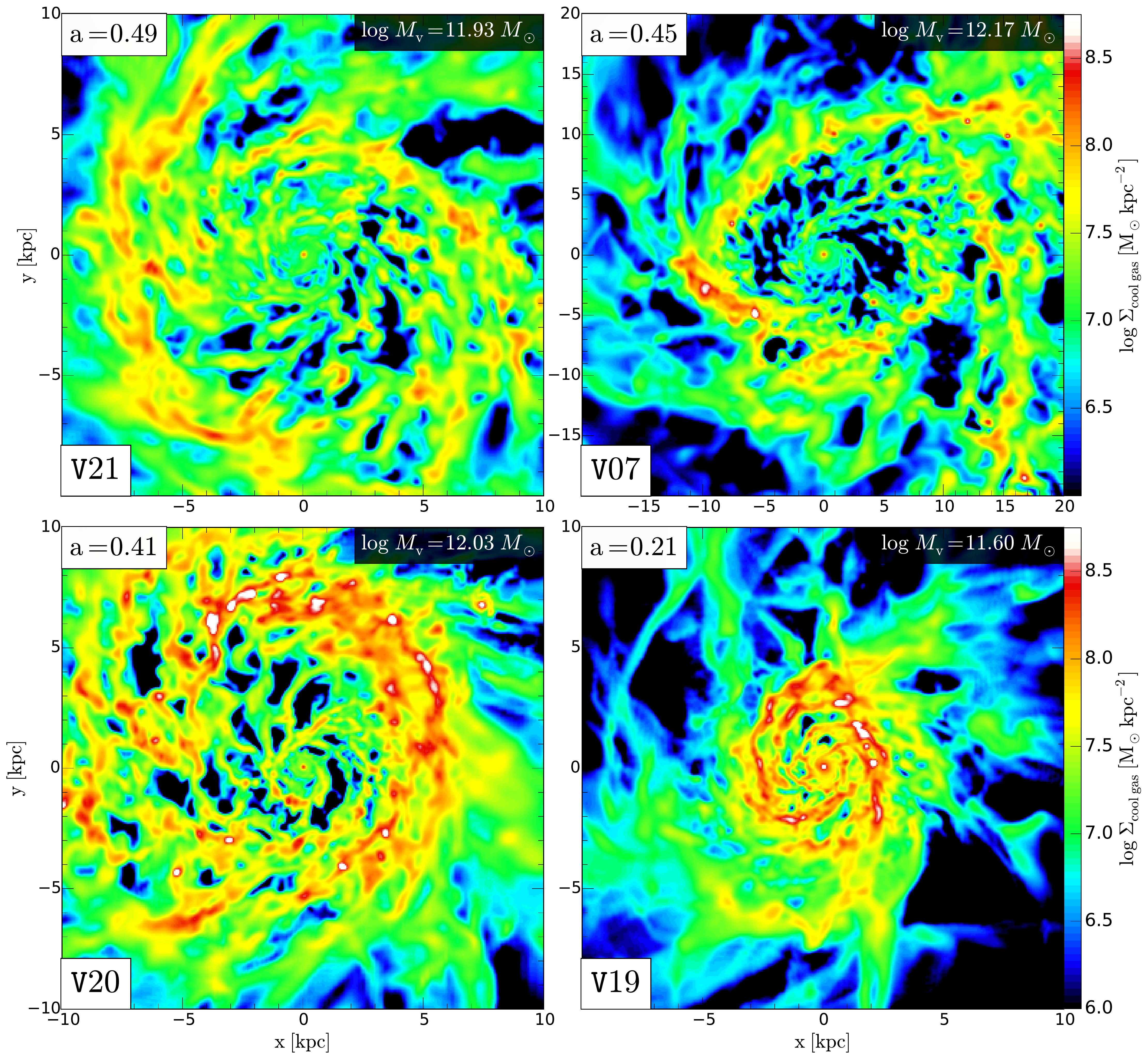}
\caption{
Post-compaction rings.
Face-on projected images of gas density in four VELA galaxies
at their post-compaction phase,
displaying pronounced, extended, clumpy, star-forming rings, 
fed by cold streams, surrounding gas-depleted central regions.
The corresponding stellar density maps show massive compact central bulges.
}
\label{fig:rings_gas}
\end{figure*}

\begin{figure*} 
\centering
\includegraphics[width=1.0\textwidth,height=0.26\textwidth]
{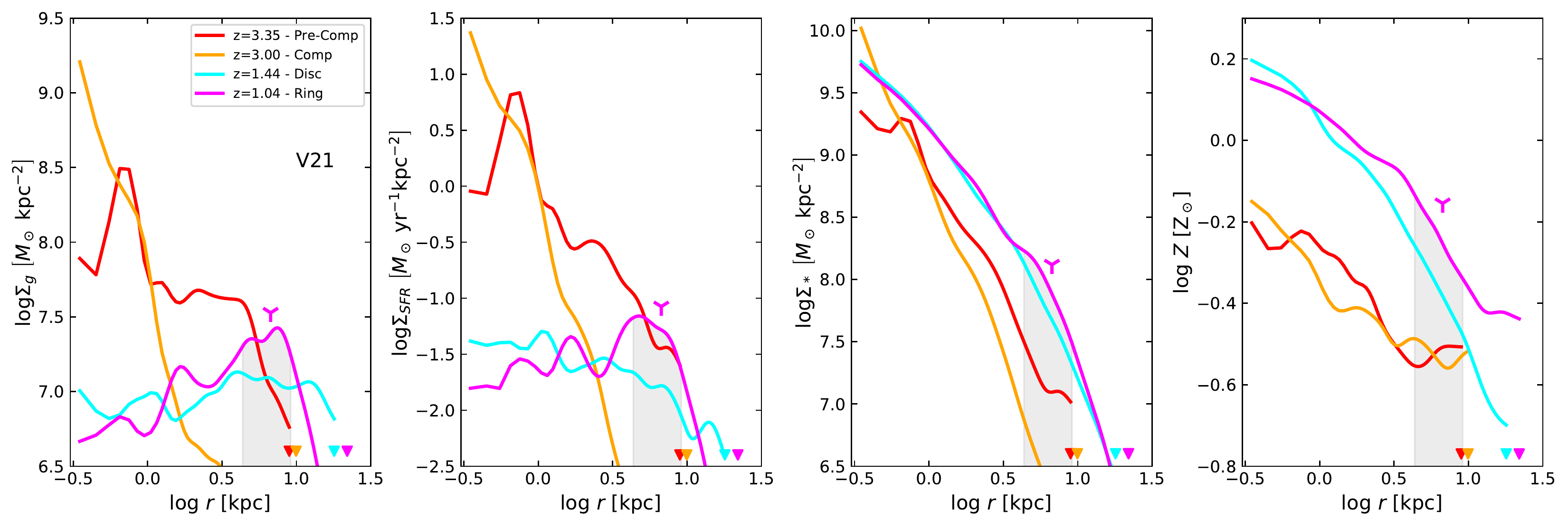}
\includegraphics[width=1.0\textwidth,height=0.26\textwidth]
{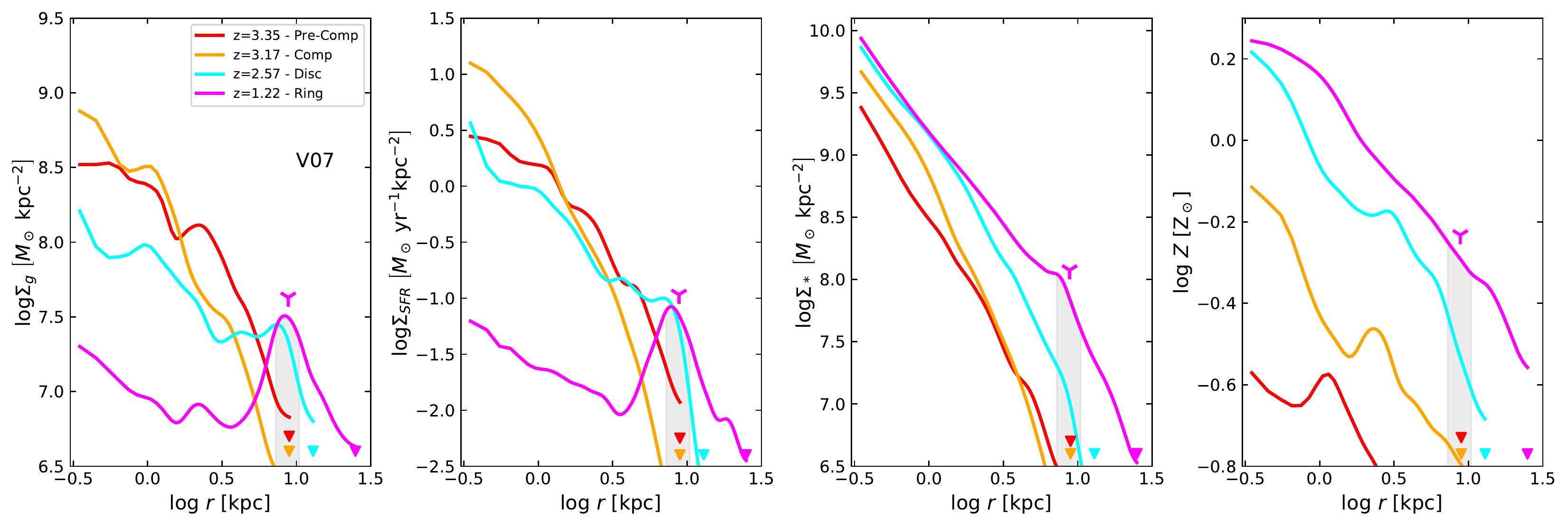}
\includegraphics[width=1.0\textwidth,height=0.26\textwidth]
{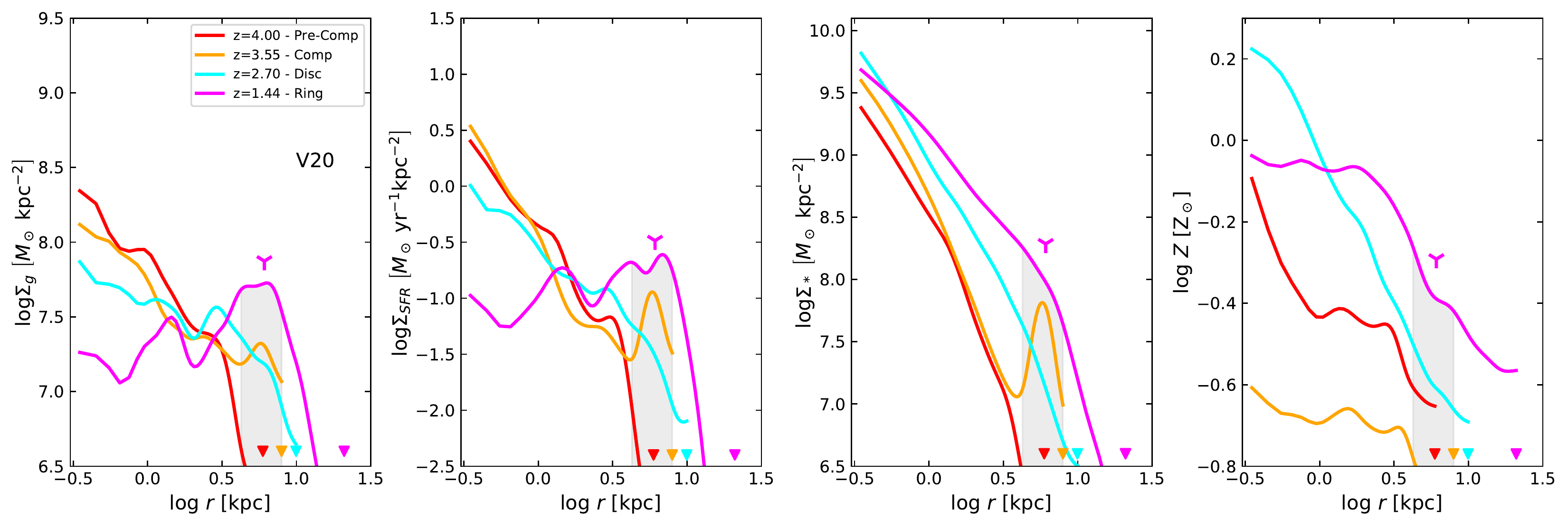}
\caption{
Evolution of profiles through compaction and ring formation.
Shown are surface-density radial profiles of gas, SFR, stars
and metallicity
for three of the VELA galaxies that develop pronounced rings.
The profiles for each galaxy are shown at the four phases of evolution 
(see \fig{mosaic_v07}),
namely pre compaction (red), during compaction (yellow), 
early post compaction (cyan) and late post compaction (magenta),
at the redshifts indicated.
The profiles are smoothed with a Gaussian of standard deviation $0.05$ in 
$\log r$.
The triangles mark the disc radii $R_{\rm d}$.
The shaded area marks the $\pm 1\sigma$ ring width in gas at the late 
post-compaction phase, and the wedge marks $r_0$.
The gas and SFR profiles show the post-compaction appearance of a ring.
It is associated with a growth in the stellar mass encompassed by the ring,
\adb{namely within $5\sdash 10\kpc$}
(note the different vertical axes for gas and stars).
The metallicity in the ring is low, reflecting the fact that it is largely
made of freshly accreted gas.
}
\label{fig:profiles_rings}
\end{figure*}

\smallskip 
The compaction events mark drastic transitions in the galaxy
structural, compositional, kinematic and other physical properties, which
translate to pronounced changes as a function of mass near the characteristic
mass for major blue nuggets
\citep{zolotov15,tacchella16_prof,tacchella16_ms}.
The compaction triggers inside-out quenching of star formation,
to be maintained by a hot CGM in massive haloes, possibly aided by AGN feedback.
This is accompanied by a structural transition from a diffuse and largely
amorphous configuration to a compact system, possibly surrounded by an
extended gas-rich ring and/or a stellar envelope.
The kinematics evolves accordingly from pressure to rotation support
(\fig{vela_V} below).
Due to the compaction, the central region turns from dark-matter dominance 
to baryon dominance, which induces a transition of global
shape from a prolate to an oblate stellar system 
\citep{ceverino15_shape,tomassetti16}
consistent with observations \citep{zhang19}.
Finally, the blue nugget marks a transition in the central black-hole growth
rate from slow to fast 
\citep{dubois15,bower17,angles17,habouzit18,dekel19_gold}, 
which induces a transition from supernova feedback to AGN feedback as the 
main source for quenching of star formation.

\smallskip 
Especially important for our purpose here is that the blue nuggets favor
a characteristic mass.
According to the simulations, minor compaction events may occur at all masses
in the history of a star-forming galaxy (SFG).
Indeed, repeated episodes of minor compactions and subsequent quenching 
attempts can explain the confinement of SFGs to a narrow Main Sequence
\citep{tacchella16_ms}.
However, the major compaction events, those that involve an order-of-magnitude
increase in central density, cause a transition from central dark-matter
dominance to baryon dominance, and trigger significant and long-lasting
quenching,
are predicted by the simulations to occur near a characteristic halo 
mass about $\Mv\!\sim\!10^{11.5}\msun$,
see \citet[][Fig.~8]{tomassetti16} and \citet[][Fig.~21]{zolotov15}.
This has been confirmed by the deep-learning study of VELA simulations
versus observed CANDELS galaxies \citep{huertas18}, which
detected a preferred stellar mass for the observed blue nuggets
near the golden mass, $\Ms \!\sim\! 10^{9.5-10}\msun$.
The significance of this finding is strengthened by the fact that
the same characteristic mass has been recovered after eliminating from the
training set the direct information concerning the mass, through the galaxy
luminosity.
One may suspect 
that the compaction events are especially pronounced in galaxies near the 
critical mass primarily due to the fact that supernova feedback, which weakens 
compactions in lower masses, becomes inefficient near and above this mass 
\citep{dekel19_gold}.
The characteristic mass for compaction events, being in the ball park of the
mass threshold for discs seen in \fig{disc_Mz_bins}, 
may indicate that the compaction
events also have a major role in the transition from non-discs to discs, to be
addressed in this paper.

\begin{figure*} 
\centering
\includegraphics[width=0.65\textwidth]
{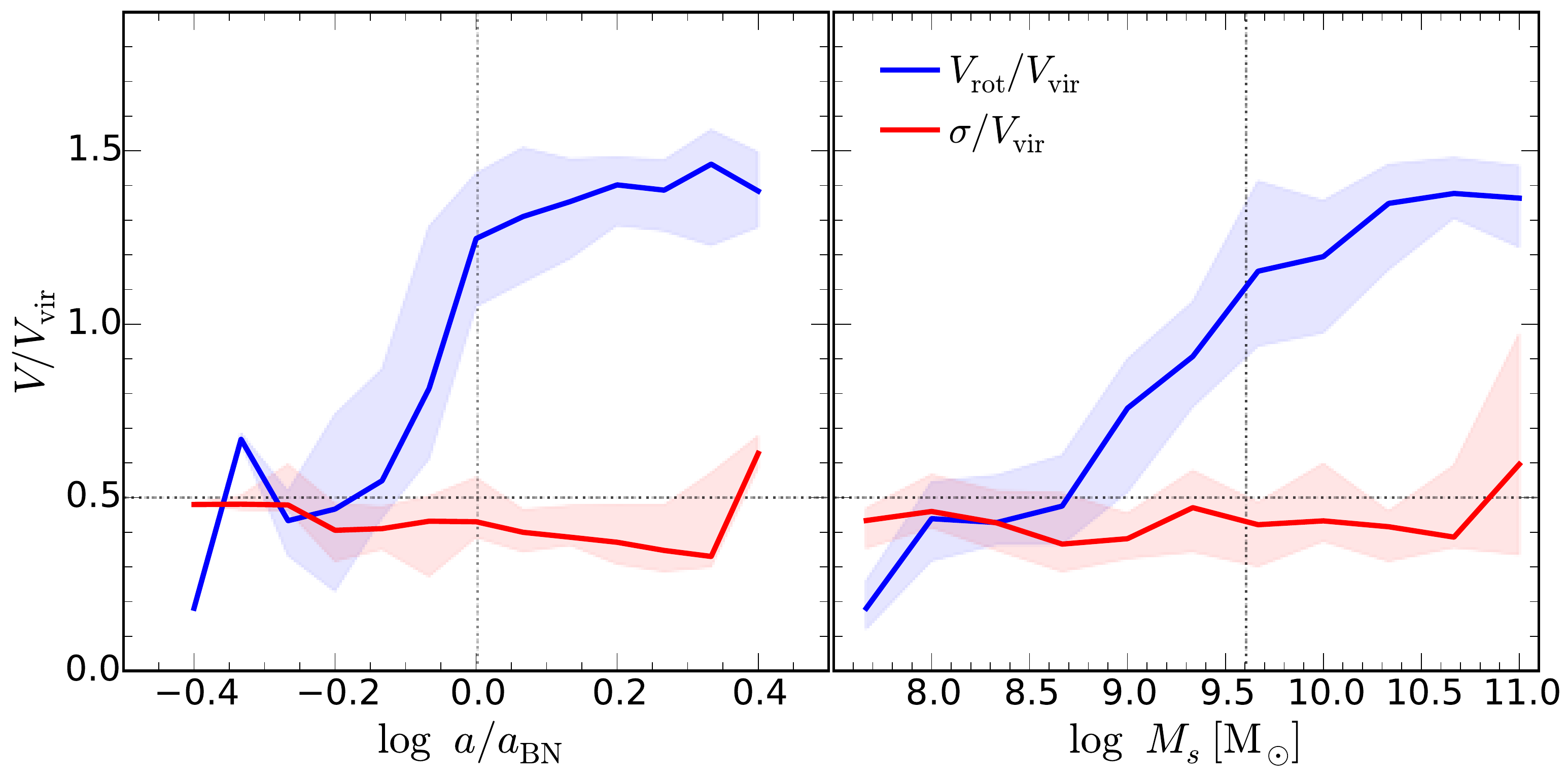}
{\includegraphics[width=0.323\textwidth,trim={0 0 0 0.15cm},clip]
{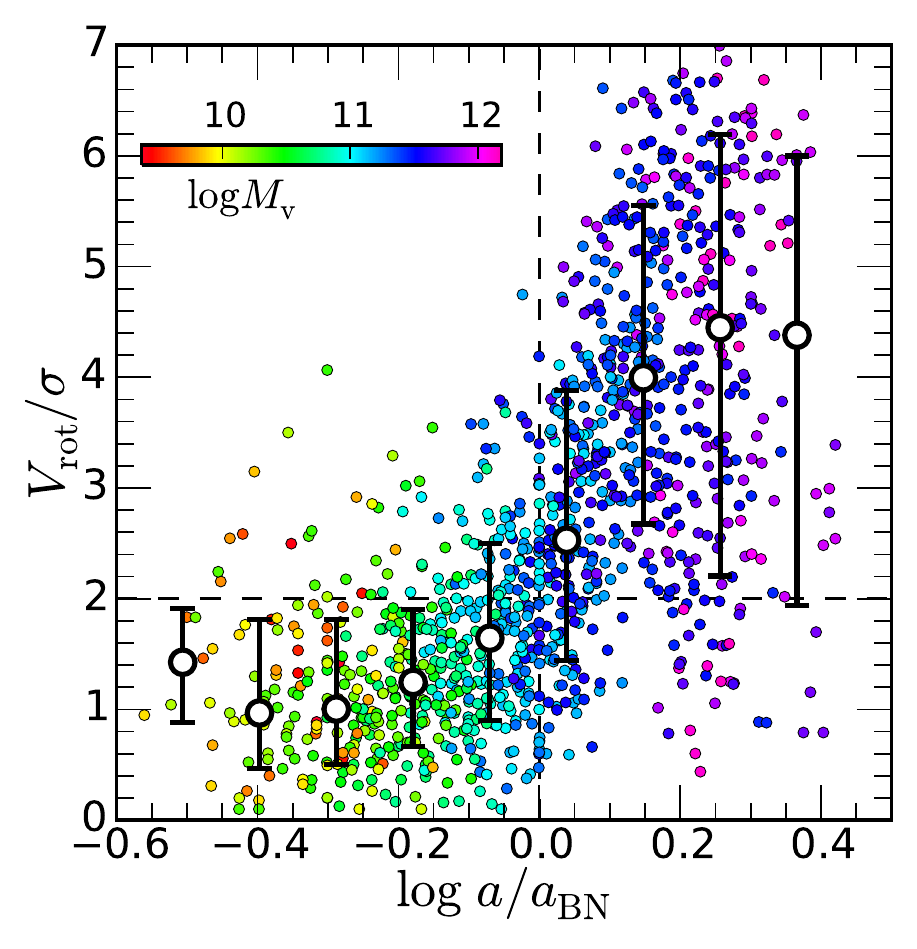}}
\caption{Kinematic evolution through compaction into rotation-supported
discs/rings.
Shown is the evolution of rotation velocity $\Vrot$ and velocity dispersion
$\sigma_r$ in VELA galaxies.
{\bf Left:}
The  median velocities (and $1\sigma$ scatter), with respect to the
virial velocity $\Vv$,
are shown as a function of time (expansion factor $a\!=\!(1+z)^{-1}$)
with respect to the blue-nugget event (left) and, quite similarly, as a
function of stellar mass $\Ms$ (right).
We see that the rotation velocity is increasing drastically during the 
compaction and blue-nugget phase, while the velocity dispersion remains 
roughly constant across this event. This argues that angular momentum is a key
to understanding the emergence of post-BN discs, and hints to extended rings 
as they naturally posses high AM.
{\bf Right:} 
The disciness measure $\Vrot/\sigma$, for all VELA snapshots, as a function of
time with respect to the BN event. The symbol color marks $\Mv$.
Symbols with error-bars are medians in bins of $a$.
We see a relatively tight correlation between the transition from non-discs 
to discs and the compaction-driven blue-nugget event where the galaxy mass is 
near a threshold of $\sim\!2\times 10^{11}\msun$, hinting that the formation of
a massive compact bulge is an important driver of disc or ring formation and
longevity.
}
\label{fig:vela_V}
\end{figure*}

\subsection{Post-compaction discs \& rings in simulations}

\subsubsection{Discs and rings about a massive bulge}

\Fig{mosaic_v07} displays the evolution of one VELA galaxy 
through the compaction and blue-nugget events and the post-compaction phases.
It shows face-on images of projected gas 
density that can serve as a proxy for the associated SFR following
the Kennicutt-Schmidt relation.
The compaction phase (top-left panel) leads to a blue nugget (top-middle)
that is characterized by the central blob of high gas density.
Immediately after (top-right), a highly turbulent rotating 
disc develops and grows in extent (bottom-left).
It shows a pronounced spiral-arm 
pattern and irregular perturbations including giant clumps.
This is a VDI phase, in which the giant clumps and
the gas between them migrate inwards 
\citep{noguchi98,bournaud07c,dsc09,cdb10,ceverino12,bournaud11,ceverino12}.
Then, the central gas is depleted into star formation and outflows 
\citep[in comparable roles,][]{zolotov15}, and an extended clumpy ring forms,
continuously fed by incoming cold streams, showing tightly wound spiral arms
and giant clumps (bottom-middle). The ring is maintained at its
extended form for several Gigayears, with no significant inward migration
(bottom-right).

\smallskip
The complementary stellar-density maps in the lower six panels
show how a compact stellar system forms following the
gas compaction process into the blue-nugget phase (top-middle), 
and how it remains massive 
and compact as it quenches to a passive red nugget. The compaction process thus
results in a massive central bulge, which soon becomes surrounded by a   
gaseous disc that develops into an extended ring. We will argue below that this
massive bulge is a key for ring longevity.

\smallskip
\Fig{rings_gas} indicates that the ring phenomenon is robust.
It shows examples of images of face-on projected gas density
in several simulated post-compaction galaxies with extended gas rings, three
at $z \sgt 1$ and one at $z \ssim 4$.
These examples will serve as our fiducial pronounced rings in the
simulations.
These cases illustrate the robustness of rings about massive bulges 
in post-compaction galaxies, above the critical
mass for major compaction events.

\smallskip
In order to explore the buildup of the rings through the different phases,
\fig{profiles_rings} shows the evolution of surface-density radial 
profiles in the disc plane for (mostly cold) gas as well as SFR, stars and 
metallicity, 
Three VELA galaxies that develop pronounced rings are shown, 
each at four phases of evolution corresponding to pre-compaction, compaction, 
early post-compaction and late post-compaction, as seen in \fig{mosaic_v07}, 
with the corresponding redshifts marked.
Inspecting the evolution of the gas profiles,
one can see the growth of gas density inside $r\!\sim\!1\kpc$ during the
compaction phase, followed by gas depletion in the inner few kpc and the
development of a long-term ring at $r\!\sim\!10\kpc$ post compaction.
The SFR density profiles roughly follow the gas density profiles, 
obeying the KS relation, showing the ring as well. 

\smallskip
The stellar profiles show the associated post-compaction growth of the stellar
mass within the sphere encompassed by the ring, which can be translated to a
decrease in the quantity $\delta_{\rm d}$ of \equ{deltad} 
that determines the inward mass transport rate via \equ{tinf} below.
The ring itself is hardly detectable in most stellar mass profile.
The metallicity is decreasing with radius, being $\sim\!0.4$dex lower in the
ring compared to the inner disc. This indicates that the ring is being built
by freshly accreted gas.

\smallskip
Based on the robust occurance of clumpy post-compaction discs and rings in the
simulations, we propose that they can be identified with 
a large fraction of the observed 
massive and extended star-forming rotating and highly turbulent ``discs" 
showing giant clumps \citep{genzel08,genzel14_rings,guo15,forster18a,guo18}.
The long-term survival of these gravitationally unstable rings, in the
simulations and in the observed galaxies, seems to be in apparent
conflict with the expected rapid inward migration of VDI discs
discussed in \se{vdi} based on \citet{dsc09}, and it thus poses a 
theoretical challenge which we address in \se{ring_toy} below.  

\subsubsection{Compaction-driven transition to rotating discs/rings}
\label{sec:V_sigma}


The kinematic transition through the blue-nugget event is of particular
relevance to our current study of extended discs and rings.
In order to see this,
\fig{vela_V} shows the transition of kinematic properties through the
blue-nugget event and, almost equivalently, through the crossing of the
threshold mass, in the VELA simulations
\citep[see similarly the evolution of spin in][Fig. G1]{jiang19_spin}.
The galaxies evolve from pressure to rotation support, with the median
$\Vrot/\sigma$ growing from near unity to about $4\!\pm\!1$.
The important fact to note is that it is the rotation velocity that is
dramatically growing during the
compaction process, from a small $\Vrot/\Vv\!=\!0.4\!\pm\!0.1$ pre compaction
to $\Vrot/\Vv\!= 1.4\!\pm\!0.1$ post compaction.
During the same period, over which the normalizing factor $\Vv$ does not vary
significantly, the velocity dispersion remains roughly constant at about
$\sigma/\Vv\!=\!0.4\!\pm\!0.1$.

\smallskip
We learn that the transition from pressure to rotation support is not because
of a significant change in the stirring of turbulence, but rather due to
an abrupt increase in the gas AM. 
This indicates that the inflowing high-AM gas is prevented from forming a
long-lived disc in the pre-compaction phase because of an efficient loss
of AM, largely due to the merger-driven flips discussed in
\se{flip} based on \citet{dekel20_flip}.
In turn, the gas seems to retain its incoming high AM in the
post-compaction phase.
This should guide our effort in \se{ring_toy} below to understand the
post-compaction survival of extended rings by means of AM exchange.

\begin{figure*} 
\includegraphics[width=0.91\textwidth]
{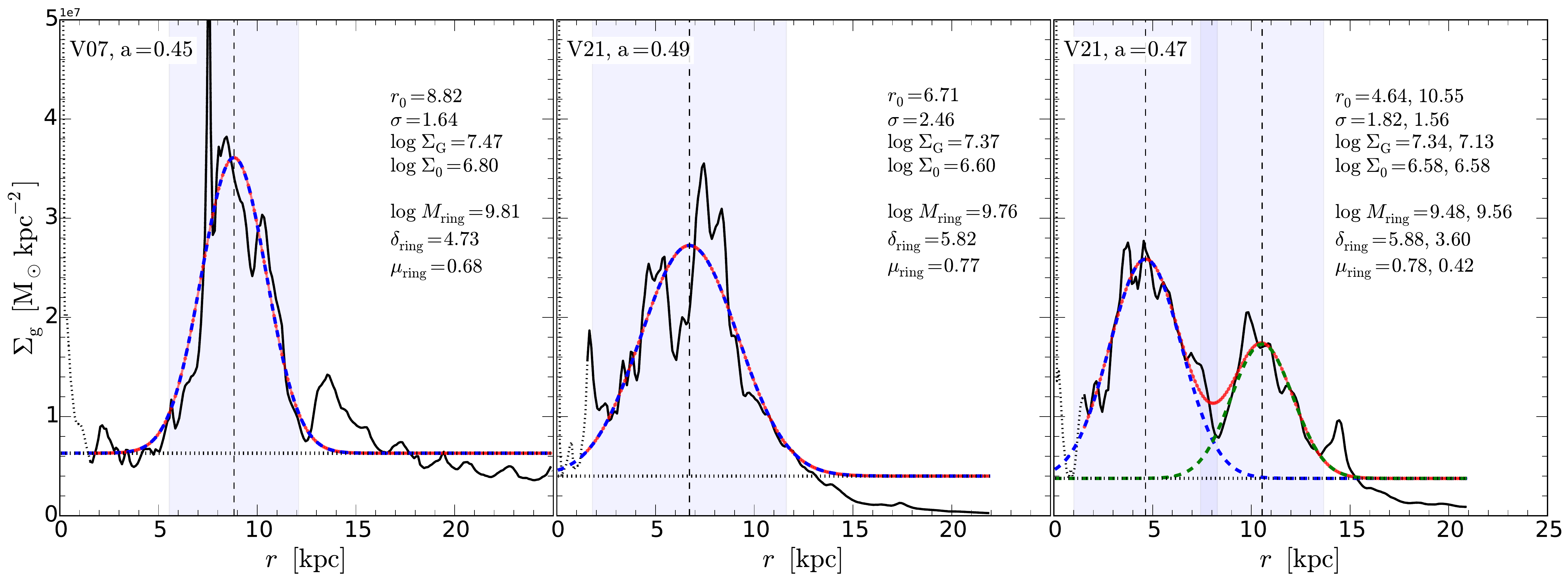}
\caption{
Measuring rings.
Three examples of gas surface brightness profiles and the best-fit Gaussian 
fitting functions, with either one or two Gaussians above a constant 
background, \equ{Gaussian}.
The four best-fit parameters are quoted for each ring.
The ring is characterized by its contrast
$\delta_{\rm ring}\!=\!\Sigma_{\rm G}/\Sigma_0$
and by the gas mass excess in the ring
$\mu_{\rm ring}\!=\!M_{\rm ring}/M_{\rm gas,tot}$, where
the ring mass is measured within $r_0\pm2\sigma$ and the total gas is inside
$r_0\!+\!2\sigma$.
In the case of two Gaussians, if one ring is three times as massive as the
other the massive one is chosen ($\sim\!10\%$ of the ringy galaxies). 
Otherwise, the two rings are combined ($\sim\!10\%$ of the ringy galaxies).
}
\label{fig:fits}
\end{figure*}

Also shown in \fig{vela_V} is the measure of disciness, $\Vrot/\sigma$,
 for all VELA snapshots as a function of the time with respect to the major 
BN event, as well as the virial mass (color).
The visual impression is that the correlation with the blue-nugget event is
rather tight, showing a clear transition from non-discs to discs near the
BN event, where the halo mass is near a threshold
of $\sim\!2\times 10^{11}\msun$.
In particular, there is only a small number of cases (except very massive
ones) where the galaxies are non-discs significantly after the blue-nugget
phase.
This is yet another possible hint that the formation of a massive central 
bulge, in the mass range where mergers are infrequent, is in most cases 
sufficient for disc or ring longevity.

\smallskip
We thus have a hint that the simulations show a correlation between the
compaction to a blue nugget and the development of an extended gas disc
or ring. 
In \se{ring_prop}, we quantify the ring properties and establish this 
correlation explicitly for the rings,
and in \se{ring_toy} we attempt to understand the origin of this correlation.

\subsection{Ring properties and correlation with nuggets}
\label{sec:ring_prop}

\subsubsection{Ring detection and properties}
\label{ring_detect}

For crude estimates of expected ring properties,
we read from the surface-density profiles of the VELA galaxies 
with pronounced rings (\fig{rings_gas} and \fig{profiles_rings})
that the galaxies of $\Ms\!\sim\!5\times 10^{10}\msun$ have 
mean gas surface densities in the ring of 
$\Sigma_{\rm gas}\! \sim 5\!\times\! 10^7\! \msun \kpc^{-2}$, 
while much of the SFR occurs in clumps of 
$\Sigma_{\rm gas}\! \gsim\! 5\!\times\! 10^8\msun \kpc^{-2}$.
In a ring of radius $r\seq 10\kpc\, r_{10}$ and width 
$\Delta r/r= 0.33\,\eta_{0.33}$,
the total gas mass in the ring is 
$M_{\rm gas} \!\sim\! 10^{10}\msun\, \eta_{0.33}\, r_{10}^2$.
The average gas number density in the ring is $n\!\sim\! 1 \cmc$.
The average SFR density in these pronounced rings is 
$\Sigma_{\rm sfr}\!\sim\!\Sigma_{\rm gas}/\tsfr$, 
which for $\tsfr\!\sim\!0.5\Gyr$ gives 
$\Sigma_{\rm sfr} \!\sim\! 0.1\msun\yr^{-1}\kpc^{-2}$.
The total SFR in the ring
is therefore SFR$\sim\!20\msun\yr^{-1} \eta_{0.33} r_{10}^2$.

\smallskip 
In order to more quantitatively measure ring properties for all the simulated 
galaxies, 
we compute for each the radial gas surface-density profile $\Sigma(r)$ 
projected onto the disc plane defined by the instantaneous angular momentum, 
and fit to it a function that captures the main
ring with a Gaussian shape in linear $\Sigma$ versus $r$, 
as described in more detail in appendix \se{app_ring_detection}.
Three examples are shown in \fig{fits} to illustrate the profiles and fits. 
In the case of a single dominant ring (left and middle panels), 
which is $\sim\!80\%$ of the cases where the galaxies have rings,
we fit a Gaussian on top of a constant background,
\be
\Sigma(r)=\Sigma_0
+\Sigma_{\rm G}\exp\left[-\frac{(r-r_0)^2}{2\sigma^2}\right]\, ,
\label{eq:Gaussian}
\ee
with four free parameters.

\smallskip
In the case where a smoothed version of the profile indicates two
well-separated rings, which occurs in $\sim\!20\%$ of the ringy galaxies,
we fit a sum of two Gaussians with the same $\Sigma_0$
(right panel). 
If one of the rings is at least three times as massive as the other, 
which happens in about half the galaxies with double rings,
we choose it as the dominant ring.
Otherwise, in $\sim\!10\%$ of the cases,
we combine the two rings into one, assigning to it the average
contrast and a combined radius, width and mass as specified in appendix 
\se{app_ring_detection}.

\smallskip
The ring is characterized by its contrast with respect to the background in its
interior,
\be
\delta_{\rm ring} = \frac{\Sigma_{\rm g}}{\Sigma_0} \, ,
\label{eq:delta_ring}
\ee
ranging from $0$ for no ring to $\delta_{\rm ring} \rightarrow \infty$ 
for an ultimate ring with an empty interior.
 
\smallskip
An alternative measure of ring strength is the mass excess, 
the ratio between gas mass in the ring and the total in the disc 
including the background,
\be
\mu_{\rm ring} = \frac{M_{\rm ring}}{M_{\rm gas,tot}} \, .
\ee
The ring mass is taken to be the mass above $\Sigma_0$ in the range 
$r_0\pm 2\sigma$, and the total mass is measured from $r\!=\!0$ to 
$r_0\!+\!2\sigma$.
The mass fraction $\mu_{\rm ring}$ thus ranges from zero for no ring 
to unity for a pure ring with otherwise no disc component.

\begin{figure} 
\includegraphics[width=0.45\textwidth]
{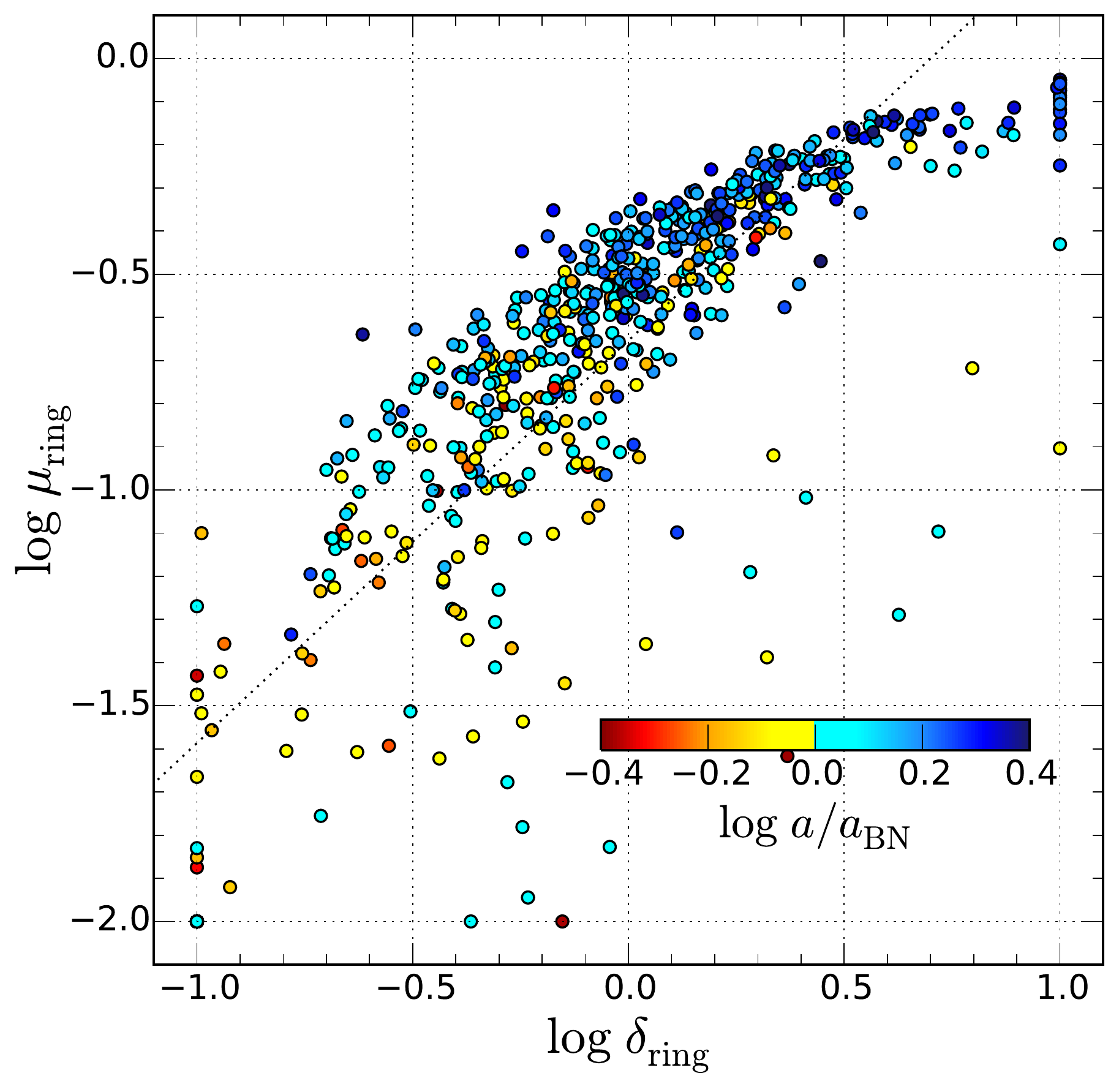}
\caption{
Ring strength.
The two measures of ring strength are compared, the mass excess 
$\mu_{\rm ring}$ and the density contrast $\delta_{\rm ring}$.
For the significant rings the two measures are tightly correlated, roughly
about the line $\log \mu_{\rm ring} = \log \delta_{\rm ring} - 0.5$.
We find $\mu_{\rm ring}$ to be more useful because
a threshold in $\mu_{\rm ring}$ implies a threshold in $\delta_{\rm ring}$,
but not vice versa.
}
\label{fig:mu_delta}
\end{figure}

\smallskip
In order to evaluate the quality of the two measures of ring strength,
\fig{mu_delta} plots the two against each other. 
For the more significant rings the two measures are tightly correlated, spread
about the line $\log \mu_{\rm ring} = \log \delta_{\rm ring} - 0.5$, except for
the very strong rings where $\delta_{\rm ring}$ becomes $\gg\!1$.
The scatter is larger for the more minor rings.
We note that a threshold in $\mu_{\rm ring}$ implies a meaningful threshold in 
$\delta_{\rm ring}$. For example 
$\log \mu_{\rm ring} \sgt -1.0, -0.7, -0.5, -0.3$ automatically 
implies $\log \delta_{\rm ring} \sgt -0.7, -0.5, -0.3, 0.1$ respectively.
On the other hand, a thresholds in $\delta_{\rm ring}$ allows a large range of 
low $\mu_{\rm ring}$ values.
We conclude that $\mu_{\rm ring}$ is a more robust measure of ring strength,
and adopt it in our analysis below.

\smallskip
\Fig{app_prop_dist} in appendix \se{app_ring_prop}
shows the distributions of certain ring properties among the simulated galaxies
with significant rings, $\mu_{\rm ring} \sgt 0.3$.
Other relevant properties will be discussed below in different specific 
contexts.

\begin{figure} 
\includegraphics[width=0.46\textwidth]
{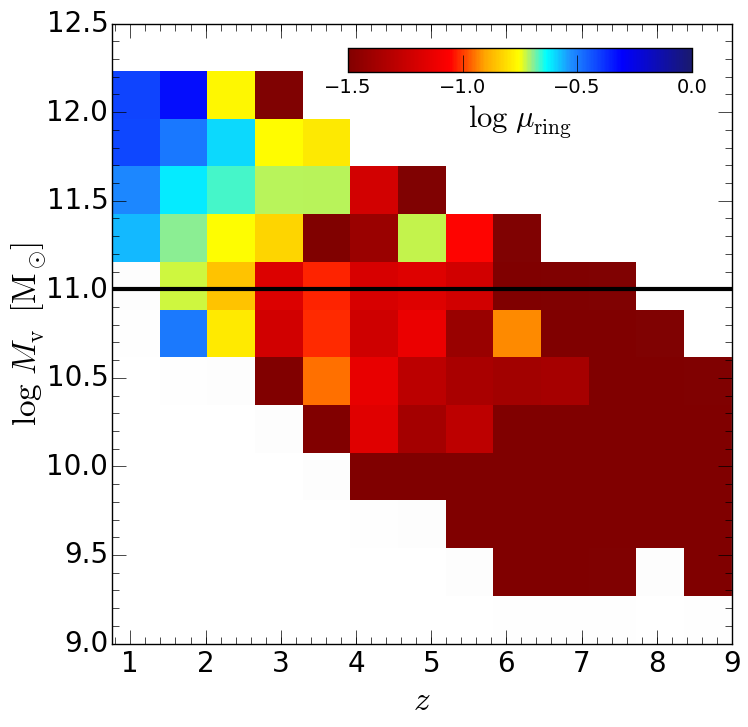}
\caption{
Ring mass excess as a function of mass and redshift.
Shown is the distribution of $\mu_{\rm ring}$, 
averaged within bins in the $\Mv\!-\!z$ plane.
The corresponding distribution of the fraction of galaxies with significant
rings is shown in \fig{delta_d_Mz_bins}.
We recall from \fig{disc_Mz_bins} that long-lived discs or rings are expected 
to dominate above a threshold mass of $\Mv \ssim 2\times 10^{11}\msun$ at all
redshifts.
We see that significant rings dominate above the threshold mass at $z\slt 4$,
while there are weaker or no rings at much lower masses and higher redshifts.
The weakness or absence of rings in intermediate-mass galaxies at very 
high redshifts indicates that a process other than spin flips also has a 
role in ring survival.
}
\label{fig:mu_Mz_bins}
\end{figure}

\begin{figure*} 
\includegraphics[width=0.463\textwidth]
{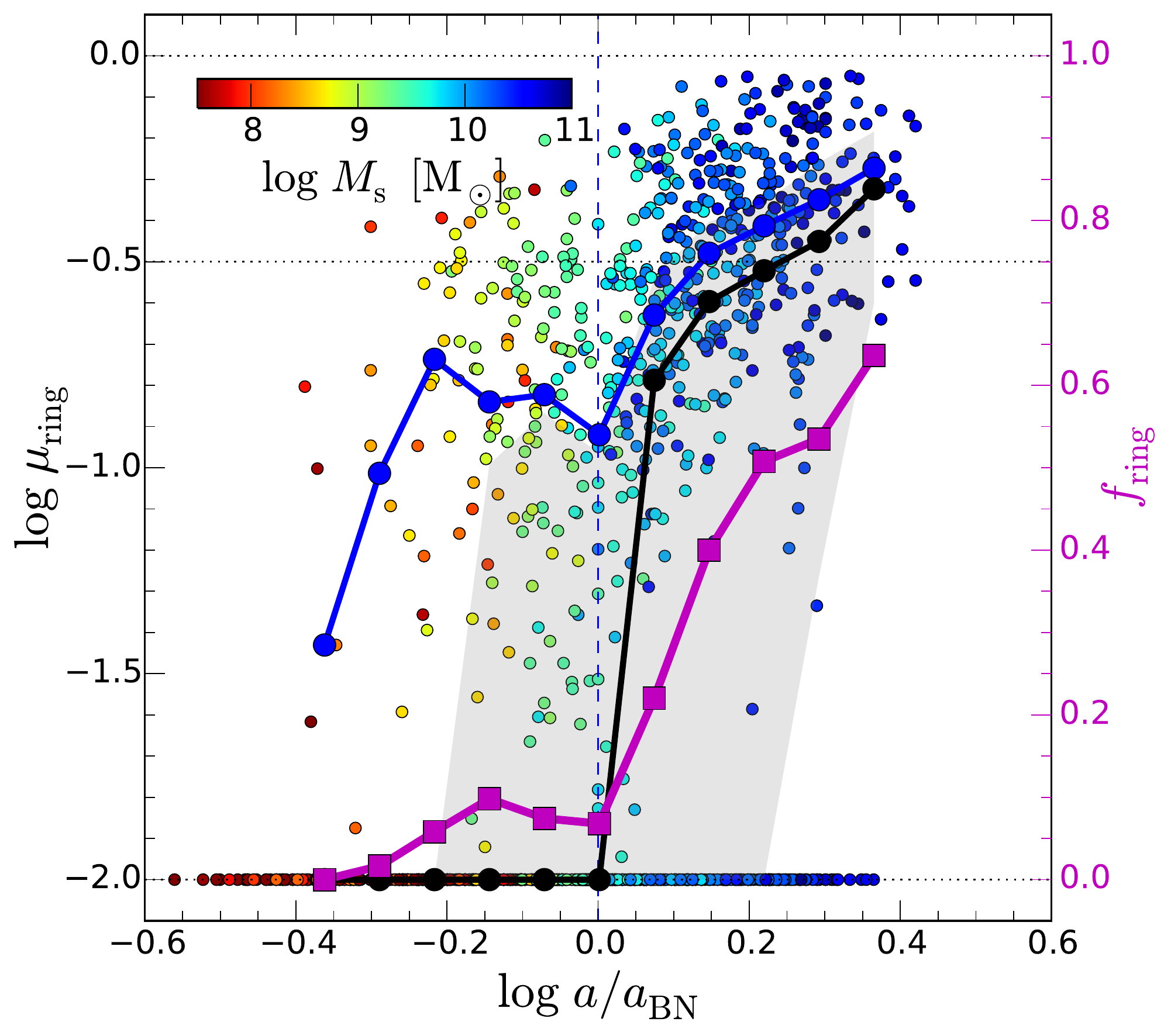}
\quad
\includegraphics[width=0.43\textwidth,trim={0 0.15cm 0 0}]
{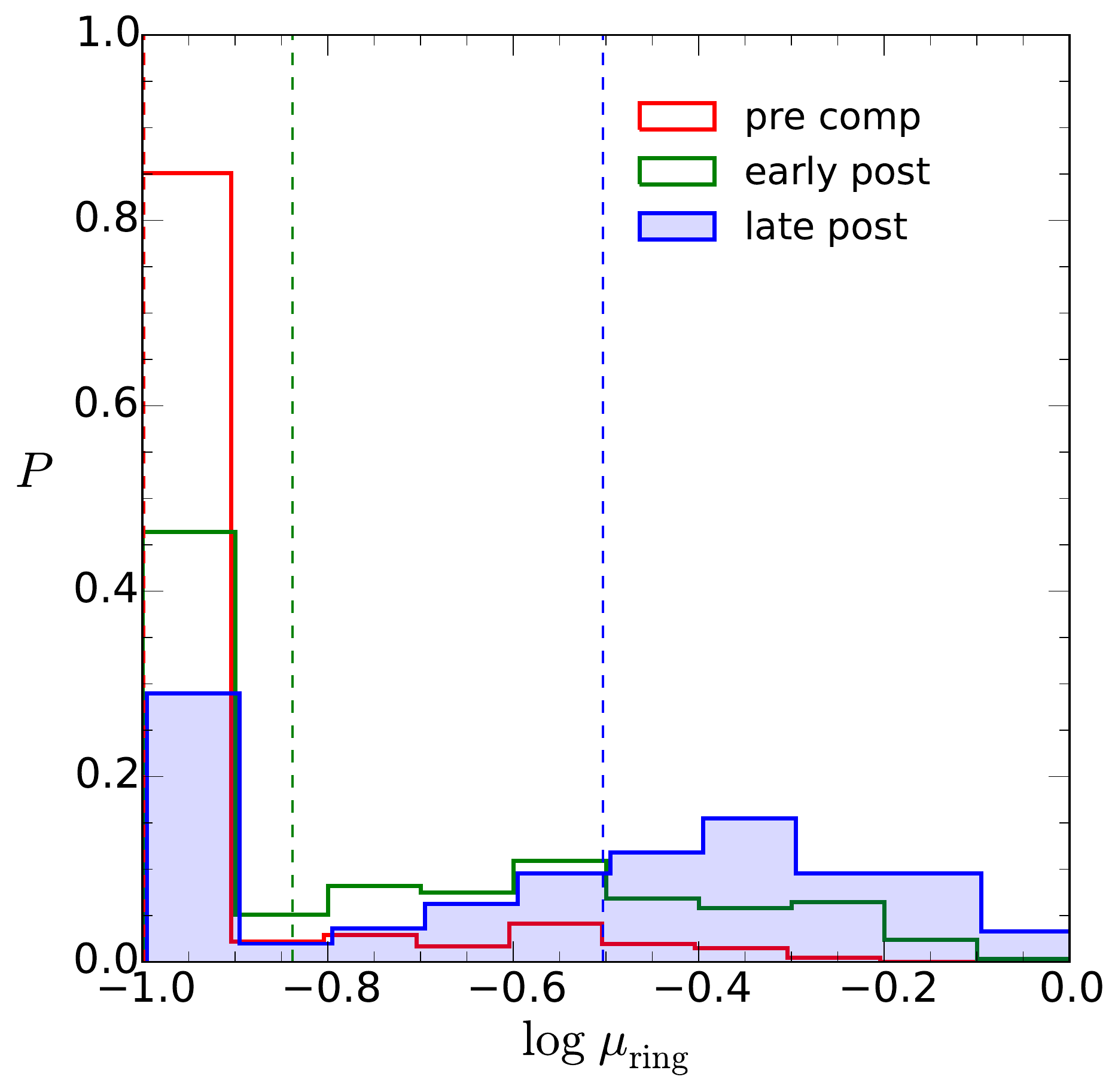}
\caption{
Ring strength and the blue-nugget event.
{\bf Left:}
Ring mass excess $\mu_{\rm ring}$ 
as a function of expansion factor $a=(1+z)^{-1}$
with respect to the blue-nugget time $a/a_{\rm BN}$ for all simulated galaxies.
Symbol color marks stellar mass, which is strongly correlated with
$a/a_{\rm BN}$.
Symbols where $\mu_{\rm ring}\!<\!0.01$ are put at 0.01.
The medians at given mass bins for all simulated galaxies are shown in black
with the 25\% and 75\% percentiles marked by the shaded area.
The medians limited to the galaxies that show rings are marked in blue.
The magenta curve and the right axes refer to the fraction of galaxies
with significant rings of $\mu_{\rm ring}\!>\!0.3$. 
We see a pronounced transition near the major compaction events to blue nuggets
at $\Ms\!\sim\!10^{10}\msun$, from a tendency to have no rings to
a dominance of high-contrast, massive rings.
The ring excess is gradually growing with galaxy mass and time with
respect to the blue-nugget phase.
{\bf Right:}
The distribution of gas ring mass excess $\mu_{\rm ring}$
in three different stages of evolution with respect to the blue-nugget event,
roughly corresponding to different mass bins with respect to
$\Ms\!\sim\!10^{10}\msun$.
Pre BN (black), early post BN (blue) and late post BN (red) refer to
$\log(a/a_{\rm BN})$ prior to $-0.05$,
between $0.00$ and $0.15$, and after $0.15$, respectively.
The medians are marked by vertical dashed lines.
Pre BN only a small fraction of the galaxies show rings, all weak,
while late post BN most of the galaxies show rings, mostly significant ones.
}
\label{fig:mu_bn}
\end{figure*}

\subsubsection{Compaction-driven rings}

The causal connection between the compaction to a blue nugget and the
presence of a ring can now be quantified using the ring properties measured 
from the simulations. 

\smallskip
Recalling that the major blue-nuggets events tend to occur at a characteristic
mass in a broad redshift range, we start by exploring the typical ring strength
as a function of mass and redshift.
For this, \fig{mu_Mz_bins} displays the ring mass excess, averaged within 
square bins in the $\Mv\sdash z$ plane.  
Shown below in \fig{delta_d_Mz_bins} is the associated distribution of 
$f_{\rm ring}$, defined as the fraction of galaxies with significant rings 
obeying $\mu_{\rm ring}\!>\!0.3$.
We recall from \fig{disc_Mz_bins} \citep[based on][]{dekel20_flip}
that $\Mv\ssim 2\times 10^{11}\msun$ is the threshold for  
long-lived discs, as discs in lower mass haloes are disrupted by merger-driven
spin flips on timescales shorter than the orbital times, as well as by
supernova feedback.
We learn that a significant fraction of the galaxies of halo masses 
$\Mv \sgt 10^{11.3}\msun$ at $z\seq 1\sdash 4$, where discs dominate,
actually have significant rings.
In this redshift range, weaker rings show up also in a fraction of the haloes 
somewhat below the threshold mass, where they are expected to be less 
long-lived.
At lower masses, $\Mv \slt 10^{10}\msun$, or at higher redshifts, $z\sgt 6$, 
there are almost no rings.
This is not surprising given that there are no long-lived discs there, as seen
in \fig{disc_Mz_bins}. 

\smallskip
An interesting feature is the absence of rings at high redshifts near
and even slightly above the threshold mass for long-lived discs, which does not
follow the insensitivity to redshift seen for discs in general
in \fig{disc_Mz_bins}.
This indicates that the formation and survival of rings is not solely 
determined by the infrequent merger-driven spin flips that provides a
necessary condition for long-term disc survival.
The galaxies develop long-lived outer rings only after they undergo certain 
stages of evolution, possibly after their major compaction events,
and, as we shall see in \se{ring_toy}, they should also obey another condition
associated with the quantity $\delta_{\rm d}$ of \equ{deltad} that 
provides a sufficient condition for long-term rings.

\smallskip
In order to better establish the causal connection between a compaction to a 
BN and the development of a ring, the left panel of \fig{mu_bn} shows the ring 
mass excess $\mu_{\rm ring}$ 
as a function of the cosmological expansion factor $a=(1+z)^{-1}$ 
with respect to the blue-nugget event at $a_{\rm BN}$, 
for all snapshots of all galaxies. 
Almost equivalently, the symbol color shows these quantities as a function of
stellar mass $\Ms$.
The distributions of these ring properties at given mass bins are indicated by
the medians and the 25\% and 75\% percentiles for all galaxies
(black, shade). The medians are also shown for 
the sub-sample of snapshots that show rings with non-vanishing
$\mu_{\rm ring}$ values (blue).         
Also shown (magenta) is the fraction of galaxies that have significant rings,
with $\mu_{\rm ring}\!>\!0.3$.

\smallskip
When inspecting the whole sample of simulated galaxy, 
we see a pronounced transition near the major compaction events to blue nuggets
at a critical mass $\Ms\!\lsim\!10^{10}\msun$. 
Prior to the BN, the vast majority of the galaxies show no rings, 
while sufficiently after the BN, most of the galaxies show rings. 
Focusing on the galaxies with rings,
we see gradual strengthening with mass, where the median mass excess  
is growing from $\mu_{\rm ring}\!\sim\!0.1$ to $0.5$.
Well above the critical mass, post blue nugget, the 
most pronounced rings approach $\mu_{\rm ring}\ssim 1$, namely pure rings with
no gas in the interior disc.
The fraction of galaxies with significant rings is rising from $\sim\!10\%$
pre BN and below the BN mass to $\sim\!65\%$ well after the BN time and
above the BN mass.
The exact quoted fractions are to be taken with a grain of salt as the VELA 
suit is not a statistically representative sample of galaxies in terms of mass 
function. 

\smallskip
To further establish the correlation between rings and the post-compaction
phase, 
the right panel of \fig{mu_bn} displays the probability distributions of ring 
mass excess in three different phases of evolution with respect to the BN
event. The pre-BN phase is defined here by $\log (a/a_{\rm BN}) \slt -0.05$,
to avoid the late compaction phase near the BN.
The early-post-BN phase is limited to $0.0\slt \log(a/a_{\rm BN})\slt 0.15$
and the late-post-BN phase is defined as $ \log(a/a_{\rm BN}) \sgt 0.15$.
These phases roughly correspond to different mass bins with respect to the
characteristic BN mass of $\Ms \ssim 10^{10}\msun$.
We read from the histograms that
pre BN less than $\sim\!20\%$ of the galaxies show rings, all weak
with a median $\mu_{\rm ring} \ssim 0.2$ for the rings.
In contrast, late post BN $\sim\!70\%$ of the galaxies show rings, mostly
significant with a median above $\mu_{\rm ring} \ssim 0.3$.
We read that $\sim\! 24\%$ of the galaxies are expected to have pronounced 
rings of $\mu_{\rm ring} \sgt 0.5$, and $\sim\! 50\%$ are significant rings
with $\mu_{\rm ring} \sgt 0.3$, while $\sim\! 30\%$ have no sign of a ring.

\smallskip
\Fig{f_Mz_bins} shows the distribution of ring fraction in the $\Mv\sdash z$
plane for significant rings of $\mu_{\rm ring} \sgt 0.3$.
\Fig{app_f_Mz_bins} in appendix \se{app_ring_prop} shows similar maps
for all rings of $\mu_{\rm ring} \sgt 0.01$ and for the pronounced rings of
$\mu_{\rm ring} \sgt 0.5$.
\Fig{app_f_Msz_bins} in appendix \se{app_ring_prop} shows the same map for 
significant rings $\mu_{\rm ring}\sgt 0.3$ but with $\Mv$ replaced by 
stellar mass $\Ms$, to make the comparison with observations more
straightforward. 
This complements the map of ring strength shown in \fig{mu_Mz_bins}.
We see that a high fraction of rings is detected above the threshold mass,
$\Mv \sgt 10^{11}\msun$, where discs survive spin flips (\fig{disc_Mz_bins}),
and at $z\slt 4$.
Focusing on massive galaxies, we read that the fraction of strong rings,
$\mu_{\rm ring} \sgt 0.5$, is $\sim\! 30\%$ at $z\ssim 1$, 
while it drops to the $\sim\! 10\%$ level at $z\seq 1.5\sdash 3.5$.
The fraction of significant rings is $\sim\! 50\%$ at $z\ssim 1$ 
and $30\sdash 40\%$ at $z\seq 2\sdash 4$.
The fraction of all rings, including marginal minor ones, 
is $\sim\! 70\%$ at $z\seq 1\sdash 3$, 
$\sim\! 50\sdash 60\%$ at $z\seq 3\sdash 4$
and $\sim\! 20\sdash 40\%$ at $z\seq 4\sdash 6$.
In comparison, according to \fig{mu_Mz_bins}, 
the average ring strength in massive galaxies is just
below $\mu_{\rm ring} \seq 0.5$ at $z \ssim 1$, and is 
$\mu_{\rm ring} \seq 0.3$ at $z \seq 1.5\sdash 4$.
These fractions and ring strengths are to be compared to observations,
where preliminary indications for qualitative agreement are discussed in 
\se{candels}.

\smallskip
The correlation of ring strength and time or mass with respect to the
blue-nugget phase, as seen in \figs{mu_Mz_bins}, 
\ref{fig:mu_bn}, and \ref{fig:f_Mz_bins},
strengthens our earlier impression that ring formation is correlated 
with the post-compaction phase of evolution, which was based on visual 
inspection of rings in gas images and profiles in the different evolution 
phases 
(\figs{mosaic_v07}, \ref{fig:rings_gas} and \ref{fig:profiles_rings}),
combined with the post-compaction appearance of discs (or rings) above the 
corresponding mass thresholds (\figs{disc_Mz_bins} and \ref{fig:vela_V}).
We next attempt to understand the origin of the post-compaction longevity 
of the rings.

\section{Ring Stabilization by a Central Mass: an analytic model}
\label{sec:ring_toy}


Having established in the simulations the ring formation and survival after 
compaction to a massive central body, and given the conflicting 
expectation for a rapid inward transport in a gaseous VDI disc summarized in 
\se{vdi},
we now proceed to an analysis that reveals the conditions for slow inward mass 
transport of an extended ring (\se{spiral}). 
\adb{Then, combined with the conditions for faster ring buildup by high-AM 
accretion and faster interior depletion by star formation, we obtain the
condition for ring formation and longevity (\se{survival}).
The longer lifetime of high-contrast rings is discussed (\se{strong}),
as well as the ring instability to clump formation (\se{toomre}).
Finally, we test certain aspects of the model using the simulations 
(\se{model_sims}).
}

\subsection{Transport in by torques from spiral structure}
\label{sec:spiral}

\subsubsection{The torque}

In order to estimate the rate of inward transport of ring material
[through AM conservation, or the angular Euler equation
\citep[eq. 6.34 of][hereafter BT]{bt08}],
we wish to compute the relative AM change in the outer disc outside radius $r$
due to torques exerted by the perturbed disc inside $r$, during one disc
orbital time.
For an order-of-magnitude estimate, we follow Chapter 6 of BT
assuming a tightly wound spiral-arm pattern in a razor-thin disc.
The $z$-component of the torque per unit mass at a position $(r,\phi)$ in the
disc plane is
\be
\tau(r,\phi) = -\frac{\pa \Phi}{\pa \phi} \, ,
\ee
where $\Phi$ is the gravitational potential, and $r$ and $\phi$ are the usual
spherical coordinates.
The relevant part of the potential exerting the torque is due to the disc, 
$\Phi_{\rm d}$, which is related to the density in the disc $\rho_{\rm d}$ via 
the Poisson equation.
The total torque on the ring outside $r$ is obtained by an integral over the
volume outside $r$,
\be
T(r)= \int_r^\infty dV\, \rho_{\rm d}\, \tau \, .
\ee
This is a simplified version of the more explicit expression in eq. 6.14 of BT.
After some algebra, the general result is 
\be
T(r)= \frac{r}{4\pi G} \int_0^{2\pi} d\phi \frac{\pa \Phi_{\rm d}}{\pa r}
\frac{\pa \Phi_{\rm d}}{\pa \phi} _{\vert_{r}} \, .
\ee

\smallskip
Next, assuming a tightly wound spiral structure with $m$ arms, 
the small pitch angle is given by 
\be
\tan \alpha = \frac{m}{|k|r} \ll 1 \, ,
\label{eq:pitch}
\ee
with $k$ the wavenumber (positive or negative for trailing or leading arms
respectively).
One assumes a thin disc, in which the spiral structure is described by a 
surface density (eq. 6.19 of BT)
\be
\Sigma(r,\phi) = \Sigma_1(r)\, \cos[m\phi+f(r)] \, ,
\ee
where $\Sigma_1(r)$ is the spiral perturbation, assumed to vary slowly with
$r$, the shape function is $f(r)$ where $df/dr=k$, the spirals are tightly 
wound $|k|\, r \!\gg\! 1$, and $m \sgt 0$.
The corresponding potential is 
\be
\Phi_{\rm d} = \Phi_1(r) \, \cos[m\phi+f(r)],
\quad \Phi_1 = -\frac{2\pi G\Sigma_1}{|k|} \, .
\ee
The torque is then (eq. 6.21 of BT)
\be
T(r) = {\rm sgn}(k) \frac{m\, r\, \Phi_1^2}{4G}  
= {\rm sgn}(k) \frac{\pi^2\, m\, r\, G\, \Sigma_1^2}{k^2} \, .
\label{eq:T1}
\ee
Thus, trailing arms ($k>0$) exert a positive torque on the outer part of the 
disc and therefore transport AM outwards (we omit the sign of the torque 
hereafter).
Using \equ{pitch}, the torque becomes
\be
T(r) = \frac{\pi^2\, G\, r^3}{m}\, \Sigma_1^2\, \tan^2\alpha \, .
\label{eq:T2}
\ee

\smallskip 
We should comment that spiral arms are indeed usually trailing.
Across the trailing spiral, the gas rotates faster than the spiral pattern
(\S 6.1.3.d of BT), namely the ring in this region is well inside the 
co-rotation radius of the spiral.  
The pattern speed for the spiral pattern may be not unique,
at different radii or at different times,
because of the varying potential due to the growth of the bulge.  
Nevertheless, the torque from the trailing spirals is transferring AM
from inside to outside the co-rotation radius and consequently gas in the 
ring flows inwards. 
\adb{We discuss these issues further in the context of rings at
resonance radii in \se{resonances}.} 

\subsubsection{Angular-momentum transport and mass inflow rate}
\label{sec:transport}

In order to compute the relative change of AM in the ring during an orbital 
time, we write for a ring at radius $r$ the torque per unit mass as
$\tau(r) = T(r)/M_{\rm r}$, where $M_{\rm r}$ is the ring gas mass,
\be
M_{\rm r} = 2 \pi\, r^2\, \eta_{\rm r}\, \Sigma_{\rm r} \, ,
\label{eq:Md}
\ee
and where $\Sigma_{\rm r}$ is the average surface density in the ring
and $\eta_{\rm r} \seq \Delta r/r$ is the relative width of the ring. 
The specific AM in the ring is $j\seq \Omega\, r^2$,
and the orbital time is $\torb \seq 2\pi\, \Omega^{-1}$,
where $\Omega$ is the circular angular velocity at $r$.
Substituting in \equ{T2}, the relative change of AM during one orbit is
\be
\frac{\Delta j}{j}_{\vert{\rm orb}} = \frac{\tau \torb}{j}
= \frac{\pi^2 G}{m\, \Omega^2\, r\, \Delta_{\rm r}\, \eta_{\rm r}} 
A_m^2 \Sigma_{\rm d} \tan^2\alpha \, ,
\label{eq:dj_1}
\ee
where $\Delta_{\rm r}\seq \Sigma_{\rm r}/\Sigma_{\rm d}$ is the ring contrast 
with respect to the disc\footnote{This is related to $\delta_{\rm ring}$ of
\equ{delta_ring} and \fig{mu_delta} 
simply by $\Delta_{\rm r}=1+\delta_{\rm ring}$.}, 
which we first assume to be $\!\gsim\! 1$,
and where the amplitude of the spiral surface-density pattern is assumed to be
a fraction $A_m$ of the axisymmetric density, 
\be
\Sigma_1(r) = A_m \Sigma(r) \, .
\label{eq:Am}
\ee
For reference, this amplitude is known to be in the range $0.15\sdash 0.60$ 
in observed spiral galaxies.

\smallskip 
It turns out that the quantity that governs the relative
AM change and simplifies the above expressions for a ring is the same mass 
ratio of cold disc to total that governs the rapid inflow of a VDI disc,
\equ{deltad}, namely 
$\delta_{\rm d} \seq {M_{\rm d}}/{M_{\rm tot}}$,
where $M_{\rm d}$ is the cold mass in the disc. 
Using \equs{deltad} and (\ref{eq:Md}), 
and approximating the circular velocity by 
$(\Omega r)^2 \seq G M_{\rm tot}(r)/r$,
the surface density in the disc can be expressed in terms of 
$\delta_{\rm d}$,
\be
\Sigma_{\rm d} = \frac{M_{\rm d}}{\pi\,r^2}
= \frac{\Omega^2 r}{\pi\, G}\, \delta_{\rm d} \, .
\label{eq:Sigma}
\ee
Inserting this in \equ{dj_1} we obtain 
\be
\frac{\Delta j}{j}_{\vert{\rm orb}} 
= \frac{\pi}{m\, \Delta_{\rm r}\, \eta_{\rm r}} 
A_m^2 \delta_{\rm d} \tan^2\alpha\, .
\label{eq:dj_2}
\ee

\smallskip  
Next, 
the pitch angle can also be related to the key variable $\delta_{\rm d}$
by appealing to the local
axi-symmetric Toomre instability, which yields a critical wavenumber for the
fastest growing mode of instability (eq. 6.65 of BT),
\be
k_{\rm crit} = \frac{\kappa^2}{2\, \pi\, G\, \Sigma}
= \frac{\psi^2}{2\, r}\, \delta_{\rm d}^{-1} \, .
\label{eq:kcrit}
\ee
Here $\kappa$ is the epicycle frequency,
$\kappa^2 \seq r\, {\dd}\Omega^2/dr \!+\! 4\Omega^2$.
This gives $\kappa \seq \psi\, \Omega$, where
$\psi \seq 1 $ for Keplerian orbits about a point mass,
$\psi \seq \sqrt{2}$ for a flat rotation curve,
$\psi \seq \sqrt{3}$ for the circular velocity of a self-gravitating uniform 
disc,
and $\psi \seq 2$ for a solid-body rotation.
The second equality made use of \equ{Sigma}.
Adopting $|k| \seq k_{\rm crit}$ in \equ{pitch} for the pitch angle, we obtain
\be
\tan\alpha = \frac{2\,m}{\psi^2}\, \delta_{\rm d} \, .
\label{eq:pitch2}
\ee
Inserting this in \equ{dj_2} we get 
\be
\frac{\Delta j}{j}_{\vert{\rm orb}} =
\frac{4\,\pi\,m}{\psi^4\,\Delta_{\rm r}\,\eta_{\rm r}}\, 
A_m^2\, \delta_{\rm d}^3 \, .
\label{eq:dj_3}
\ee

\smallskip 
The rate at which AM is transported out is actually driven by the sum of the
gravitational torques computed above and the advective current of AM.
The advective transport rate
is given by the same expression as in \equ{T1} times a factor of order
unity or smaller, explicitly [$v_{\rm s}^2\, |k|/(\pi G \Sigma) - 1]$,
where $v_{\rm s}$ is the speed of sound (BT eq. 6.81, based on appendix J).
Thus, the advective transport is generally comparable in magnitude to or
smaller than the transport by gravitational torques. We therefore crudely
multiply the AM exchange rate of \equ{dj_3} by a factor of two.

\smallskip 
The inverse of $\Delta j/j$ in an orbital time is the desired timescale for 
the ring mass to be transported inwards with respect to the orbital time, 
\be
\tinf \sim \frac{\psi^4}{8\,\pi\,m} 
\frac{\Delta_{\rm r}\, \eta_{\rm r}}{A_m^2}\, 
\delta_{\rm d}^{-3} \, \torb \, .
\label{eq:tinf1}
\ee
With the fiducial values 
$\psi \seq \sqrt{2}$, $m\seq 2$, $\Delta_{\rm r}\seq 1$, $A_m\seq 0.5$, 
and $\eta_{\rm r} \seq 0.5$
we finally obtain
\be
\tinf \sim 5.89\, \delta_{{\rm d},0.3}^{-3}\, \torb \, .
\label{eq:tinf}
\ee
We learn that, for a fixed value of $\eta_{\rm r}$ that does not strongly
depend on $\delta_{\rm d}$,
the inward mass transport rate is very sensitive to the value of 
$\delta_{\rm d}$. A value near unity implies a rapid inflow, while for
$\delta_{\rm d} \!\ll\! 1$ the inflow rate is very slow.

\smallskip 
With $\delta_{\rm d} \!\lsim\! 1$, e.g., corresponding to a bulge-less very
gas-rich disc in radii where it dominates over the dark matter, \equ{tinf}
indicates a significant AM loss corresponding to inward mass transport
in a few orbital times, as estimated in \citet{dsc09} for a VDI disc,
\equ{vdi1} above.
In this case, the pitch angle is not necessarily very small, 
and the spiral pattern can cover a large fraction of the disc with a 
significant radial component.
In contrast, with $\delta_{\rm d} \!\ll\! 1$,
\equ{pitch2} indicates that the pitch angle is very small, 
making the above calculation valid and
practically confining the spiral arms to a long-lived ring,
with a negligible mass transport rate.

\smallskip 
A low value of $\delta_{\rm d}$
is expected when $M_{\rm tot}$ is dominated either by a central massive bulge
and/or by a large dark-matter mass within the ring radius.
The former is inevitable after a compaction event. 
The latter is likely when the dark-matter halo is cuspy and when
the incoming streams enter with a large impact parameter and therefore form 
an extended ring that encompasses a large dark-matter mass,
as may be expected at late times.
A small $\delta_{\rm d}$ is also expected when the gas fraction in the
accretion is low, as expected at late times.

\subsection{The conditions for long-lived rings}
\label{sec:survival}

The formation and fate of the disc and the ring is determined by the interplay 
between three timescales.
First,
an extended ring originates from high-AM spiraling-in cold streams  
(\se{ring_formation}), which accrete mass on a timescale $\tacc$.
Second,
the ring gas is transported inwards toward filling up the interior disc 
by torques from the perturbed disc on a timescale $\tinf$, 
computed in \se{spiral}. 
Third,
this disc gas is depleted into stars and outflows on a timescale 
$\tsfr$.
We consider the 
two conditions for ring formation and longevity to be
\be
\tacc < \tinf \ \ {\rm and} \ \ \tsfr < \tinf \, .
\label{eq:conditions}
\ee
When either of these conditions is violated, the ring does not survive.
When $\tacc \sgt \tinf$, the ring is evacuated into the disc
before it is replenished, and when $\tsfr \sgt \tinf$,
the disc remains gaseous, both leading to a gas disc. 
These conditions for a disc are
expected to be fulfilled for high values of $\delta_{\rm d}$.
On the other hand, a ring would form and survive if 
$\tacc \slt \tinf$, namely the ring is replenished before it is
transported in, and if $\tsfr \slt \tinf$, causing 
the interior disc to be depleted of gas. 
These conditions for a ring are expected to be valid
for low values of $\delta_{\rm d}$. 
We next quantify these conditions, assuming for simplicity an
EdS cosmology, approximately valid at $z \sgt 1$.
 
\smallskip
The orbital time of the extended ring is on average a few percent of the 
Hubble time $t_{\rm hub}$ for all galaxy masses 
\citep[e.g. following][]{dekel13},
\be
\torb \sim 0.088\, \lambda_{0.1}\, t_{\rm hub} 
            \sim 1.5 \Gyr\, (1+z)^{-3/2} \, . 
\label{eq:torb}
\ee
Here $\lambda=R_{\rm ring}/\Rv$ is the contraction factor from the virial
radius to the extended gas ring, here in units of $0.1$, which we adopt as our
fiducial value.

\smallskip  
The accretion timescale, the inverse of the specific accretion rate, is on
average \citep{dekel13}
\be
\tacc \sim 30\, \Gyr\, (1+z)^{-5/2} 
            \sim 20\, \torb\, (1+z)^{-1} \, ,
\label{eq:tacc}
\ee
with a negligible mass dependence across the range of massive galaxies.
The second equality is based on \equ{torb}.
Comparing \equ{tacc} and \equ{tinf} for the transport time, 
we obtain net {\it ring replenishment}, $\tacc \slt \tinf$, for
\be
\delta_{\rm d} < 0.20\, (1+z)^{1/3} \, .
\label{eq:rep}
\ee
The redshift dependence is weak, e.g., $\delta_{\rm c} \slt 0.29$ at $z\seq 2$. 

\smallskip 
For the depletion time,
it is common to assume that the gas turns into stars on a timescale
\be
\tsfr \sim \epsilon_{\rm ff}^{-1} t_{\rm ff}
            \sim 5\, \torb \, ,
\label{eq:tsfr}
\ee
where $\epsilon_{\rm ff}$ is the efficiency of SFR in a free-fall time
and $t_{\rm ff}$ is the free-fall time in the star-forming regions 
\citep[e.g.][]{kdm12}.
We adopted above the observed standard value of $\epsilon_{\rm ff} \sim 0.01$
\citep[e.g.][]{krum17},
and $t_{\rm ff} \ssim 0.3\, t_{\rm dyn} \ssim 0.05\, \torb$. 
The latter is assuming that stars form in clumps 
that are denser than the mean density of baryons in the ring by a factor of 
$\sim\!10$ \citep{ceverino12}.\footnote{A similar timescale for SFR is obtained 
in the VELA simulations.
Using \tab{sample}, the ratio of gas mass to SFR at $z\!=\!2$ is typically
$\tsfr\! \sim\! 1\Gyr$. 
With $\torb\! \sim\! 0.29\Gyr$ at
that redshift from \equ{torb}, the SFR timescale is consistent with
\equ{tsfr}.}
Comparing \equs{tsfr} and (\ref{eq:tinf}), we obtain net inner 
{\it disc depletion}, $\tsfr \slt \tinf$, for
\be
\delta_{\rm d} < 0.32 \, .
\label{eq:dep}
\ee 

\smallskip 
From \equs{rep} and (\ref{eq:dep}) we learn that the two conditions
are in the same ball park, though at $z \slt 3$ the ring
replenishment condition is somewhat more demanding, while at higher redshifts 
the disc depletion condition is a little more demanding. 
We conclude that ring formation and survival is crudely expected below  
$\delta_{\rm d} \ssim 0.2$, give or take a factor $\sim\!2$ due to
uncertainties in the values of the fiducial parameters.

\smallskip 
\adb{
We note that
since the ring is primarily a gaseous phenomenon, the ring would not form and
survive once the gas fraction in the disc and ring is too low. This implies a
lower bound for $\delta_{\rm d}$ in long-lived ring galaxies, 
when its low value is driven by a low gas fraction. 
This lower bound is estimated using the simulations to be 
$\delta_{\rm d} \ssim 0.03$ (\se{model_sims}).
}

\smallskip
As the ring develops, once $\delta_{\rm d}$ becomes smaller than the threshold, 
the ring becomes over-dense with respect to the disc,
$\Delta_{\rm r} \sgt 1$. Then in \equ{tinf} the inflow timescale becomes longer
accordingly so the ring can continue to grow in a runaway process.

\subsection{Long-lived high-contrast rings}
\label{sec:strong}

The estimate of $\tinf$ in \se{transport} was valid for the stage of
transition from disc to ring, namely where the disc is still gas rich 
with a spiral structure that exerts torques on the outer ring, namely when
the ring contrast $\Delta_{\rm r}$ is not much larger than unity.
Once the ring develops and is long lived, under $\delta_{\rm d} \!\ll\! 1$,
it can become dominant over the disc. To evaluate the inflow rate in this
situation, we now consider the limiting case of a pure ring,  
in which the spiral structure exerts torques on other parts of the ring. 
$M_{\rm d}$ of the previous analysis is now replaced by $M_{\rm r}$,
$\Sigma_{\rm d}$ in \equ{Sigma} is replaced by 
$\Sigma_{\rm r} \seq (2\eta_{\rm r})^{-1} \Sigma_{\rm d}$,
while $\Delta_{\rm r}$ is unity.
Now $\Delta j/j$ in \equ{dj_2} and  $\tan\alpha$ in \equ{pitch2}
are both divided by $2\eta_{\rm r}$, so $\Delta j/j$ in \equ{dj_3} 
is divided by $(2\eta_{\rm r})^3$.
For otherwise the fiducial values of the parameters,
and in particular for the value of $\eta_{\rm r}$ fixed at $0.5$,
the inflow timescale is the same as it was in \equ{tinf}.

\smallskip
However, in the case of a tightly wound strong ring, one may relate the
relative ring width to the pitch angle, assuming that the width is the
``wavelength" of the spiral arm, namely the radial distance between the parts
of the arm separated by $2\pi$,
\be
\eta_{\rm r} = \frac{\Delta r}{r} = \frac{2\pi/k}{r}
= \frac{2\pi}{m}\, \tan\alpha
= \frac{2\pi}{\psi^2} \frac{1}{\eta_{\rm r}}\, \delta_{\rm d} \, ,
\ee
where we used \equs{pitch} and (\ref{eq:pitch2}).
Solving for $\eta_{\rm r}$ we obtain
\be
\eta_{\rm r} = \frac{(2\pi)^{1/2}}{\psi}\, \delta_{\rm d}^{1/2} \, .
\label{eq:eta_deltad}
\ee
Then from \equ{dj_3} divided by $(2\eta_{\rm r})^3$ we obtain
\be
\frac{\Delta j}{j}_{\vert{\rm orb}} =
\frac{m}{8\pi}\,A_m^2\,\delta_{\rm d} \, .
\ee
After multiplying by two for the advective contribution,
the inflow timescale becomes
\be
\tinf = \frac{4\pi}{m\, A_m^2}\, \delta_{\rm d}^{-1}\, \torb
= 84\, \delta_{{\rm d},0.3}^{-1}\, \torb \, ,
\label{eq:tinf_b}
\ee
where the fiducial values of $m=2$ and $A_m=0.5$ were assumed.
This is well longer than the Hubble time and
the accretion and depletion timescales, implying that once a high-contrast
ring forms, it is expected to live for long.

\begin{figure} 
\centering
\includegraphics[width=0.46\textwidth]
{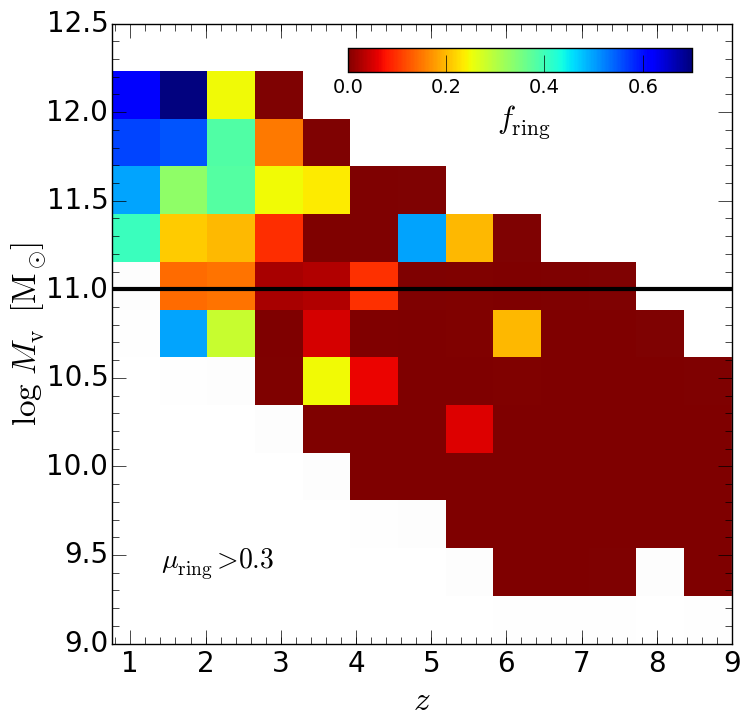}
\caption{
Fraction of significant rings.
Shown is the 2D distribution of the fraction of rings 
with $\mu_{\rm ring}\!>\!0.3$ in the $\Mv\sdash z$ plane.
Similar maps for all rings with $\mu_{\rm ring}\!>\!0.01$ and
for pronounced rings with $\mu_{\rm ring}\!>\!0.5$ are shown in
\fig{app_f_Mz_bins} in appendix \se{app_ring_prop}.
This complements the distribution of ring strength in \fig{mu_Mz_bins}.
We see that a high fraction of rings is detected above the threshold mass,
$\Mv \sgt 10^{11}\msun$, where discs survive spin flips (\fig{disc_Mz_bins}),
and at $z\slt 4$.
}
\label{fig:f_Mz_bins}
\end{figure}

\begin{figure} 
\centering
\includegraphics[width=0.46\textwidth]
{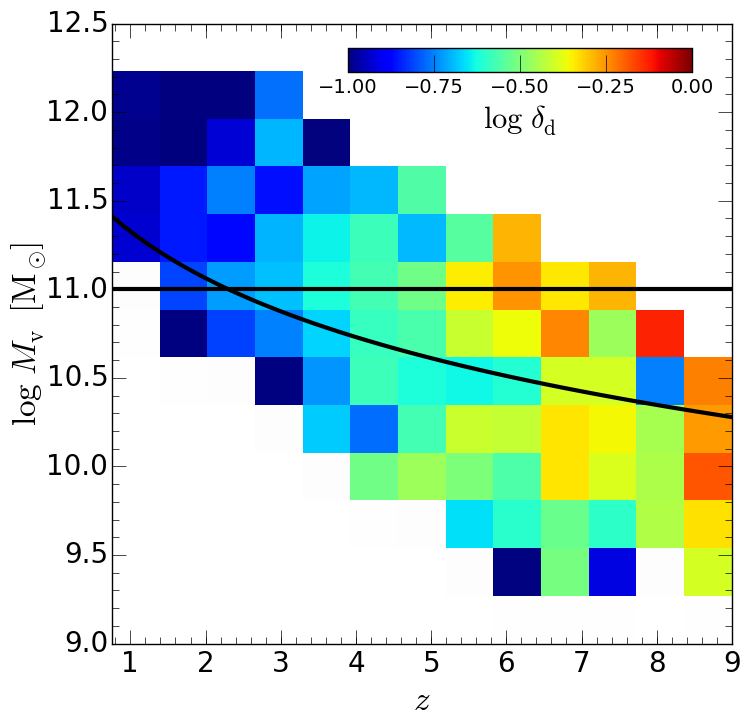}
\caption{
Analytic model versus simulations in the $\Mv\sdash z$ plane.
Shown is the average $\delta_{\rm d}\seq \Md/M_{\rm tot}$ 
in bins within the plane.
This quantity is expected to govern the inward mass transport rate
and thus tell where rings are expected in this plane,
to be compared to the ring fraction map in \fig{f_Mz_bins}.
We see that a high fraction of rings is detected above the threshold mass,
$\Mv \sgt 10^{11}\msun$, where discs survive spin flips (\fig{disc_Mz_bins}),
and at $z\slt 4$.
This regime is indeed where on average $\delta_{\rm d} \slt 0.3$,
as predicted in \equs{rep} and (\ref{eq:dep}).
The redshift dependence of $\delta_{\rm d}$, also reflected in the ring
fraction in \fig{f_Mz_bins}, is largely due to the general increase of gas 
fraction with redshift.
Below the threshold mass $\delta_{\rm d}$ is not too meaningful because
the galaxies are dominated by non-discs (\fig{disc_Mz_bins}).
The low values of $\delta_{\rm d}$ at low masses could be due to gas removal
by supernova feedback below the critical potential well of $\Vv\ssim 100\kms$
\citep[black curve][]{ds86}, but they do not lead to long-term rings because
of spin flips.
}
\label{fig:delta_d_Mz_bins}
\end{figure}

\begin{figure} 
\vskip 0.17cm
\includegraphics[width=0.49\textwidth]
{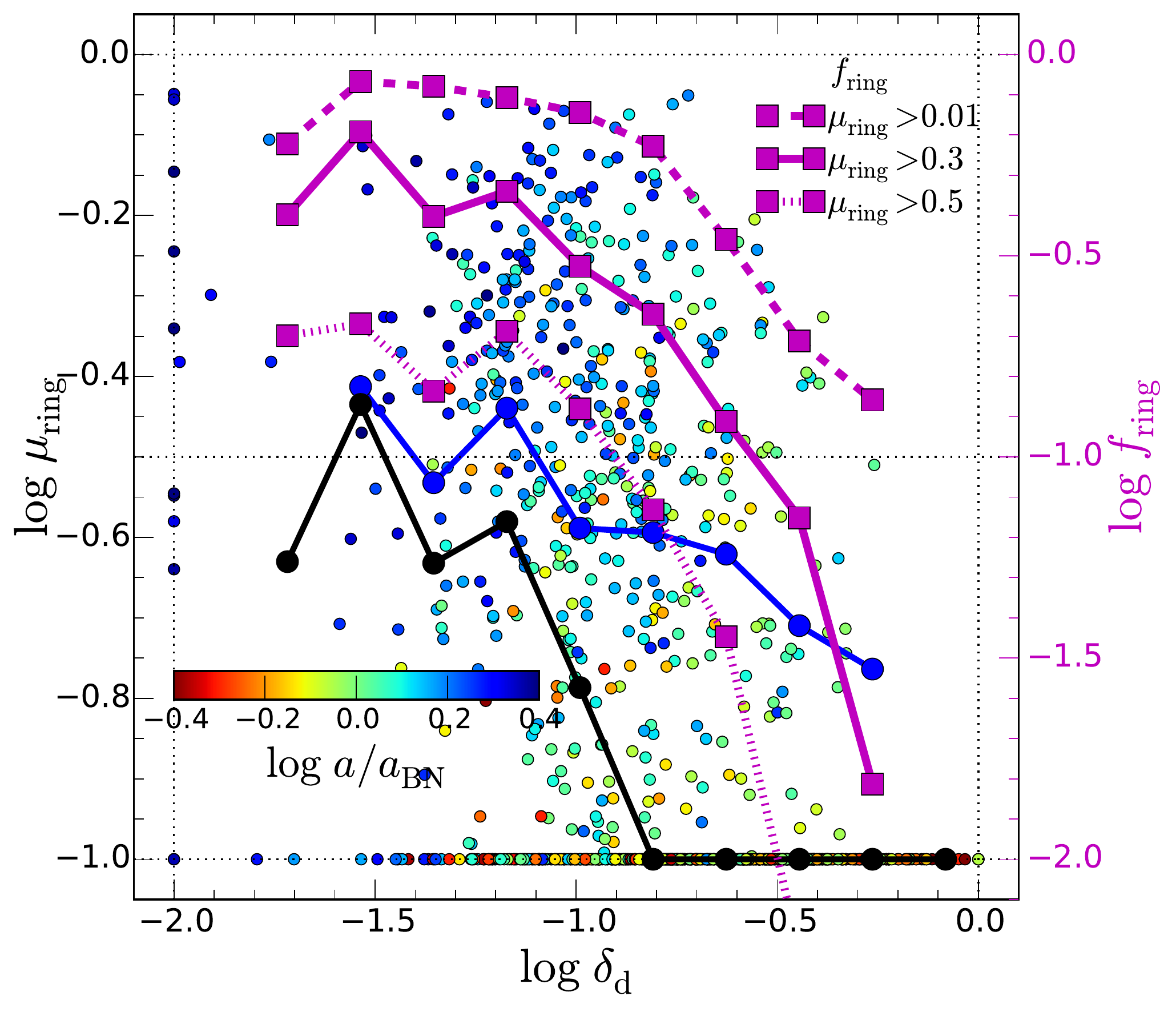}
\caption{
Analytic model versus simulations - ring strength and fraction.
Shown for each galaxy is the ring gas mass excess $\mu_{\rm ring}$
versus $\delta_{\rm d}$, the cold-to-total mass ratio interior
to the ring, predicted to control the ring inward mass transport.
The medians of $\mu_{\rm ring}$ in bins of $\delta_{\rm d}$
are shown for all galaxies (black, shade) and for the galaxies
with rings (blue).
Symbol color marks time with respect to the BN event (correlated with
mass with respect to the characteristic BN mass).
Also shown is the fraction of galaxies with rings of 
$\mu_{\rm ring}\sgt 0.01,\,0.3,\,0.5$ 
\adb{(dashed, solid, dotted lines respectively)} 
in bins of $\delta_{\rm d}$ (magenta, right axis and label).
We see an anti-correlation between ring strength and $\delta_{\rm d}$ as
predicted by the analytic model. 
The fraction of rings with $\mu_{\rm ring}\sgt 0.3$ ranges from 
$\sim\!64\%$ at $\delta_{\rm d}\ssim 0.03$ 
to $\sim\!9\%$ at $\delta_{\rm d}\ssim 0.5$.
A fraction of 50\% in all rings is obtained near $\delta_{\rm d}\ssim 0.2$,
in general agreement with the model prediction, \equs{rep} and (\ref{eq:dep}).
The fraction of strong rings with $\mu_{\rm ring}\sgt 0.5$ is $\sim\!20\%$,
obtained for $\delta_{\rm d} \!\leq\! 0.1$, 
and $\sim\!10\%$ at $\delta_{\rm d}\!\simeq\!0.15$
The $\delta_{\rm d}$ dependence of the ring fraction is expected to be similar 
to that of $\tinf$, and, indeed, the inverse linear relation 
$\tinf\! \prop\! \delta_{\rm d}^{-1}$ of \equ{tinf_b},
predicted for significant rings, is roughly reproduced overall.
The slight steepening of $f_{\rm ring}$ at higher $\delta_{\rm d}$, 
and the absence of rings at $\delta_{\rm d} \sgt 0.5$, 
are consistent with the prediction for weaker rings in \equ{tinf}.
The flattening of $f_{\rm ring}$ at low $\delta_{\rm d}$ is consistent 
with saturation of the ring population when $\tinf$ is very long compared 
to all other timescales.
}
\label{fig:ring_delta}
\end{figure}

\begin{figure} 
\centering
\includegraphics[width=0.44\textwidth]
{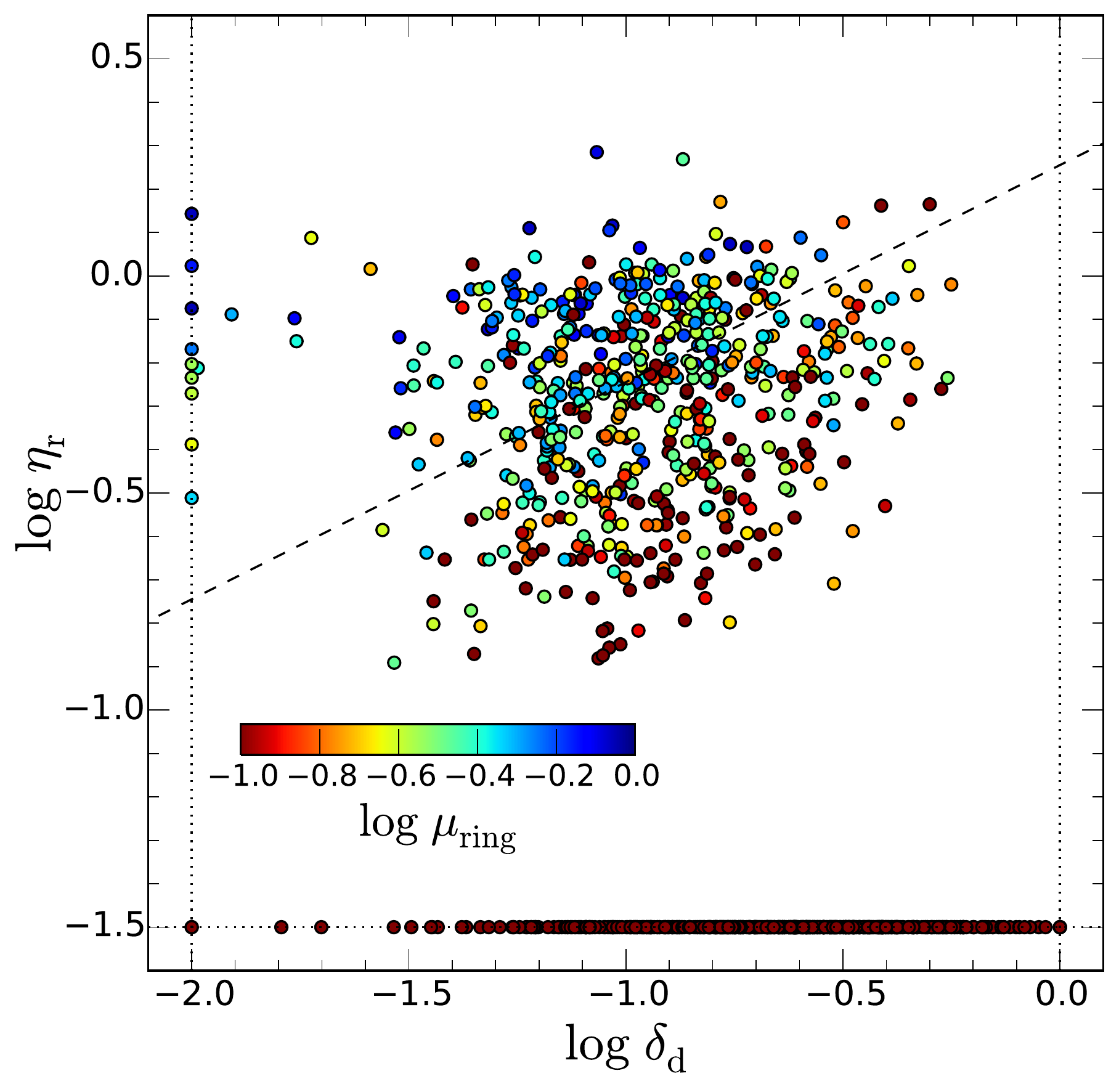}
\caption{
Analytic model versus simulations - ring thickness.
Shown is 
the relative ring width $\eta_{\rm r}\seq \Delta r/r$ versus $\delta_{\rm d}$
for all simulated galaxies.
The symbol color represents ring contrast.
Low values outside the box are put near the box axes (with many points
accumulated at $log \delta_{\rm d} \seq -2$ and $\log \eta_{\rm r}\seq -1.5$).
We see that these quantities are \adb{marginally} correlated. 
For high-contrast rings \adb{(blue and cyan)}
the correlation \adb{and the normalization are} consistent with the relation
$\eta_{\rm r} \ssim 1.8\,\delta_{\rm d}^{1/2}$
predicted in \equ{eta_deltad} (dashed line).
}
\label{fig:eta_deltad}
\end{figure}

\subsection{Ring Toomre instability to clump formation}
\label{sec:toomre}

The rings in the simulations, as seen in \fig{rings_gas}, are clumpy and star 
forming, as observed \citep[e.g.][]{genzel14_rings}, 
with the clump properties analyzed in \citet{mandelker14,mandelker17}.
This indicates that the rings are gravitationally unstable.
Can this be consistent with the rings being stable against inward mass 
transport?

\smallskip
The Toomre $Q$ parameter can also be expressed in terms of the cold-to-total
mass ratio $\delta_{\rm d}$ \citep{dsc09}.
\adb{
For a high-contrast ring at $r$, with a relative width $\eta_{\rm r}$, mean
cold surface density $\Sigma_{\rm r}$, and cold mass 
$M_{\rm r} \seq 2\,\eta_{\rm r}\,\pi\,r^2\,\Sigma_{\rm r}$,
we obtain with $V^2 \seq G\,M_{\rm tot}(r)/r$ 
\be
Q = \frac{\kappa \sigma}{\pi G \Sigma_{\rm r}}
=2\,\eta_{\rm r}\,\psi\,\delta_{\rm d}^{-1} \left( \frac{\sigma}{V} \right) \, ,
\label{eq:Q}
\ee
where $\psi\ssim 1\sdash2$ (mentioned in the context of \equ{kcrit}).
Here $\delta_{\rm d}$ refers in practice to the cold mass in the ring versus
the total mass encompassed by the ring.
With $\psi\ssim \sqrt{2}$ and $\eta_{\rm r}\ssim 0.5$,
this implies that a ring with $\sigma/V \ssim 0.2$ (\fig{vela_V})
could be Toomre unstable with $Q \!\ssim\! 1$ 
if $\delta_{\rm d} \ssim 0.2$.
}
This is indeed in the regime of long-lived rings based on \equs{rep} and
(\ref{eq:dep}), and certainly so in the high-contrast-rings case, \equ{tinf_b}. 
Based on \figs{delta_d_Mz_bins} and \ref{fig:ring_delta}, 
a significant fraction of the rings have $\delta_{\rm d} \ssim 0.2$,
especially at $z\ssim 2\sdash 5$.

\smallskip
In fact, it has been shown using the simulations that clumps may form even with 
$Q\ssim 2\sdash 3$ as a result of an excess of compressive modes of turbulence 
due to compressive tides during mergers or flybys \citep{inoue16}.
This would allow clumpy rings even when $\delta_{\rm d}\!\lsim\!0.1$.
\adb{
The ring could thus fragment to giant clumps and form stars while it is
deep in the regime where it is stable against inward mass transport,
provided that it is above the threshold for a gaseous ring, 
$\delta_{\rm d} \ssim 0.03$ (\se{model_sims}).
}

\smallskip
At the same time, the value of the $Q$ parameter in the disc region 
encompassed by the ring can be much higher than unity, leading to 
``morphological quenching" \citep{martig09}. 
This is because $\delta_{\rm d}$ in this region is low,
partly due to the high central mass and partly due to the low density of gas, 
which has been depleted into stars and outflows.

\subsection{Model versus simulations via the cold fraction}
\label{sec:model_sims}

We turn to the simulations again, now for the variable 
$\delta_{\rm d} \seq M_{\rm d}/M_{\rm tot}$
that is predicted to control the ring formation and longevity according to
the analytic model derived in the previous subsections.
First, \fig{delta_d_Mz_bins} shows in bins within the $\Mv \!-\! z$ plane
the averaged values of $\delta_{\rm d}$ for all galaxies.
The cold mass $M_{\rm d}$ is computed in the disc within $\Rd$
including the gas and young stars,
with the latter typically contributing about $20\sdash 25\%$ of the cold mass.
It is divided by the total mass including gas, stars and dark matter
in the sphere interior to the ring radius $r_0$.
In the absence of a ring, the total mass is computed within the median
ring radius in the galaxies with rings, $r_0 \!\simeq\! 0.5\,\Rd$
(\fig{app_prop_dist} in appendix \se{app_ring_prop}).

\smallskip
This map of $\delta_{\rm d}$ is to be compared to the maps of $f_{\rm ring}$
in \fig{f_Mz_bins} for rings of different strengths,
and to the distribution of ring strength in \fig{mu_Mz_bins}.
We learned there that a high fraction of rings is detected above the galaxy
threshold mass, $\Mv \sgt 10^{11}\msun$, where discs or rings
survive merger-driven spin flips according to \fig{disc_Mz_bins} 
and \citet{dekel20_flip}, and typically at $z\slt 4$.
This ring-dominated range in the $\Mv\sdash z$ diagram is indeed where on 
average $\delta_{\rm d} \slt 0.3$,
consistent with the analytic predictions in \equs{rep} and (\ref{eq:dep}).
These low values of $\delta_{\rm d}$ are largely due to the post-compaction 
massive bulges that appear above a similar threshold mass. 

\smallskip 
Near the threshold mass we see an increase of $\delta_{\rm d}$ with redshift
roughly as $\delta_{\rm d} \prop (1+z)$.
This mostly reflects the general increase of gas fraction with redshift.
We learn from \equs{tinf} and (\ref{eq:tacc}) that the quantity that
characterizes a pronounced ring is expected to depend on redshift as
$\tinf/\tacc \ssim 0.3\,(1+z)\,\delta_{{\rm d},0.3}^{-3}$,
so it is predicted to decrease with redshift as $\prop\!(1+z)^{-2}$.
This explains the absence of rings at high redshifts even above the mass
threshold for discs, as seen in the distribution of $f_{\rm ring}$.

\smallskip 
Well below the threshold mass $\delta_{\rm d}$ is not too meaningful for 
ring survival because these galaxies are dominated by irregular non-disc 
gas configurations (\fig{disc_Mz_bins}),
as discs/rings are frequently disrupted by merger-driven spin flips
\citep[][Fig.~4]{dekel20_flip}.
The relatively low values of $\delta_{\rm d}\ssim 0.1\sdash 0.2$ at low masses
and moderate redshifts partly reflect little gas, possibly due to gas removal 
by supernova feedback.
Indeed, the upper limit for effective supernova feedback at $\Vv\ssim 100\kms$
\citep{ds86}, marked in the figure by a black curve, roughly matches 
the upper limit for the region of low $\delta_{\rm d}$ values, 
with a similar redshift dependence of $\Mv$.
Another contribution for the low $\delta_{\rm d}$ values in this regime
may come from the pre-compaction central dominance of dark matter 
\citep{tomassetti16}.
However, these low values of $\delta_{\rm d}$ do not lead to long-lived rings
because they are disrupted by merger-driven spin flips (\fig{disc_Mz_bins}).

\smallskip
To further explore the match of the analytic model with the simulations,
\fig{ring_delta} shows more explicitly the relation between the ring strength
and the variable $\delta_{\rm d}$ that governs the model.
Shown for each galaxy is the ring gas mass excess $\mu_{\rm ring}$ 
versus $\delta_{\rm d}$, as well as the  
median values of $\mu_{\rm ring}$ in bins of $\delta_{\rm d}$,
for all galaxies (black, shade) and limited to the galaxies with rings (blue).
Most interesting are the fractions of galaxies with rings 
of $\mu_{\rm ring} \sgt 0.01, 0.3$ or $0.5$,
shown in bins of $\delta_{\rm d}$ 
(magenta symbols and lines, as labeled, right axis).
We see an anti-correlation between ring strength and $\delta_{\rm d}$ as
predicted by the analytic model.
The fraction of rings with $\mu_{\rm ring}\sgt 0.3$ ranges from
$\sim\!64\%$ at $\delta_{\rm d}\ssim 0.03$
to $\sim\!9\%$ at $\delta_{\rm d}\ssim 0.5$.
It is encouraging to note that for all rings ($\mu_{\rm ring}\sgt 0.01$)
the fraction is $f_{\rm ring} \seq 0.5$ near $\delta_{\rm d}\ssim 0.2$,
in general agreement with the model prediction in \equs{rep} and (\ref{eq:dep}).
We see that the fraction of strong rings with $\mu_{\rm ring}\sgt 0.5$ 
is $\sim\!20\%$, obtained for $\delta_{\rm d} \!\leq\! 0.1$.
This fraction is $\sim\!10\%$ at $\delta_{\rm d}\!\simeq\!0.15$.
\adb{We see an indication for a decrease in ring fraction for 
$\delta_{\rm d}\slt 0.03$, as expected for galaxies with a very low gas 
fraction (\se{survival}).
}

\smallskip
The way $f_{\rm ring}$ depends on $\delta_{\rm d}$ may be expected to crudely 
resemble that of $\tinf$. 
Indeed, the inverse linear relation
$\tinf\! \prop\! \delta_{\rm d}^{-1}$ predicted in \equ{tinf_b}
for strong rings is crudely reproduced overall.
The steepening at higher values of $\delta_{\rm d}$,
and in particular the absence of rings at $\delta_{\rm d} \sgt 0.5$,  
are consistent with the prediction in \equ{tinf} for rings in their earlier 
growth phase.
The flattening of the ring fraction at very low $\delta_{\rm d}$ values
is consistent with saturation of the ring population when $\tinf$ 
is very long compared to the Hubble time and all other timescales.

\smallskip
Another way to test the validity of the analytic model is via the ring width.
\Fig{eta_deltad} shows 
the relative ring width $\eta_{\rm r}\seq \Delta r/r$ versus $\delta_{\rm d}$
for all simulated galaxies. The width is deduced from the Gaussian fit,
$\eta_{\rm r} \seq 2\sigma/r_0$.
Also shown by color is the ring contrast.
We see a correlation between $\eta_{\rm r}$ and $\delta_{\rm d}$,
which for high-contrast rings is well modeled by the relation
$\eta_{\rm r} \ssim 1.8\,\delta_{\rm d}^{1/2}$ predicted in \equ{eta_deltad}.

\smallskip
We conclude that the survival of rings about massive central masses, as seen 
in the simulations post compaction, \fig{mu_bn}, 
can be understood in terms of the analytic model, \equ{rep} and (\ref{eq:dep})
as well as \equs{tinf_b} with \equ{eta_deltad}.

\section{The central body of ring galaxies}

\subsection{Baryons versus dark matter}
\label{sec:BD}

\begin{figure} 
\centering
\includegraphics[width=0.46\textwidth,trim={{0.02\textwidth} 0 0 0}]
{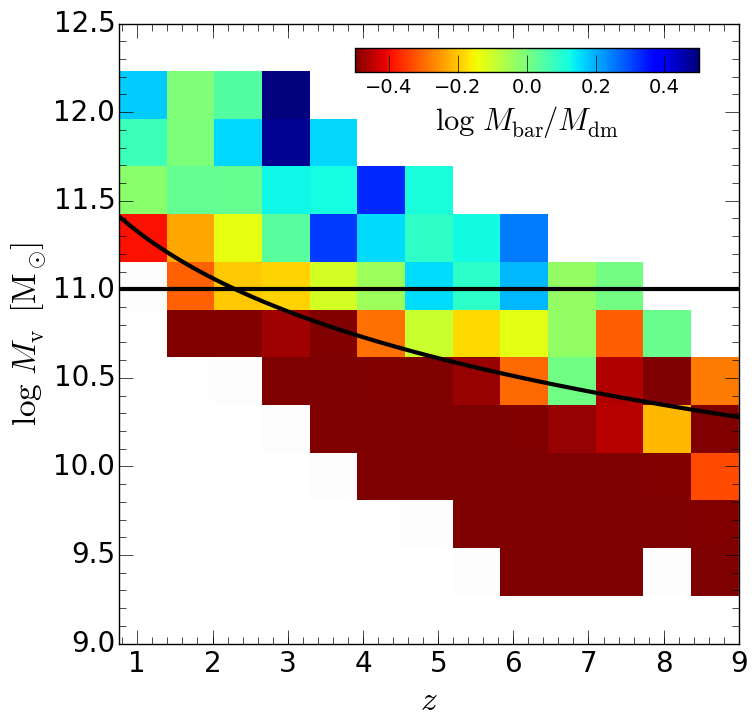}
\caption{
Nature of the central mass.
Shown in bins within the $\Mv\sdash z$ plane
is the average of the baryon to dark-matter ratio within the ring radius.
For no-rings the radius is chosen to be $r_0 \seq 0.5\,\Rd$, the typical ring
radius.
The dark matter dominates below the characteristic mass of 
$\sim\!  10^{11}\msun$, while the baryons dominate above it,
reflecting the transition due to the major compaction event.
Nevertheless, even post compaction the two components are comparable.
}
\label{fig:BD_2D}
\end{figure}

\begin{figure*} 
\centering
\includegraphics[width=0.44\textwidth,trim={{0.02\textwidth} 0 0 0}]
{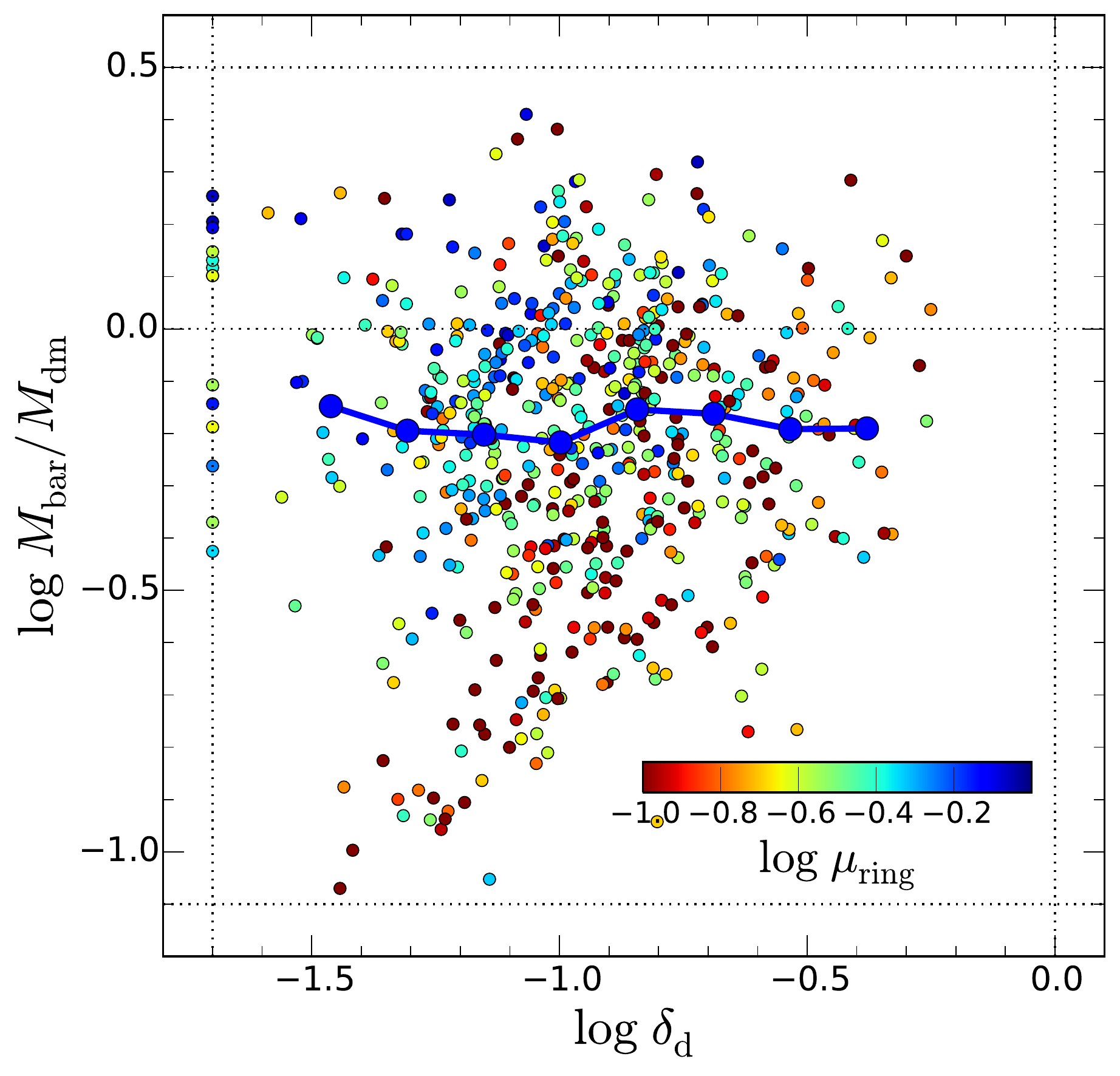}
\quad
\includegraphics[width=0.44\textwidth,trim={{0.02\textwidth} 0 0 0}]
{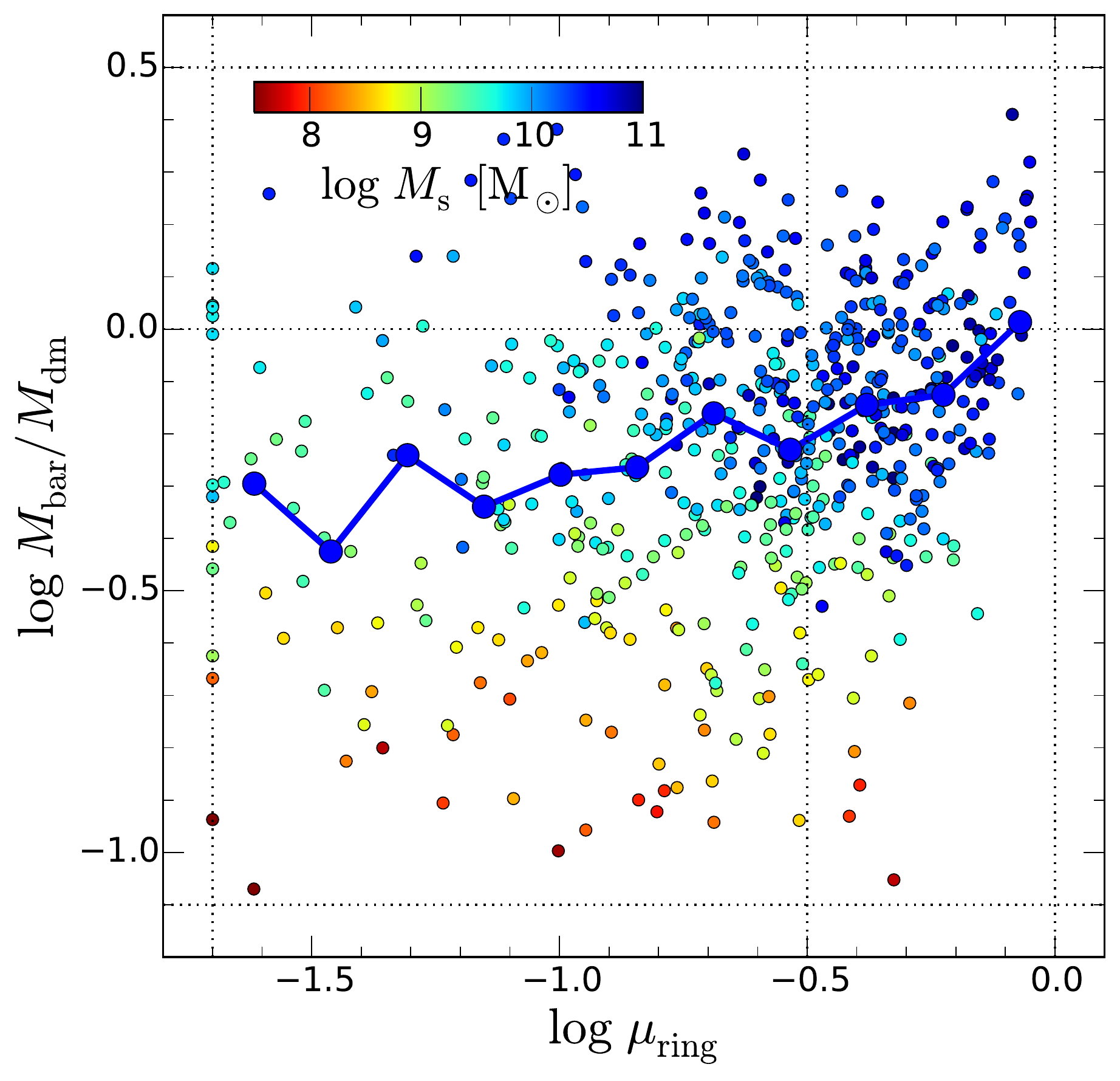}
\caption{
Baryons versus dark matter interior to the rings.
Shown is 
the baryon-to-DM ratio versus $\delta_{\rm d}$ (left) and $\mu_{\rm ring}$
(right) for all galaxies with rings, with the median marked (blue). 
Low $\delta_{\rm d}$ values (e.g. $\slt 0.3$) and the associated
significant rings (e.g. $\mu \sgt 0.1$ in this case) are obtained for a range of
ratios, from baryon dominance by a factor $\times\!2.5$ to DM dominance
by a similar factor for massive galaxies, and by $\times\! 10$ for low-mass
galaxies.  
In the median there is slightly more DM than baryons.
}
\label{fig:BD_dd_mu}
\end{figure*}

It would be interesting to figure out the contributions of the different
mass components to the central mass that determines the low values of
$\delta_{\rm d}$ and thus leads to long-lived rings. 
First,
\fig{BD_2D} shows the distribution of B/DM, the mass ratio of baryons to dark
matter interior to the ring. In the case of a ring the masses are computed
within the ring radius $r_0$, and in the case of no-ring they are measured
within $0.5\!\Rd$, the average ring radius when there is a ring.
Shown is the average of B/DM in bins within the $\Mv\sdash z$ plane.
We see that the dark matter tends to dominate below the threshold mass near
$\Mv \ssim 10^{11}\msun$ while the baryons tend to dominate above it.
This reflects the major compactions to blue nuggets near this characteristic
mass. 
We note that in the regime that tends to populate rings, 
namely above the threshold mass and at $z\slt 1$, 
the average contributions of baryons and dark matter are comparable. 
We learn that the average post-compaction baryon dominance is typically
marginal, by a factor of order unity and $\times\!2.5$ at most. 

\smallskip
To find out how dark matter may contribute to low values of $\delta_{\rm d}$
and the associated high ring strength, 
\fig{BD_dd_mu} shows B/DM versus $\delta_{\rm d}$ and versus $\mu_{\rm ring}$
for all the simulated galaxies which show rings.
We learn that the low values of $\delta_{\rm d} \slt 0.3$, which are supposed
to lead to rings, as well as the significant rings themselves, 
with $\mu_{\rm ring} \sgt 0.1$, say, could be associated with central bodies
that are either dominated by baryons or by dark matter.  
Typically the contributions of the two components are comparable,
with a slight preference for the dark matter. 
However, for massive galaxies B/DM ranges from $0.4$ to $2.5$. 
This implies that long-lived rings could appear even in cases where the
central mass is dominated by dark matter with a negligible bulge.
This is to be compared to observations.
As mentioned in \se{Halpha}, both cases of baryon dominance and dark-matter
dominance are detected \citep{genzel20}.

\smallskip
The important contribution of the dark matter to the mass interior to the ring
even after a major wet compaction event
is partly because the wet compaction of gas induces an adiabatic
contraction of stars and dark matter, and partly because the ring radius is
significantly larger then the $1\kpc$, or the effective radius $\Re$, 
within which the post-compaction baryons are much more dominant.

\subsection{Nuggets: naked versus ringy, blue versus red}
\label{sec:nuggets}

\begin{figure*} 
\centering
\includegraphics[width=0.44\textwidth,trim={{0.02\textwidth} 0 0 0}]
{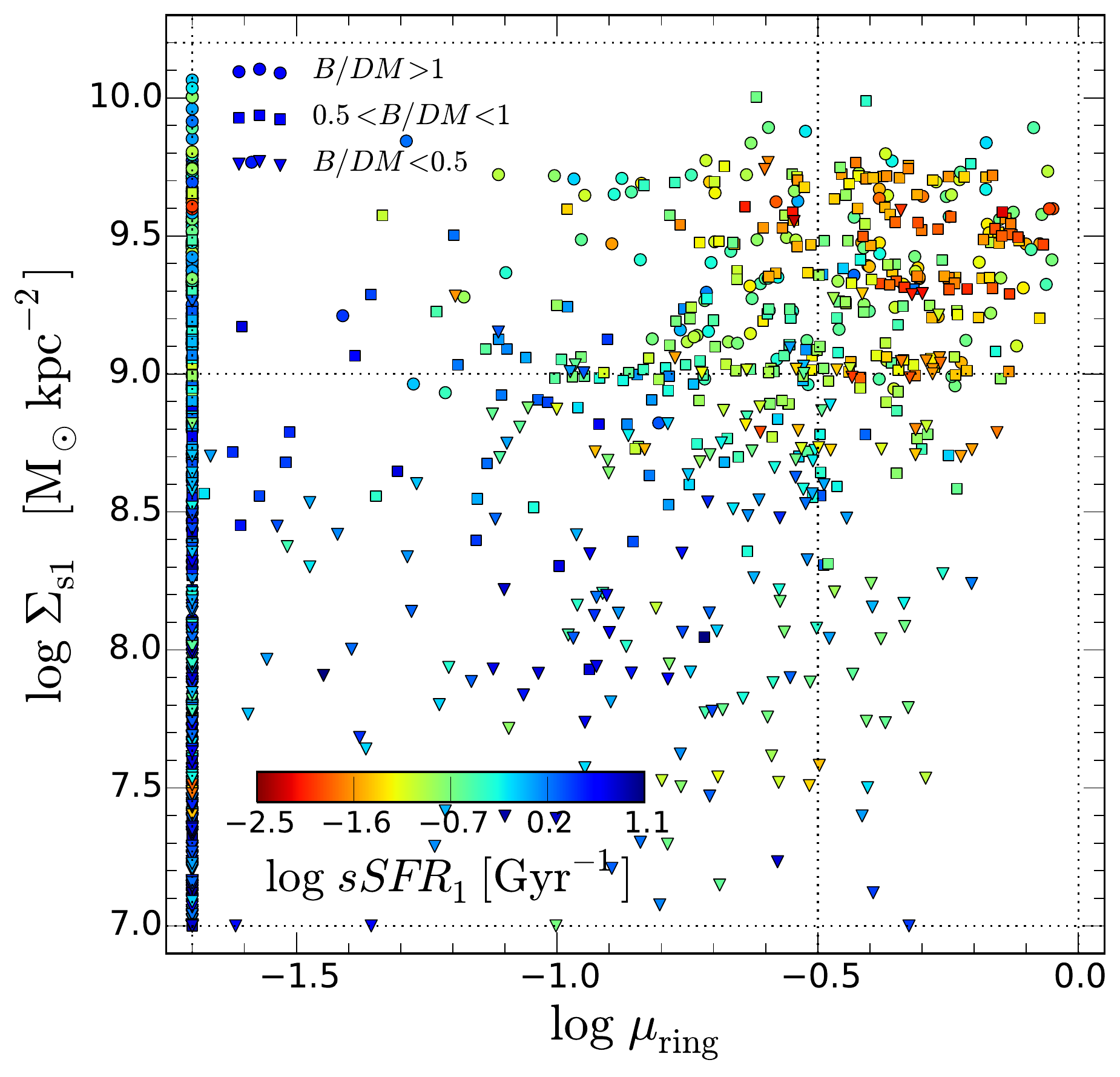}
\quad
\includegraphics[width=0.44\textwidth,trim={{0.02\textwidth} 0 0 0}]
{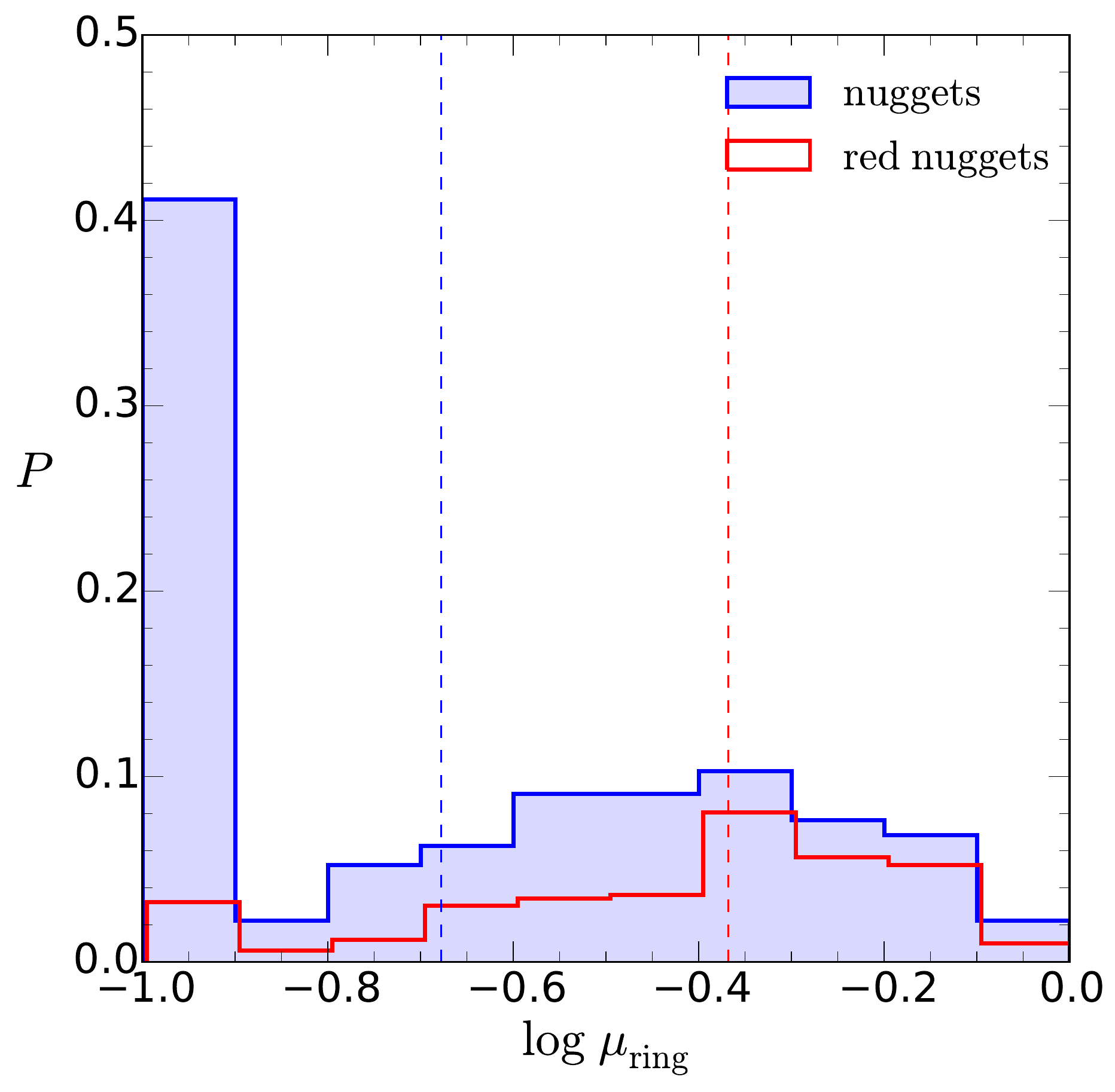}
\caption{ 
Nuggets and rings. 
{\bf Left:}
Stellar surface density within the inner $1\kpc$, $\Sigma_{{\rm s}1}$, 
versus ring gas mass excess, $\mu_{\rm ring}$,
for all simulated galaxies.
Nuggets are characterized by $\Sigma_{{\rm s}1} \sgt 10^9\msun\kpc^{-2}$. 
Color marks the sSFR within $1\kpc$,
with blue and red nuggets separated at 
$\log {\rm sSFR}_1/\Gyr^{-1} \ssim -1$.
{\bf Right:}
The probability distribution of $\mu_{\rm ring}$ for nuggets,
obeying $\Sigma_{{\rm s}1} \sgt 10^9\msun\kpc^{-2}$ (blue, shaded). 
The fraction of red nuggets, $\log {\rm sSFR}_1 \slt -1$, is marked by 
the red histogram, with the difference between the blue and red histograms 
referring to the blue nuggets.
We see on the left that pronounced rings tend to surround nuggets, 
while a significant fraction of the weaker rings have no central nuggets.
Among the nuggets, on the right, $\sim\! 40\%$ have significant rings of
$\mu_{\rm ring} \sgt 0.3$, while $\sim\! 40\%$ are naked with no rings.
The nuggets inside rings are roughly half blue and half red nuggets. 
}
\label{fig:nuggets_mu}
\end{figure*}

It is interesting to address the interplay between the strength of the ring
and the nature of the mass in the central $1\kpc$.
\Fig{nuggets_mu} refers to this in terms of being a nugget, 
and if so whether it is an early star-forming blue nugget or a late quenched 
red nugget.
We learn from the left panel that significant rings, e.g., 
$\mu_{\rm ring}\sgt 0.3$, predominantly surround nuggets at their centers,
defined by $\Sigma_{{\rm s}1} \sgt 10^9\msun\kpc^{-2}$ (see \fig{compaction}).
A larger fraction of the weaker rings have no central nuggets (though they may
still have large masses interior to the ring radius, which is typically
several kpc, leading to a small $\delta_{\rm d}$). 
These no-nuggets are typically star-forming, pre compaction and DM dominated.

\smallskip
Focusing on the galaxies with central nuggets,
the right panel of \fig{nuggets_mu} shows the distribution of 
$\mu_{\rm ring}$ for the nuggets only.
We learn that $\sim\! 40\%$ of the nuggets are naked,
while $\sim\! 60\%$ are surrounded by rings, with $\sim\! 40\%$ having
significant rings of $\mu_{\rm ring} \sgt 0.3$.

\smallskip
The red histogram refers to the fraction of red nuggets among the nuggets,
defined by $\log {\rm sSFR}_1/\Gyr^{-1} \slt -1$.
We learn that among the significant rings, the nuggets are divided 
roughly half and half between quenched red nuggets and 
star-forming blue nuggets, which tend to appear later and earlier after the 
compaction phase respectively.

\smallskip
These predicted fractions of rings with no central nuggets, 
of naked nuggets, and of blue versus red nuggets in
galaxies that show rings, are to be compared to observations
(see Ji et al., in prep., and a preliminary discussion in \se{comp_obs}).

\subsection{Torques by a prolate central body}
\label{sec:prolate_sum}

To complement the analytic model for disruptive mass transport,
we note that it may be aided by torques exerted by a central 
body, stars or dark matter, if it deviates from cylindrical symmetry. 
Indeed, as can be seen in \fig{app_shape},
the VELA simulated galaxies tend to be triaxial, prolate pre compaction
and oblate post compaction, showing a transition about the critical mass
for blue nuggets \citep{ceverino15_shape,tomassetti16}. 
A similar transition has been deduced for the shapes of observed CANDELS
galaxies \citep{vanderwel14_shape,zhang19}.
The prolate pre-compaction bulge may exert non-negligible torques 
that could possibly disrupt the disc. 
In appendix \se{app_prolate}, we learn from a very crude estimate that an
extreme prolate central body could have a non-negligible 
effect on the survival of a disc around it.

\smallskip 
The central bulge, which tends toward an oblate shape (\fig{app_shape}),
does not exert torques on masses orbiting in the major plane of the
oblate system, but it does exert torques off this plane.
An extreme oblate system, namely a uniform disc, yields values of
$\Delta j/j \ssim  0.1$ per quadrant of a circular orbit 
in a plane perpendicular to the disc and
close to it \citep[][Figure 16]{danovich15}. 
This implies that the post-compaction central oblate body, above the critical
mass for blue nuggets, is not expected to significantly affect
the AM of the disc and disrupt it.

\begin{figure*} 
\centering
\includegraphics[width=1.01\textwidth,trim={{0.02\textwidth} 0 0 0}]
{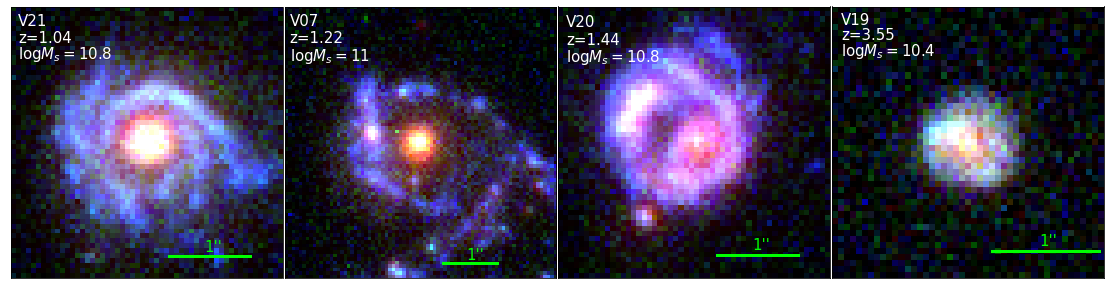}
\caption{
Mock images of simulated rings.
Shown are 
three-color rgb images showing blue rings about red massive bulges
as ``observed" from the four simulated galaxies seen in \fig{rings_gas}.
The corresponding mock images in the three filters F606W, F850LP and F160W
are shown in \fig{app_mock_4x4} of appendix \se{app_more_figures}.
Dust is incorporated using Sunrise
and the galaxy is observed face-on through the HST filters using the HST
resolution and the noise corresponding to the CANDELS survey in the deep
GOODS-S field.
}
\label{fig:mock_rgb}
\end{figure*}

\begin{figure} 
\centering
\includegraphics[width=0.49\textwidth,trim={{0.02\textwidth} 0 0 0}]
{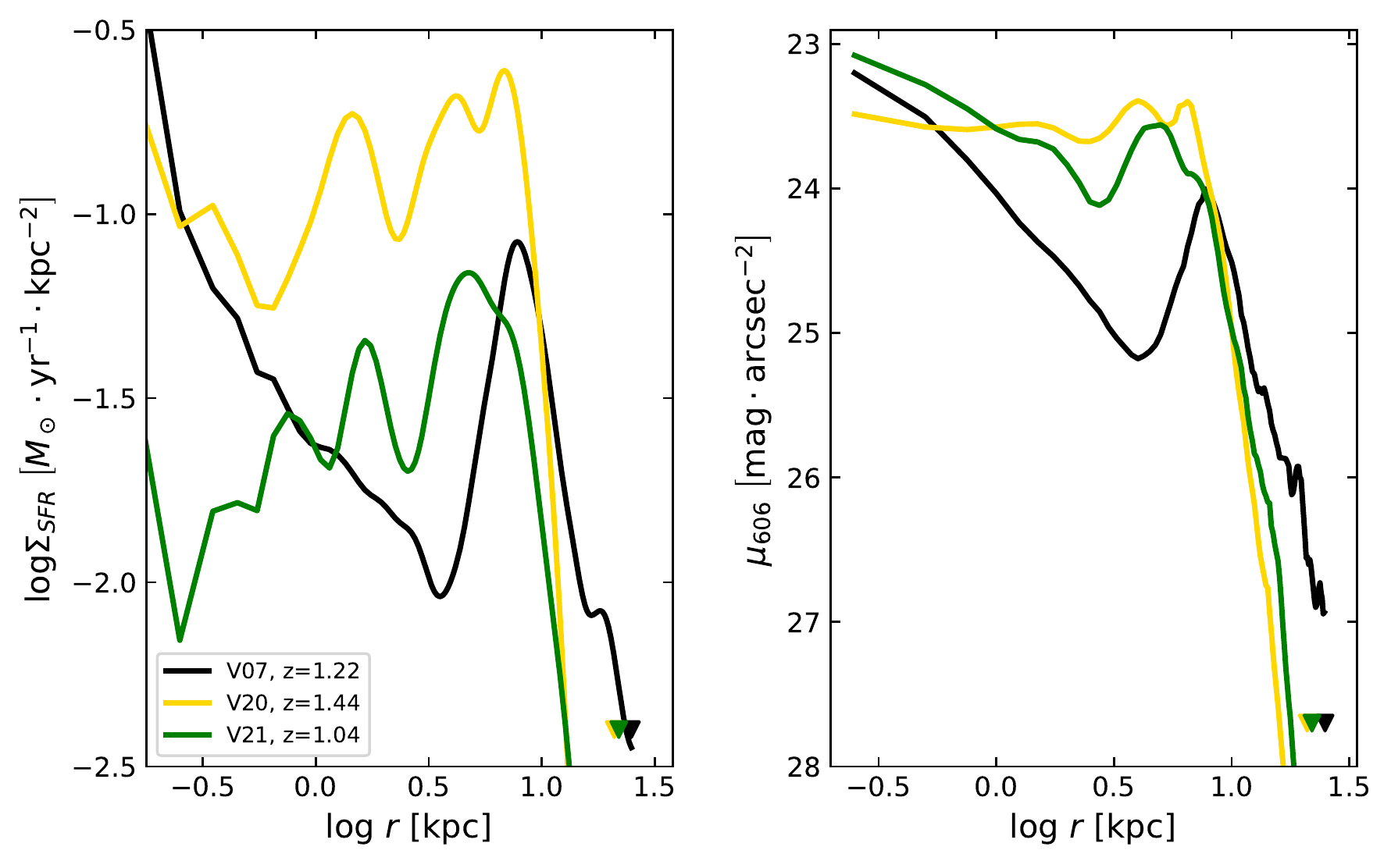}
\caption{
Rings in SFR versus luminosity from mock images of simulated galaxies.
Shown are profiles of three example simulated galaxies with pronounced rings.
{\bf Left:} Face-on projected SFR surface density profiles from the simulations.
{\bf Right:} Light density profiles in the F606W filter from the mock
images shown in \fig{mock_rgb}.
The ring SFR density of $\sim\! 0.1 \msun \yr^{-1} \kpc^{-2}$ shows as
$\sim\! 24\, {\rm mag}\, {\rm arcsec}^{-2}$, indicating weak dust extinction,
but the predicted contrast between the ring and the interior is significantly
smaller.
}
\label{fig:mock_profiles}
\end{figure}

\begin{figure*} 
\includegraphics[width=0.33\textwidth,trim={0.cm 0.2cm 5.7cm 0.0cm},clip]
{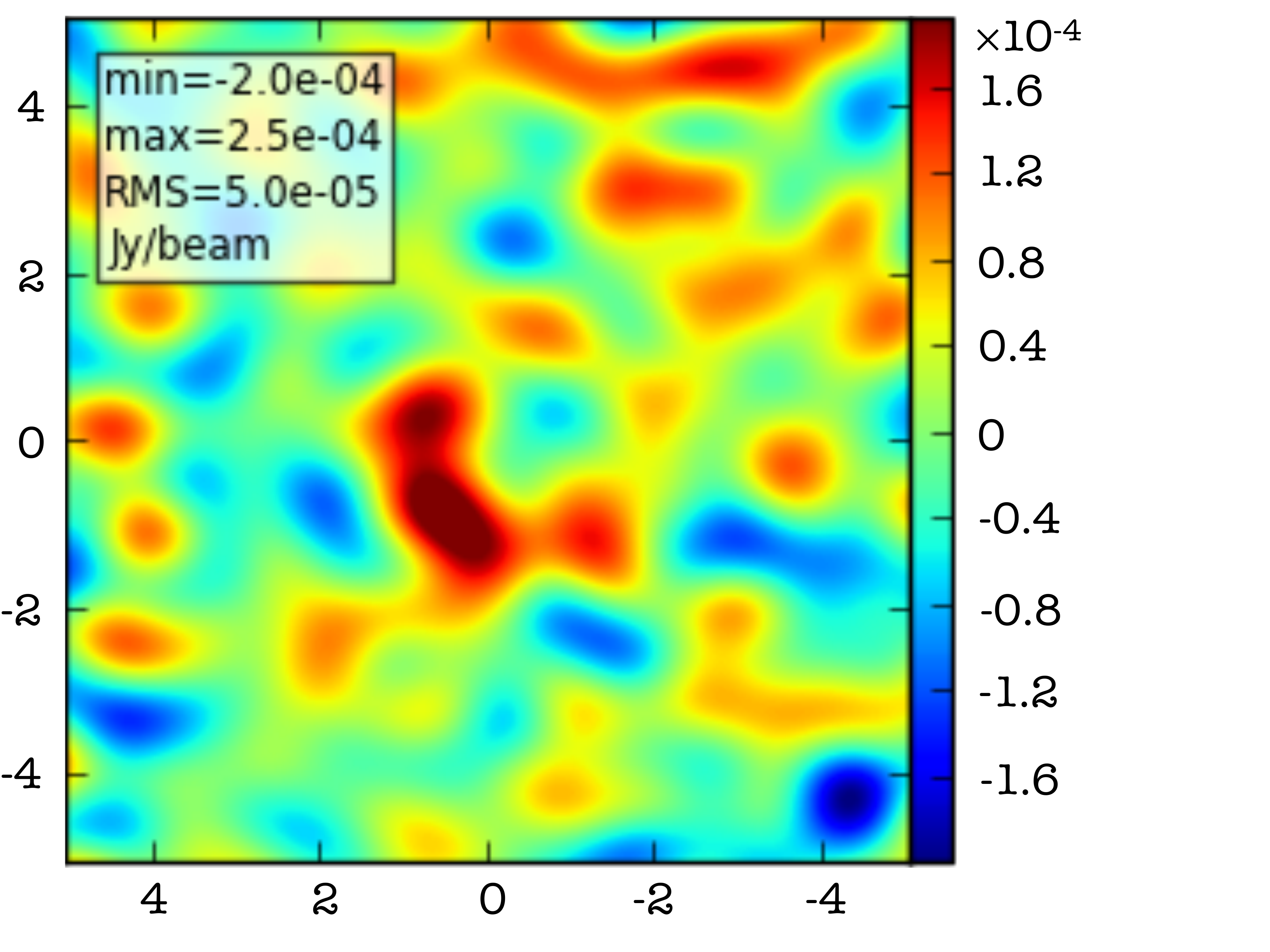}
\hfill
\includegraphics[width=0.33\textwidth,trim={0.cm 0.2cm 5.7cm 0.0cm},clip]
{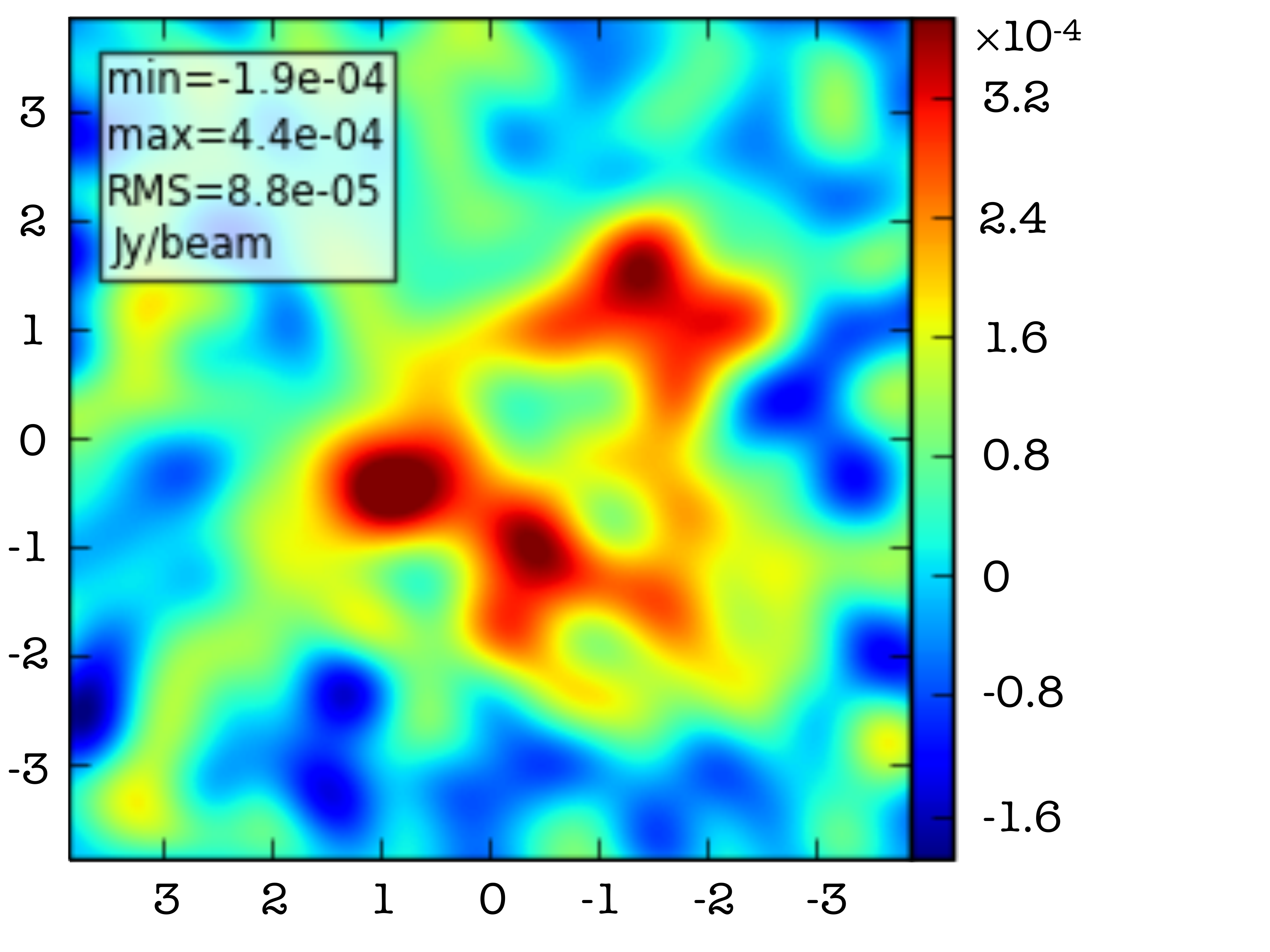}
\hfill
\includegraphics[width=0.33\textwidth,trim={0.cm 0.2cm 5.7cm 0.0cm},clip]
{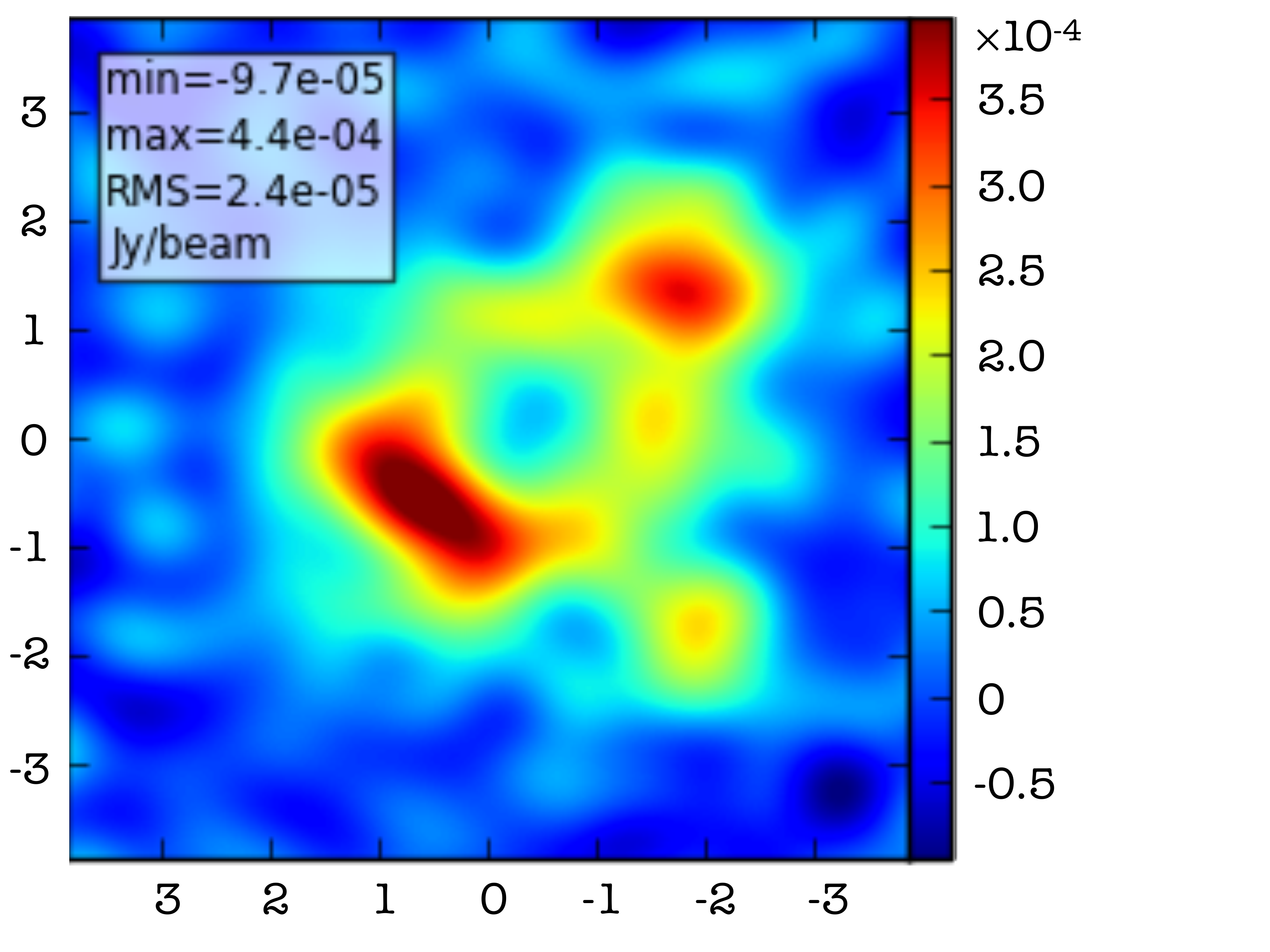}
\caption{
Mock ALMA images of a simulated ring.
Shown are mock CO(2-1) ALMA observations of the V07 simulated gas ring
(\fig{rings_gas}). Color is in Jy per beam.
{\bf Left:} 10h of observation with the galaxy at $z\!=\!1$.
{\bf Middle:} 2h at $z\!=\!0.5$.
{\bf Right:} 10h at $z\!=\!0.5$.
The simulations assumed the 12m array in Cycle 5 configuration 3
together with the Atacama Compact Array (ACA), yielding a spatial resolution
of $\sim$1.5 arcsec (12 kpc) and $\sim$1 arcsec (6 kpc) at $z\!=\!1$ and
$z\!=\!0.5$ respectively. The channel width was chosen to be
$\rm 67 ~\rm km~s^{-1}$ in order to resolve the line width expected from the
velocity dispersion (FWHM $200~\rm km ~s^{-1}$) with 3 channels.
A typical precipitable water vapor level of 0.6 mm was assumed.
The color-map is in Jy/beam and distances are in arcsec.
The simulations were carried out with the \texttt{simalma} task of
\texttt{CASA} \citep{CASA07}.
}
\label{fig:alma}
\end{figure*}

\section{Preliminary Comparison to Observations}
\label{sec:comp_obs}

\subsection{Mock observations of simulated rings}
\label{sec:predictions}

\subsubsection{Crude estimates} 

In order to crudely estimate H$\alpha$ fluxes for the simulated star-forming
rings, we use our estimated SFR surface densities, averaged in the pronounced 
rings and in their giant clumps, of
$\Sigma_{\rm SFR}\!\sim\!(0.1\!-\!1) \msun\yr^{-1}\kpc^{-2}$.
We then assume 
(1)
a conversion to H$\alpha$ luminosity of
$L_{H\alpha} (\ergs) \!=\! 1.8\times 10^{41} {\rm SFR} (\msun\yr^{-1})$
\citep{kennicutt09}, 
(2) an effective screen optical depth of 
$\tau_{\rm H\alpha}\! =\! 0.73\, A_{\rm V}$ 
\citep{calzetti00,genzel13}, 
and
(3) a typical visible extinction of $A_{\rm v}=1$ 
\citep[e.g.][]{forster11a, forster11b, freundlich13}. 
This yields, for sources at $z\!=\!1$, H$\alpha$ fluxes of 
$(1\!-\!10) \times 10^{-16} \ergs\cm^{-2}\,{\rm arcsec}^{-2}$. 
Such values are in the ball park of those detected 
in resolved H$\alpha$ observations at $z\!\sim\! 1$ 
\citep[e.g.][]{nelson12, contini12, genzel13}. 
Simulated rings traced by $H_\alpha$ are also seen in \citet{ceverino16_disc}.

For crude estimates relevant for HST imaging, we note that 
for a source at $z \ssim 1$ the F606W band of HST falls near $3000\AA$
and thus traces the rest-frame UV luminosity, which can serve as a proxy for 
star formation in the gaseous rings. 
Using the relation between UV luminosity and SFR based on 
\citet{kennicutt98_araa},
corrected for a \citet{chabrier03} initial mass function by a 
factor of $\times 1.8$, the SFR surface densities in the rings and the clumps,
$\Sigma_{\rm SFR}\!\sim\! (0.1\!-\!1)\msun\yr^{-1}\kpc^{-2}$ respectively,
yield crudely estimated UV magnitudes of 
$(23.4\!-\!20.9)~\rm mag~arcsec^{-2}$ without taking extinction into account. 
The fainter magnitudes may be more representative of
the average surface brightness in the most pronounced rings in the simulations.
Dust extinction in the UV is expected to lie between zero and 3 
magnitudes \citep[e.g., ][]{buat96, freundlich13}.

\subsubsection{Mock ``Candelized" Images}
\label{sec:mock_images}

In order to make more quantitative observable predictions for rings based on 
the simulated galaxies, we use mock 2D images that are generated for each 
given VELA simulated galaxy at a given time
to mimic HST CANDELS images (CANDELization).
As described in \citet{snyder15} and \citep{simons19}
(see MAST archive.stsci.edu/prepds/vela/), 
the stellar light is propagated through 
dust using the code \textsc{sunrise}.
A spectral energy distribution (SED) is assigned to every star 
particle based on its mass, age and metallicity. 
The dust density is assumed to be proportional to the metal density
with a given ratio and grain-size distribution.
\textsc{Sunrise} then performs dust radiative transfer using a Monte Carlo 
ray-tracing technique.  As each emitted multi-wavelength ray  
encounters dust mass, its energy is probabilistically absorbed or scattered 
until it exits the grid or enters the viewing aperture,
selected here to provide a face-on view of the gas disc. 
The output of this process is the SED at each pixel. 
Raw mock images are created by integrating the SED in each pixel over the 
spectral response functions of the CANDELS ACS and WFC3 
R and IR filters ($F606W$, $F850LP$ and $F160W$)
in the observer frame given the redshift of the galaxy.   
These correspond at $z\!\sim\!1\!-\!2$ to rest-frame UV, B to U, and R to V,
respectively.
Thus, the first two are sensitive to young stars while the third refers to
to old stars.
The images are then convolved with the corresponding HST PSF at a given 
wavelength. 
Noise is added following \citet{mantha19},
including random real noise stamps from the CANDELS data, to ensure
that the galaxies are simulated at the depth of the real GOODS-S data and 
that the correlated noise from the HST pipeline is reproduced

\smallskip
\Fig{mock_rgb} shows the resultant mock rgb images for the four example
VELA galaxies with rings, whose gas densities are shown in \fig{rings_gas} and
the corresponding profiles in \fig{profiles_rings}.
\Fig{app_mock_4x4} in appendix \se{app_more_figures}
further shows the images of the same simulated galaxies in each filter 
separately.
These are all post-compaction galaxies with $\Ms \ge 10^{10.4}\msun$.
At $z\!\sim\!1\!-\!1.5$ the images show extended blue rings encompassing 
massive red bulges.
The rings could be described as tightly-wound spiral arms, which are sometimes
not concentric about the bulge, and they tend to show giant clumps.  
With the added noise the ring structure becomes less obvious in the 
rgb picture of the $z\!=\!3.55$ galaxy, though it is seen pretty clearly
in the F606W filter (\fig{app_mock_4x4}).

In order to make the predictions a bit more quantitative,
\fig{mock_profiles} shows the SFR surface density profiles of the three VELA
galaxies with clear rings at $z\!\sim\!1\!-\!1.5$. 
These are compared to the light density profiles in the F606W filter from 
the mock images shown in \fig{mock_rgb}.
The ring SFR density of $\sim\! 0.1 \msun \yr^{-1} \kpc^{-2}$ shows as 
a surface brightness of $\mu_{606} \sim\! 24\, {\rm mag}\, {\rm arcsec}^{-2}$, 
indicating only weak dust extinction.
However, the predicted contrast in light between the ring and the interior is 
significantly smaller in the mock images.
These are consistent with the crude estimates made at the beginning of this
section.

\begin{figure*} 
\centering
\includegraphics[width=1.01\textwidth,trim={{0.02\textwidth} 0 0 0}]
{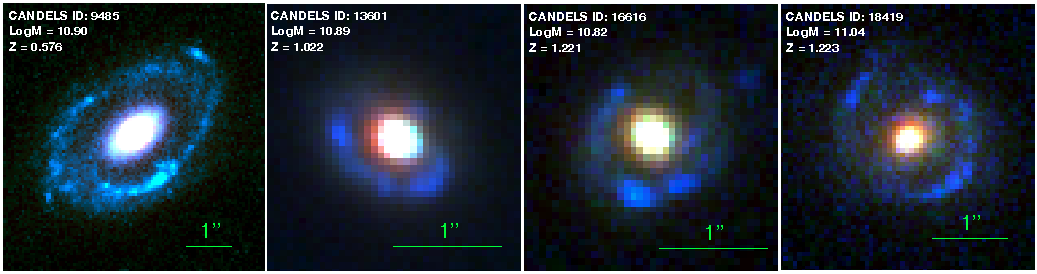}
\caption{
Observed rings.
Shown are three-color rgb images from the deepest, GOODS fields of CANDLES,
displaying extended blue rings about red massive bulges.
The corresponding images in the three filters F606W, F850LP and F160W
are shown in \fig{app_obs_4x4} of appendix \se{app_more_figures}.
The average F606W surface brightness in the rings is
$24.6$, $24.7$, $25.6$, $25.5 \,{\rm mag}\,{\rm arcsec}^{-2}$ in rings id
9485, 13601, 16616, 18419 respectively.
The identified rings with bulges are a non-negligible fraction of the galaxies
with masses $\Ms\!>\!10^{10.5}\msun$ at $z\seq 0.5\!-\!1.2$
(Ji, Giavalisco et al., in prep.).
}
\label{fig:obs_rgb}
\end{figure*}

\subsubsection{Predictions for ALMA CO emission}
\label{sec:alma}

The gas rings could be observable near the peak of star 
formation through their CO line emission.
In particular, the CO(2-1) line at $\nu_{\rm rest}\seq 230.538~{\rm GHz}$ 
is the lowest-excitation line observable with NOEMA and ALMA at $z\!\lsim\!2$. 
We assume (a) the Galactic value for the CO(1-0) conversion factor of molecular 
gas mass to luminosity, 
$\alpha_{\rm CO} \seq 4.36~\rm M_\odot/(K~km~s^{-1}~pc^2)$, 
(b) a conservative CO(2-1)/CO(1-0) line ratio $r_{21} \seq 0.77$ 
\citep[e.g., ][]{daddi15}, 
and (c) the standard relation of \citet{solomon97} for converting
intrinsic CO(2-1) luminosity into integrated flux 
\citep[cf. also][]{freundlich19_PHIBSS2}. 
Then, the typical ring and clump molecular gas 
surface densities of $\Sigma_{\rm ring}\seq 5\stimes 10^7$ and 
$5\stimes 10^8~\rm M_\odot~kpc^{-2}$ 
respectively (assuming that the ring is dominated by molecular gas)
yield integrated CO(2-1) line fluxes of 
$0.04$ and $0.4\,\rm Jy\,km\,s^{-1}\,arcsec^{-2}$ at $z \seq 1$, and  
$0.10$ and $1.0\,\rm Jy\,km\,s^{-1}\,arcsec^{-2}$ if the galaxy is put at 
$z \seq 0.5$. 

\smallskip
We estimate that at $z \seq 1$, practical ALMA observation times per 
galaxy of $<\!10$h would enable a detection of the dense clumps of such rings,
but probably not the lower-density 
regions between the clumps. ALMA would need about $50$h to detect the mean 
surface density of the rings with integrated SNR$=6$, assuming a line width of 
$200\kms$ resolved by three channels. 
In order to visualize the detectability of rings by ALMA,
\fig{alma} presents simulated CO(2-1) ALMA observations of the gas ring of V07 
shown in \fig{rings_gas}. We see that if the galaxy is put at $z\!=\!1$, 
parts of its ring would be traceable with 10h of ALMA but with a low signal to 
noise. 
At $z\!=\!0.5$, the ring is detectable using 2h of ALMA (60h of NOEMA),
and can be mapped with a higher signal-to-noise with 10h of ALMA.

\subsection{Observed rings}
\label{sec:obs}

\subsubsection{In H-alpha}
\label{sec:Halpha}

\citet{genzel14_rings} detected a significant fraction of massive galaxies 
at $z\ssim 2$ with extended $H_\alpha$ star-forming, rotating rings, 
most of which surrounding a central massive stellar bulge.
This is consistent with the proposed formation and stabilization of the rings 
by central masses and the presence of post-compaction rings 
in our simulations.
Then \citet{genzel20} and Genzel et al. (in prep.), following \citet{genzel17}, 
utilizing rotation curves for 40 massive star-forming galaxies at
$z\!=\!0.6\!-\!2.7$ based on data from the 3D-HST/KMOS$^{\rm 3D}$/SINFONI/NOEMA 
surveys,
verify the common existence of extended gas rings. 
They report that some of the rings surround massive compact bulges, 
typically with little dark-matter mass within one or a few effective radii,
while other rings surround less massive bulges but preferably with higher 
central dark-matter masses \citep{genzel20}.  
This is qualitatively consistent with the model in \se{ring_toy} addressing
the crucial role of a massive central body, 
which could either be a post-compaction massive bulge and/or a centrally 
dominant massive dark-matter halo.
It is also qualitatively consistent with our findings in the simulations
(\se{BD}, \fig{BD_dd_mu}) that the mass interior to significant rings 
in massive galaxies range from baryon dominance by up to a factor $\times\!2.5$ 
to dark-matter dominance by up to a similar factor or more.

\smallskip
In terms of the ring properties,
in \citet{genzel14_rings} the reported gas densities in the rings 
(e.g. their Fig.~23)
are $\Sigma_{\rm gas} \!\sim\! 10^{8.6-9.0} \msun\kpc^{-2}$. 
This is higher than the average in the simulated rings and similar to the 
peak densities in the giant clumps within the simulated rings.
The corresponding SFR densities deduced from the observed rings are
$\Sigma_{\rm SFR}\!\sim\! (1\!-\!2) \msun\yr^{-1}\kpc^{-2}$.
This is again higher than the average across the simulated rings and
comparable to the SFR densities in the simulated giant clumps.
This difference may be partly due to the fact that the observed galaxies of
$\Ms\!\sim\!10^{10.0-11.5}\msun$ are systematically more massive than the
simulated galaxies with significant rings where $\Ms \ssim 10^{9.5-11.0}$
(\fig{app_prop_dist} in appendix \se{app_ring_properties}).
It may also reflect the imperfection of the VELA-3 suite of simulations used 
here, which tend to underestimate the gas fractions  
(\fig{app_prop_dist}). 
This is largely due to the relatively weak feedback incorporated, 
which leads to overestimated SFR at high redshifts.

\smallskip
\adb{
Using AO-SINFONI $H_\alpha$ spectroscopy as well as HST WFC3 multi-wavelength 
imaging for 22 $z\ssim 2.2$ star forming galaxies, \citet{tacchella15} revealed 
an inside-out quenching process, with the massive galaxies showing a pronounced
peak in their stacked sSFR profile at $5\sdash10\kpc$ (their Fig.~1), 
indicating rings, surrounding massive quenched bulges.
Then, analyzing dust obscuration in ten such galaxies, 
\citet{tacchella18} detected rings in the dust-corrected surface-density
SFR profiles (their Fig.~8), making sure that the SFR rings are real and not
an artifact of missing SFR in the observed central regions. 
The corrected SFR surface densities are in the range
$0.1\sdash 1\msun\kpc^{-2}$, similar to the simulation results. 
}

\subsubsection{In HST images}
\label{sec:candels}

\adb{
In a pioneering study,
\citet{elmegreen06_ring} showed HST-ACS $V_{606}$-band images of 24 
galaxies in the GEMS and GOODS surveys, in the redshift range $0.4\sdash 1.4$.
They detected 9 rings and 15 partial rings, each containing a few giant clumps
and surrounding a massive bulge, 
with no obvious bars or grand-design spiral structure.  
Several observed ``chain" galaxies, showing similar clumps,
were identified as the edge-on analogs of the more face-on ring galaxies.
}

\smallskip
The general impression from crude visual inspections of the more recent
CANDELS-HST galaxies used to be that they are not showing rings, 
somehow ignoring the \citet{elmegreen06_ring} results and
leading to a common notion that blue and red nuggets tend to be ``naked".
Naked nuggets were indeed reported at $z\ssim 1\sdash 2$
\citep[e.g.][]{williams14,lee18}.
This impression might have emerged from rest-frame optical images, focusing on
old stars, rather than rest-frame UV that reflects star formation.
In contrast, deeper CANDELS images focusing on rest-frame UV 
did show hints for star-forming rings around massive 
bulges (e.g., J.~Dunlop, private communication). 
Recall that in \se{nuggets}, we found in the simulations that 
while $\sim\!40\%$ of the nuggets are expected to be totally naked, another 
$\sim\!40\%$ of the nuggets are expected to have significant rings of 
$\mu_{\rm ring}\sgt 0.3$.

\smallskip
Indeed, an ongoing search in the deeper GOODS fields
(Ji, Giavalisco, Dekel et al., in prep.)
reveals many star-forming rings about massive bulges.
The rings are visually identified in the F606W bandpass, 
which is deep enough and relatively sensitive to young stars at $z\!\sim\!1$,
and the images are then studied in the complementary  
F850LP and F160W filters,
the latter capturing older stars.
Our preliminary inspections indicate that, among the galaxies of 
$\Ms\!>\!10^{10.5}\msun$ at $z\!=\!0.5\!-\!1.2$,
a non-negligible fraction of order $\sim\! 10\%$ clearly show blue
star-forming, clumpy rings, typically surrounding a massive bulge, 
which is either star forming or quenched.
Clearly, this detected fraction of rings is a far lower limit, limited to 
galaxies of low inclinations, sufficiently extended rings, high-contrast rings
and bright enough rings that are more easily detected at $z\!\lsim\! 1$. 
Furthermore, a large fraction of the galaxies were eliminated from the 
ring search based on a pronounced spiral structure, while our theoretical 
understanding in \se{ring_toy} is that the spiral structure is intimately 
linked to the presence of a ring.

\smallskip
In our simulations, we read from \fig{mu_bn} that 
$\sim\! 24\%$ of the galaxies are expected to have pronounced rings of
$\mu_{\rm ring} \sgt 0.5$.
Focusing on massive galaxies at $z\ssim 1$, we read from the colors in
\fig{app_f_Mz_bins} in the appendix 
that the fraction relevant to the range of masses and
redshifts where the observations were analyzed is closer to $\sim\! 30\%$.
In the range $z\seq 1.4\sdash 2$, this fraction is reduced to $\sim\! 10\%$.
According to \fig{mu_Mz_bins},
the average ring strength in massive galaxies at $z\ssim 1$ is indeed just
below $\mu_{\rm ring} \seq 0.5$. 
Given the underestimate in the observational detection of rings, the numbers
in the simulations and the observations may be in the same ball park.

\smallskip
Visualization of such observed rings is provided by \fig{obs_rgb}, 
which displays four preliminary example rgb images of CANDELS 
galaxies showing rings, 
with masses $\Ms\!>\!10^{10.8}\msun$ at $z\!=\!0.58\!-\!1.22$
as marked in the figure.
\Fig{app_obs_4x4} in appendix \se{app_more_figures}
shows the same galaxies in the three filters separately.
These observed images show rings that qualitatively resemble the 
mock images from the simulated galaxies shown in \fig{mock_rgb}.
The average F606W surface brightness in the rings is 
$24.6$, $24.7$, $25.6$, $25.5 \,{\rm mag}\,{\rm arcsec}^{-2}$ in rings id 
9485, 13601, 16616, 18419 respectively.
These are in the ball park of the 
$\sim\!24\,{\rm mag}\,{\rm arcsec}^{-2}$ of the most pronounced mock rings
shown in \fig{mock_profiles}, given the uncertainties in the simulations,
their mock images and the way the ring surface brightness is estimated
as well as in the surface brightness deduced for the observed rings.
Both the simulations and the observations show massive bulges, 
though two of the observed bulges are blue
while all the four displayed simulated bulges are redder.
We recall from \se{nuggets} that among the significant simulated rings 
with central nuggets about one half are red nuggets,
consistent with what is indicated observationally.
 
\smallskip
These pictures of observed high-redshift rings in CANDELS are just a sneak 
preview of a detailed analysis.
The challenge to be addressed is to evaluate the effect of dust on 
the appearance of rings and bulges in these images.
If these rings are real and not artifacts of dust absorption in the inner
regions, they would be qualitatively
compatible with the $H_\alpha$ observations of Genzel et al. and
along our theoretical understanding  
of extended star-forming gas rings about blue nuggets or red nuggets.
Complementary spectroscopic studies would explore the ring kinematics 
and dynamics.

\begin{figure*} 
\centering
\includegraphics[width=1.01\textwidth,trim={{0.02\textwidth} 0 0 0}]
{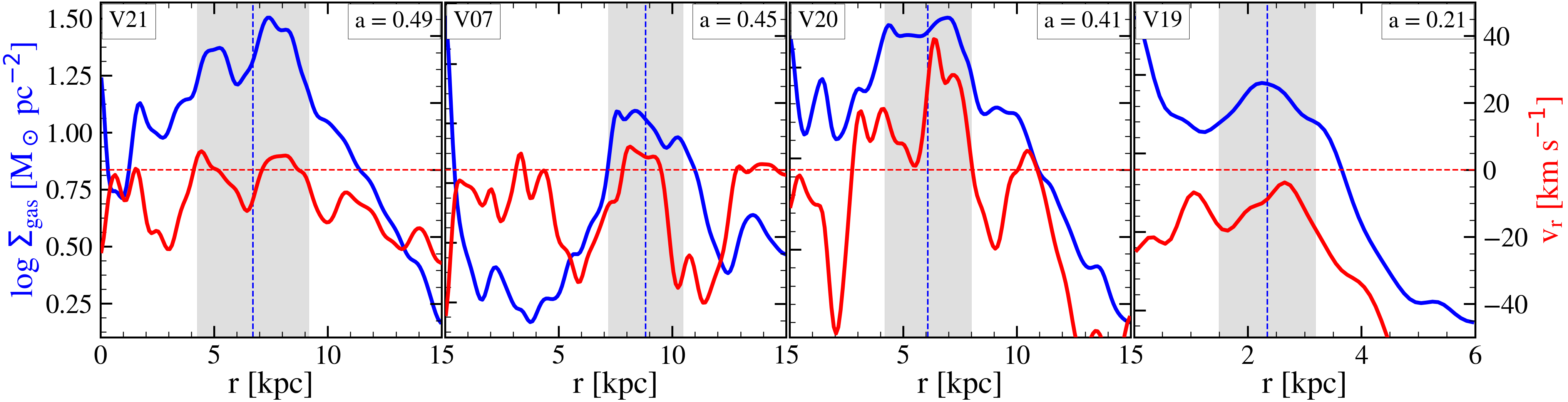}
\caption{
\adb{
Testing whether the rings may reside at resonant radii.
Shown (in red) is the radial profile of radial velocity in the disc plane,
mass-weighted averaged over circular rings, for the four cases of strong
rings in the VELA simulations.
Shown in comparison (in blue) is the gas surface density profile,
with the ring radius and width of its main body marked by a vertical line and
a shaded area.
In all four cases, we see inward averaged radial velocity both interior
and exterior to the ring, indicating that the rings do not reside
at CR or at the OLR.}
}
\label{fig:vr}
\end{figure*}

\subsubsection{At low redshifts}

\smallskip
Interestingly, \citet{salim12} found that most low-redshift ``Green-Valley" 
galaxies, at the early stages of their quenching process,  
or S0 type galaxies,
consist of massive quenched bulges surrounded by star-forming rings that
are seen in rest-frame UV.
Even closer to home, M31 and the Sombrero galaxy are known to show very 
pronounced dusty rings in IR surrounding a massive stellar bulge and disc.
While our simulations refer more explicitly to high-redshift, stream-fed
gaseous, un-relaxed galaxies in which the conditions for ring formation 
may be different from the more secular $z\!=\!0$ galaxies, 
we note that the high-$z$ rings tend to appear 
after the major compaction events, which are the early stages of quenching, 
namely in the Green Valley \citep{tacchella16_ms}. 
Our general analysis in \se{ring_toy} of ring stabilization by a central 
mass through a low $\delta_{\rm d}$, and the considerations based on $\tinf$
versus $\tacc$ and $\tsfr$, 
may be relevant with some modifications also for 
explaining the longevity of the low-redshift rings. 
\adb{
Some of the low-redshift rings may be the descendants of rings that formed 
after an earlier compaction event and remained in the Green Valley due to
the continuous star formation in the long-lived rings.
}
\adb{
This is alongside the traditional modeling of rings at resonances in secular 
discs or as outcomes of head-on collisions.  
}

\section{\adb{Discussion: Other Mechanisms for ring formation}}
\label{sec:discussion}

We discussed here a picture where the high-redshift rings are long-lived
dynamic features that are fed by high-AM cold gas streams, where their
torque-driven mass transport inwards is suppressed by a massive central body
and where the gas is depleted efficiently from the interior disc
by star formation and outflows.
Low redshift rings are commonly assumed to be resonant features in secular
evolution, or they are proposed to be formed by a collision with another galaxy
\citep[rings types O and P respectively,][]{few86}. 
We discuss here the potential applicability of these ring-forming
mechanisms to the high-redshift rings that we see in our simulations.

\subsection{\adb{Rings at Resonances in Secular Discs}}
\label{sec:resonances}

In our analysis in \se{ring_toy}, we assumed that the high-redshift discs,
which are continuously fed by intense streams and are growing massive bulges,
are subject to significant variations of the gravitational potential
and are not in their secular phase yet.
We further assumed that if the spiral structure has a semi-well-defined
pattern speed $\Omega_{\rm p}$, the gas angular velocity in the main disc
is larger, $\Omega(r) \sgt \Omega_{\rm p}$, namely the main disc is inside the
co-rotation radius (CR).
This is typically the case in low-$z$ grand-spiral galaxies, where the arms
are trailing and the gas is indicated to cross the arms from the inside, where
it is compressed and it forms stars. This is based, e.g., on the observation
that the IR light, characteristic of the first Myr of star formation,
is emitted from the inner, back side the arms, while the UV, associated with
stars of $\sim\!100\Myr$, tends to be emitted from the arms and their outer,
front side.
Inside CR, the torques by trailing arms induce AM transport
outwards, associated with mass transport inwards (\S 6.1.3.d of BT).

\smallskip
On the other hand,
open spiral structure, and especially strong bars, in relaxed discs undergoing
secular evolution at low redshifts,
where there is a constant pattern speed $\Omega_{\rm p}$,
may be responsible for rings of gas at resonance radii
\citep[e.g.][]{devaucouleurs59,buta96,buta17a,buta17b}. %
Is it possible that the rings we see in the high-z simulated galaxies reside
in resonance radii despite their unrelaxed nature?
One such resonance is at CR, where $\Omega(r)=\Omega_{\rm p}$,
and where no torques are acting to induce mass transport inward or outward.
More common at $z\seq 0$ is a ring at the Outer Lindblad Resonance (OLR),
at a radius outside co-rotation where $m\,[\Omega(r)-\Omega_{\rm p}]=-\kappa$
with the epicyclic frequency $\kappa^2=r\,d\Omega^2/dr +4\,\Omega^2$.
The torques by the spiral structure outside CR induce mass transport outwards,
pushing the gas to accumulate in the OLR.
The radial velocities interior to the ring and exterior to it
can thus indicate whether the rings may be residing in a resonance radius.

\smallskip
In order to address this distinguishing feature,
\fig{vr} shows the radial profile of radial velocity in the disc plane,
mass-weighted averaged over circular rings, for our four fiducial
cases of strong rings in the VELA simulations.
Shown in comparison is the gas surface density profile, with the ring
radius and width of its main body marked.
In all four cases, we see inward averaged radial velocity both interior
and exterior to the ring. Naturally, when inspecting the 2D maps of radial
velocity, there are some angular directions where an outward velocity is
detected interior or exterior to the ring. This is as expected from a rather
perturbed, turbulent disc.
The radial velocities roughly range from $-100\kms$ to $+50\kms$ (where the
general rotation velocity is $\sim\!300\kms$).
We interpret the fact that the average radial velocity is inwards
both interior and exterior to the ring as an indication for the rings not
to reside at resonant radii, neither at CR nor at the OLR.

\smallskip
The simulated galaxies at high redshift typically do not show massive bars or
grand-design spiral structure.
This is consistent with the observed low abundace of bars at high redshifts.
According to an analysis of the COSMOS field by \citet{sheth08},
while about $30\%$ of the $z=0$ disc galaxies show strong bars ($65\%$ show
any bars), this fraction drops to $10\%$ ($20\%$) already at $z\seq 0.84$.
Indeed, \citet{elmegreen06_ring} report the absence of obvious bars or open
spiral structure in the ring galaxies that they were the first to identify
in the GOODS fields at $z= 0.4\sdash 1.4$.
This implies that the most likely drivers of resonance rings, namely bars or
open spirals, are missing in the high-redshift ring galaxies.
Nevertheless,
possible tidal effects from a msssive companion, or oval distortions,
as alternatives to a bar in driving resonances \citep{buta96},
are not strictly ruled out in all cases.

\smallskip
Additional evidence not supporting the possibility that the simulated
high-redshift rings are associated with resonant radii is provided by the
facts that %
(a) the metallicity in the simulated rings is lower than in the
interior disc as seen in \fig{profiles_rings}, hinting for an external origin
of the gas,
(b) the rings are actually seen to be fed by inflowing cold streams, and
(c) the rings predominantly consist of gas as opposed to stars.

\smallskip
We can report yet another preliminary test for being inside the CR.
For discs at the early post-compaction phase, where one could identify
spiral structure, we plotted the spatial distribution of very young stars of
$0-60\Myr$ (assumed to be responsible for IR emission) versus that of somewhat
less young stars of $80-120\Myr$ (assumed to dominate the UV emission).
We find visual evidence
for the younger population to lie inside the arms, indicating that the gas is
crossing the trailing arm from the inside as it forms stars, which is
another evidence for being inside CR, similar to the cases of grand spirals at
low redshift.
A further, more detailed study of the high-z rings in the context of the
pattern speed and the resonant radii is beyond the scope of the current paper
and is deferred to a future study.

\subsection{\adb{Collisional Ring Galaxies}}
\label{sec:collisons}

Rings could be nearly symmetrical density waves that are driven into a disc as
a result of a bulls-eye collision with another galaxy, as reviewed by
\citet{appleton96} following the original proposals
\citep{lynds76,theys77, toomre78}.
The prototypical low-redshift example is the Cartwheel galaxy, where a suspect
perturbing galaxy is possibly identified in the image.
Ring galaxies in cosmological simulations have been widely studied in the
context of collisions
\citep[e.g.][]{donghia08,snyder15_rings,renaud18,elagali18}. %
The collision scenario for ring formation is supposed to be associated  %
with inside-out ring buildup from stars and gas, an outward ring velocity, %
and a relatively short ring lifetime of a few hundred $\Myr$ %
\citep[e.g.][]{mapelli08a,mapelli08b}. %
Recently, a collision has been proposed as a possible interpretation for an
observed ring
galaxy at $z\seq 2.19$ \citep{yuan20}, largely based on marginal evidence
for an outward radial velocity of the ring.

\smallskip
Most of the ring galaxies seen in our simulations at high redshifts are not
likely to be driven by collisions for several reasons, as follows.
First, the simulated rings typically do not show outward radial velocities.
In this respect galaxy V20 shown in \fig{vr} at $a=0.41$ is an exception,
with the ring moving outward at $\leq\! 40\kms$.
Instead, the rings (including V20) show inward velocities interior and exterior
to the ring, as seen in \fig{vr}, while a collision origin would imply outward
velocities also interior to the ring.
Second, the metallicity in the rings is lower than in the disc interior to the
rings, as seen in \fig{profiles_rings},
indicating that they do not emerge from the inner disc but are rather
fed by external fresh accretion.
Indeed, the rings are typically seen to be associated with visible incoming
cold streams.
Third, the simulated rings are primarily gaseous, while the collisional
rings are expected to contain a significant component of stars as well.
Finally, some of the rings are long lived, beyond the relatively short
lifetimes expected for collisional rings.

\smallskip
Still, since we have demonstrated that the longevity of the rings in our
simulations is associated with compaction-driven central masses,
and since a significant fraction of the compaction events are actually
trigerred by mergers, our proposed scenario is also associated with mergers,
even if in an indirect way. This calls for a further study of the interplay
between the scenario discussed here and the collision scenario for ring
formation.

\subsection{\adb{The effect of AGN feedback on rings}}
\label{sec:agn}

The robustness of our results based on the VELA simulations should be examined
by analyzing rings in other cosmological simulations, based on different codes
with different resolutions, and different subgrid recipes for processes such as
star formation and feedback. 

\smallskip
In particular,
since the current VELA simulations on which our ring analysis is based  
do not include black holes and the associated AGN feedback, one may worry 
whether AGN feedback could damage the rings and possibly eliminate them 
altogether. Indeed, strong AGN activity is predicted and observed in
star-forming galaxies above the golden mass, where extended rings form
\citep[e.g.][]{forster18b,dekel19_gold}. 

\smallskip
In an ongoing work, we are searching for and analyzing high-redshift rings 
in cosmological simulations that do include AGN feedback.
For example, the TNG simulations that do incorporate rather strong
AGN feedback would be appropriate for such an exploration of the AGN effects
on the rings. We can report that a preliminary visual inspection of images of 
gas and SFR surface density of TNG galaxies at $z \ssim 1 \sdash 2$ indicates 
a large abundance of rings, which seem to largely resemble the VELA rings 
studied here. 
This indicates that the rings survive the AGN feedback, possibly because
the ejected AGN energy and momentum are largely collimated in the polar 
disc and ring direction.

\smallskip
The significant abundance of observed rings above the golden mass
at $z \ssim 1-2$, as summarized in \se{obs}, by itself provides evidence for 
the survivability of the rings under AGN feedback.
The theoretical and observational aspects of the interplay between AGN activity
and the rings are to be quantitatively explored.

\section{Conclusion}
\label{sec:conc}

In \citet{dekel20_flip} we argued, analytically and using simulations,
that galactic gas discs are likely to be long-lived only in dark-matter 
haloes of 
mass above a threshold of $\sim\!2 \times 10^{11}\msun$, corresponding to a 
stellar mass of $\sim\!10^9\msun$, with only little dependence on redshift.
In haloes of lower masses, the gas does not tend to settle into an extended
long-lived rotating disc, as several different mechanisms act to drastically
change the angular momentum and thus disrupt the disc.
First, the AM is predicted to flip on a timescale shorter than 
the orbital timescale due to mergers associated with a pattern-change in the
cosmic-web streams that feed the galaxy with AM.
Second, in this pre-compaction low-mass regime \citep[e.g.][]{zolotov15},
violent disc instability exerts torques that drive AM out and
mass in, thus making the disc contract in a few orbital times
\citep[e.g.][]{dsc09}.
Third, in this regime the central dark-matter and stellar system tend 
to be prolate \citep[e.g.][]{tomassetti16}
and thus capable of producing torques that reduce the AM
of the incoming new gas.

\smallskip 
Furthermore,
supernova feedback is expected to have a major role in disrupting discs below 
the critical
mass. This emerges from a simple energetic argument that yields an upper limit 
of $\Vv\!\sim\!100\kms$ for the dark-matter halo virial velocity 
(i.e. potential well) within which supernova feedback from a burst of star 
formation could be effective in significantly heating up the gas \citep{ds86}.
Supernova feedback stirs up turbulence that puffs up the disc
and it suppresses the supply of new gas with
high AM \citep{tollet19}, 
possibly even ejecting high-AM gas from the disc outskirts.
As argued in section 3 of \citet{dekel20_flip},
supernova feedback determines the mass dependence of the
stellar-to-halo mass ratio that enters into the merger rate and thus affects
the frequency of disc disruption by spin flips.
Finally,
supernova feedback has a major role in confining the major compaction events
to near or above the golden mass \citep{dekel19_gold}, 
and, as shown in the current paper, these compaction events
are responsible for the formation and longevity of extended rings.  

\smallskip 
Above this golden mass, the disruptive mergers are less frequent and are not
necessarily associated with a change in the pattern of the feeding streams, 
allowing the discs to survive for several orbital times.
In parallel, the effects of supernova feedback are reduced due to the depth of
the halo potential well.

\smallskip 
The main issue addressed in this paper is the post-compaction formation of 
long-lived rings above a critical mass, similar to the golden mass
for supernova feedback and merger-driven disc flips.
We showed using the simulations that in the post-compaction regime, 
typically after $z\ssim 4$,
the inflowing high-AM streams from the cosmic web settle into 
extended discs that evolve into long-lived rings.
Using measures of ring strength in each simulated galaxy, such as contrast and
mass fraction, we quantified the tendency of the rings to appear after 
the major compaction events and above the corresponding mass threshold, 
and showed that their strength is growing with time and mass with respect to 
the blue-nugget phase.

\smallskip 
In order to understand the ring longevity, we have worked out the torques 
exerted by a tightly wound spiral structure on the disc outskirts. 
We found that the timescale for inward mass transport for a ring of
constant relative width is roughly 
$\tinf \ssim 6\, \delta_{{\rm d},0.3}^{-3}\, \torb$, 
and the spiral pitch angle is given by $\tan\alpha \ssim \delta_{\rm d}$,
where $\delta_{\rm d}$ is the cold-to-total mass ratio interior to the ring.
By comparing this to the timescales for external accretion and interior SFR, 
$\tacc$ and $\tsfr$, requiring ring replenishment $\tinf \sgt \tacc$ and
depletion of the interior, $\tinf \sgt \tsfr$,
we learned that a ring forms and survives when $\delta_{\rm d} \slt 0.3$.
The required low values of $\delta_{\rm d}$ are most naturally due  
to the post-compaction massive bulge.
A similar extended long-lived ring would appear about a massive dark-matter 
dominated central region, which could be another reason for a reduced 
$\delta_{\rm d}$.
\adb{There is a lower bound on the values of $\delta_{\rm d}$ that allow
rings when the low values are driven by a low gas fraction,
because the ring phenomenon is primarily gaseous.}
Once the ring develops a high contrast, 
the inward transport rate becomes longer than the Hubble time and all other 
relevant timescales. The ring remains in tact but it gradually weakens due to
the weakening accretion rate with cosmological time and the gradual 
ring depletion into stars.
The long-lived ring could be Toomre unstable, with giant clumps forming stars,
as long as it is fed by high-AM cold gas streams.

\smallskip 
\adb{
The high-$z$ rings seen in our simulations are unlikely to be associated
with resonant radii as in secular discs at low redshifts because the high-$z$
galaxies are not in a secular phase, they show no bars or open spirals,
the radial velocities interior and exterior to the ring tend to be inwards,
the ring gas is indicated to come from external accretion based on its
low metallicity and the association with inflowing streams, and
the rings are predominantly gaseous, not stellar.
The simulated rings are also not likely to originate from collisions a la the
Cartwheel galaxy based on the very same properties, including the fact that
in most cases the rings are not moving outward, and they live for more than a 
few hundred $\Myr$. %
}

\smallskip
In order to allow first crude comparisons of the simulated rings-about-bulges
to observations, we generated mock images from ringy simulated galaxies  
that mimic multi-color HST images in CANDELS deep fields including dust
extinction.
The pronounced rings at $r\!\sim\!10\kpc$ are expected to form stars at a 
surface density of $\Sigma_{\rm SFR}\!\sim\!(0.1-1)\msun\yr^{-1}\kpc^{-2}$.
This corresponds at $z\ssim 1$ to an average surface brightness of 
$\sim\! 24\,{\rm mag}\,{\rm arcsec}^{-2}$ in the F606W filter,
corresponding to rest-frame UV, with weak dust extinction.
We also showed mock ALMA images of CO(2-1) emission, indicating that
$z\!\sim\!1$ rings would be detectable but at a low signal to noise 
with 10h of ALMA observations. A ring at $z\!\sim\!0.5$ would be detectable 
at a higher signal to noise even with a few-hour exposure. 

\smallskip
Observational studies including H$\alpha$ kinematics 
\citep[][and work in progress]{genzel14_rings,genzel17}
show gaseous star-forming clumpy rings around massive bulges or dark-matter 
dominated centers in a significant fraction of $z\! \sim\! 1\sdash 2$ galaxies 
above the threshold mass. 
This is qualitatively consistent with our theoretical understanding that a 
massive central mass is expected to support an extended ring for long times. 
It is also along the lines of the prediction of major compaction events
that generate massive bulges typically above a similar threshold mass.
In our simulations we find that for massive galaxies with pronounced
rings, the baryon-to-DM ratio interior to the ring ranges from 0.4 to 2.5.

\smallskip
We provided a sneak preview of an ongoing study of rings
in the deepest fields of the HST-CANDELS multi-color imaging survey
(Ji et al., in preparation).
The sample galaxies shown qualitatively resemble the mock images from the 
simulations, with star-forming clumpy rings about massive bulges.
Our preliminary results indicate that, indeed, when observed deep enough,
a non-negligible fraction of the galaxies of $\Ms\!>\!10^{10.5}\msun$ at
$z\!\sim\!0.5\!-\!3$ show blue rings about massive bulges.

\smallskip
In our simulations, strong rings typically have nuggets in their $1\kpc$
centers, of which roughly half are star-forming blue nuggets and the other half
are quenched red nuggets. Among the nuggets, about half are naked, and the
other half are surrounded by significant rings. There are preliminary
indications that these predictions are in the ball park of the observed ring
and nuggets populations, but this is a subject for future studies.

\section*{Acknowledgments}

We acknowledge Greg Snyder and Raymond Simons for the CANDELIZED mock images.
We thank Francoise Combes, Jim Dunlop, Sandy Faber, Reinhard Genzel, 
David Koo, Gary Mamon, Christophe Pichon, Samir Salim and Sandro Tacchella
for stimulating interactions.  
This work was partly supported by the grants 
Germany-Israel GIF I-1341-303.7/2016, Germany-Israel DIP STE1869/2-1
GE625/17-1, I-CORE Program of the PBC/ISF 1829/12, 
US-Israel BSF 2014-273, and NSF AST-1405962.
The cosmological VELA simulations were performed at the National Energy
Research Scientific Computing Center (NERSC) at Lawrence Berkeley National
Laboratory, and at NASA Advanced Supercomputing (NAS) at NASA Ames Research
Center. Development and analysis have been performed in the astro cluster at
HU.

\bibliographystyle{mn2e}
\bibliography{ring}

\vfill\eject
\appendix

\section{The VELA Cosmological Simulations}
\label{sec:app_vela}

The VELA suite consists of hydro-cosmological simulations of 34 moderately
massive galaxies. Full details are presented in \citet{ceverino14,zolotov15}.
This suite has been used to study central issues in the evolution of galaxies
at high redshifts, including compaction to blue nuggets and the trigger of
quenching
\citep{zolotov15,tacchella16_ms,tacchella16_prof},
evolution of global shape
\citep{ceverino15_shape,tomassetti16},
violent disc instability \citep{mandelker14,mandelker17},
OVI in the CGM \citep{roca19},
and galaxy size and AM \citep{jiang19_spin}.
Additional analysis of the same suite of simulations are discussed in
\citet{moody14,snyder15}.
In this appendix we give an overview of the key aspects of the simulations
and their limitations.

\subsection{The Cosmological Simulations}

The VELA simulations make use of the Adaptive Refinement Tree (ART) code
\citep{krav97,krav03,ceverino09}, which accurately follows the
evolution of a gravitating N-body system and the Eulerian gas dynamics using
an adaptive mesh refinement approach. The adaptive mesh refinement best
physical 
resolution is $17-35\pc$ at all times, which is achieved at densities of
$\sim10^{-4}-10^3\cmc$.
Beside gravity and hydrodynamics, the code incorporates physical process
relevant for galaxy formation such as gas cooling by atomic hydrogen and
helium, metal and molecular hydrogen cooling, photoionization heating by the
UV background with partial self-shielding, star formation, stellar mass loss,
metal enrichment of the ISM and stellar feedback. Supernovae and stellar winds
are implemented by local injection of thermal energy as described in
\citet{ceverino09,cdb10} and \citet{ceverino12}. Radiation-pressure
stellar feedback is implemented at a moderate level, following
\citet{dekel13}, as described in \citet{ceverino14}.

\smallskip
Cooling and heating rates are tabulated for a given gas density, temperature,
metallicity and UV background based on the CLOUDY code \citep{ferland98},
assuming a slab of thickness $1\kpc$. A uniform UV background based on the
redshift-dependent \citet{haardt96} model is assumed, except at gas densities
higher than $0.1\cmc$, where a substantially suppressed UV
background is used
($5.9\times10^6 \erg\, {\rm s}^{-1} \cms\, {\rm Hz}^{-1}$)
in order to mimic the partial self-shielding of dense gas, allowing dense gas
to cool down to temperatures of $\sim300$K. The assumed equation of
state is that of an ideal mono-atomic gas. Artificial fragmentation on the cell
size is prevented by introducing a pressure floor, which ensures that the Jeans
scale is resolved by at least 7 cells \citep[see][]{cdb10}.

\smallskip
Star particles form in timesteps of $5 \Myr$ in cells where the gas density
exceeds a threshold of $1~\cmc$ and the temperatures is below
$10^4$K. Most stars ($>90\%$) end up forming at temperatures well
below $10^3$K, and more than half of the stars form near
$300$K in cells where the gas density is higher than $10~\cmc$.
The code implements a stochastic star-formation where a star particle with a
mass of $42\%$ of the gas mass forms with a probability
$P=(\rhog/10^3\cmc)^{1/2}$ but not higher than $0.2$.
This corresponds to a local SFR that crudely mimics
$\drhos \epsf \rhog/\tff$ with $\epsf \sim 0.02$.
A stellar initial mass function of \citet{chabrier03} is assumed.

\smallskip
Thermal feedback that mimics the energy release from stellar winds and
supernova explosions s incorporated as a constant heating rate over
the $40~\Myr$ following star formation.
A velocity kick of $\sim10\kms$ is applied
to $30~\%$ of the newly formed stellar particles -- this enables SN explosions
in lower density regions where the cooling may not overcome the heating without
implementing an artificial shutdown of cooling \citep{ceverino09}.
The code also incorporates the later effects of Type Ia supernova and
stellar mass loss, and it follows the metal enrichment of the ISM.

\smallskip
Radiation pressure is incorporated through the addition of a non-thermal
pressure term to the total gas pressure in regions where ionizing photons
from massive stars are produced and may be trapped. This ionizing radiation
injects momentum in the cells neighbouring massive star particles younger than
$5\Myr$, and whose column density exceeds
$10^{21}\cms$, isotropically pressurizing the star-forming
regions \citep[see more details in][]{agertz13,ceverino14}.

\begin{table*}
\centering
\begin{tabular}{@{}ccccccccccccc}
\multicolumn{13}{c}{{\bf Properties of the VELA galaxies}} \\
\hline
Galaxy & $\Mv$ & $\Ms$ & $\Mg$ & SFR & $\Re$ & $\Rd$ & $\Hd$ & $V_{\rm rot}$ & $\sigma$ & $e$ & $f$ & $a_{\rm fin}$\\
 & $10^{12}\msun$ & $10^{10}\msun$ & $10^{10}\msun$ & $\msun/\yr$ & kpc & kpc & kpc & km/s & km/s & & &\\

\hline
\hline
01 & 0.16 & 0.20 & 0.14 & 0.93 & 2.64 & 5.15 & 2.57 & 66.0 & 50.3 & 0.72 & 0.97 & 0.50 \\
02 & 0.13 & 0.16 & 0.12 & 1.81 & 1.43 & 6.37 & 3.57 & 71.6 & 39.1 & 0.81 & 0.98 & 0.50 \\
03 & 0.14 & 0.38 & 0.08 & 1.41 & 3.67 & 5.21 & 2.34 & 78.3 & 59.7 & 0.75 & 0.96 & 0.50 \\
04 & 0.12 & 0.08 & 0.08 & 1.73 & 0.45 & 5.71 & 2.79 & 23.3 & 54.3 & 0.96 & 0.88 & 0.50 \\
05 & 0.07 & 0.07 & 0.05 & 1.81 & 0.38 & 5.36 & 1.98 & 60.4 & 32.8 & 0.94 & 0.75 & 0.50 \\
06 & 0.55 & 2.14 & 0.33 & 1.05 & 20.60 & 2.53 & 0.42 & 221.1 & 48.1 & 0.56 & 1.00 & 0.37 \\
07 & 0.90 & 5.75 & 0.79 & 2.85 & 18.13 & 12.59 & 2.06 & 285.5 & 71.4 & 0.85 & 1.00 & 0.54 \\
08 & 0.28 & 0.35 & 0.15 & 0.74 & 5.70 & 4.03 & 1.53 & 91.6 & 48.5 & 0.80 & 0.96 & 0.57 \\
09 & 0.27 & 1.03 & 0.29 & 1.74 & 3.57 & 7.34 & 2.12 & 152.9 & 36.0 & 0.99 & 0.85 & 0.40 \\
10 & 0.13 & 0.60 & 0.13 & 0.46 & 3.20 & 4.51 & 1.19 & 137.1 & 40.9 & 0.50 & 0.99 & 0.56 \\
11 & 0.27 & 0.76 & 0.33 & 2.14 & 8.94 & 8.34 & 5.08 & 121.3 & 71.1 & 0.90 & 0.80 & 0.46 \\
12 & 0.27 & 1.95 & 0.20 & 1.13 & 2.70 & 6.53 & 1.72 & 181.2 & 43.7 & 0.97 & 0.78 & 0.44 \\
13 & 0.31 & 0.57 & 0.35 & 2.48 & 4.48 & 9.74 & 4.75 & 131.7 & 41.3 & 0.97 & 0.88 & 0.40 \\
14 & 0.36 & 1.26 & 0.44 & 0.32 & 23.31 & 1.10 & 0.14 & 213.9 & 82.8 & 0.43 & 0.98 & 0.41 \\
15 & 0.12 & 0.51 & 0.08 & 1.07 & 1.35 & 6.26 & 1.08 & 110.2 & 37.9 & 0.80 & 0.98 & 0.56 \\
*16* & 0.50 & 4.09 & 0.50 & 0.61 & 18.46 & 6.05 & 0.99 & 269.4 & 104.8 & 0.37 &
0.98 & *0.24* \\
*17* & 1.13 & 8.48 & 1.11 & 1.36 & 61.37 & 7.70 & 1.10 & 288.6 & 180.6 & 0.43 &
0.99 & *0.31* \\
*19* & 0.88 & 4.49 & 0.57 & 1.22 & 40.46 & 1.55 & 0.12 & 257.2 & 91.9 & 0.70 &
0.99 & *0.29* \\
20 & 0.53 & 3.59 & 0.35 & 1.72 & 5.55 & 9.57 & 2.75 & 235.0 & 62.0 & 0.78 & 1.00 & 0.44 \\
21 & 0.62 & 4.05 & 0.43 & 1.73 & 7.89 & 9.48 & 1.18 & 261.6 & 42.9 & 0.52 & 1.00 & 0.50 \\
22 & 0.49 & 4.40 & 0.25 & 1.31 & 12.00 & 4.70 & 0.40 & 285.6 & 50.3 & 0.48 & 1.00 & 0.50 \\
23 & 0.15 & 0.76 & 0.13 & 1.16 & 3.06 & 6.28 & 1.54 & 133.0 & 49.2 & 0.78 & 0.99 & 0.50 \\
24 & 0.28 & 0.88 & 0.25 & 1.68 & 3.88 & 7.29 & 1.95 & 131.5 & 42.0 & 0.99 & 0.97 & 0.48 \\
25 & 0.22 & 0.69 & 0.08 & 0.73 & 2.29 & 5.70 & 0.82 & 93.9 & 71.3 & 0.80 & 0.99 & 0.50 \\
26 & 0.36 & 1.58 & 0.26 & 0.74 & 9.36 & 5.42 & 1.30 & 179.6 & 65.0 & 0.74 & 1.00 & 0.50 \\
27 & 0.33 & 0.71 & 0.29 & 1.98 & 6.10 & 9.16 & 4.97 & 122.8 & 60.4 & 0.25 & 0.98 & 0.50 \\
28 & 0.20 & 0.18 & 0.21 & 2.32 & 5.54 & 5.66 & 2.97 & 37.3 & 84.6 & 0.92 & 0.63 & 0.50 \\
29 & 0.52 & 2.29 & 0.32 & 1.89 & 16.83 & 7.46 & 0.97 & 185.6 & 108.4 & 0.96 & 0.91 & 0.50 \\
30 & 0.31 & 1.57 & 0.24 & 1.43 & 2.97 & 9.32 & 1.67 & 192.1 & 37.3 & 0.68 & 1.00 & 0.34 \\
*31* & 0.23 & 0.78 & 0.13 & 0.43 & 15.26 & 4.19 & 0.96 & 195.4 & 48.1 & 0.82 &
0.99 & *0.19* \\
32 & 0.59 & 2.66 & 0.43 & 2.58 & 14.86 & 4.98 & 1.06 & 195.4 & 56.4 & 0.84 & 1.00 & 0.33 \\
33 & 0.83 & 4.81 & 0.44 & 1.23 & 32.68 & 4.59 & 0.88 & 262.7 & 114.2 & 0.49 & 0.95 & 0.39 \\
34 & 0.52 & 1.57 & 0.44 & 1.84 & 14.47 & 5.29 & 1.87 & 156.9 & 70.8 & 0.29 & 1.00 & 0.35 \\
*35* & 0.23 & 0.56 & 0.25 & 0.33 & 22.93 & 1.13 & 0.30 & 204.4 & 40.4 & - & - &
*0.22* \\

\hline
\end{tabular}
\caption{Relevant global properties of the VELA 3 galaxies.
The quantities are quoted at $z=2$ ($a=0.33$) 
or at the final timestep $a_{\rm fin}$ when it is $<0.33$ 
(marked by stars).
$\Mv$ is the total virial mass.
The following four quantities are measured within $0.1\Rv$:
$\Ms$ is the stellar mass,
$\Mg$ is the gas mass,
SFR is the star formation rate,
and $\Re$ is the half-stellar-mass radius.
The disc outer cylindrical volume, as defined in \citet{mandelker14}, 
is given by $\Rd$ and $\Hd$, the disc radius and half height that contain 85\% 
of the gas mass within $0.15\,\Rv$.
$V_{\rm rot}$ and $\sigma$ are the rotation velocity and the radial velocity 
dispersion of the gas.
$e$ and $f$ are the shape parameters of the gas distribution, 
representing the ``elongation" and ``flattening" as defined in \citet{tomassetti16}.
$a_{\rm fin}$ is the expansion factor at the last output.
}
\label{tab:sample}
\end{table*}

\smallskip
The initial conditions for the simulations are based on dark-matter haloes that
were drawn from dissipationless N-body simulations at lower resolution in
cosmological boxes of $15-60\Mpc$. The $\Lambda$CDM cosmological model was
assumed with the WMAP5 values of the cosmological parameters,
$\omm=0.27$, $\oml=0.73$, $\omb=0.045$, $h=0.7$ and
$\sigma_8=0.82$ \citep{komatsu09}. Each halo was selected to have a
given virial mass at $z = 1$ and no ongoing major merger at $z=1$.
This latter criterion eliminated less than $10~\%$ of the haloes, those
that tend to be in a dense, proto-cluster environment at $z\sim1$.
The virial masses at $z=1$ were chosen to be in the range
$\Mv=2\times10^{11}-2\times10^{12}~M_{\odot}$, about a
median of $4.6\times10^{11}~M_{\odot}$. If left in isolation, the median mass
at $z=0$ was intended to be $\sim10^{12}~M_{\odot}$.

\smallskip 
The VELA cosmological simulations are state-of-the-art in terms
of high-resolution adaptive mesh refinement hydrodynamics and the treatment of
key physical processes at the subgrid level.
In particular, they trace the cosmological streams that feed galaxies at high
redshift, including mergers and smooth flows, and they resolve the violent disc
instability that governs high-$z$ disc evolution and bulge formation
\citep{cdb10,ceverino12,ceverino15_e,mandelker14}.
Like in other simulations, the treatments of star formation and feedback
processes are rather simplified. The code may assume a realistic SFR efficiency
per free fall time on the grid scale
but it does not follow in detail the formation of
molecules and the effect of metallicity on SFR.
The feedback is treated in a crude way, where the resolution does not allow
the capture of the Sedov-Taylor phase of supernova bubbles.
The radiative stellar feedback assumed
no infrared trapping, in the spirit of low trapping advocated by
\citet{dk13} based on \citet{krum_thom13},
which makes the radiative feedback weaker than in other simulations
that assume more significant trapping \citep{murray10,hopkins12b}.
Finally, AGN feedback, and feedback associated with cosmic rays and magnetic
fields, are not yet implemented. Nevertheless, as shown in
\citet{ceverino14}, the star formation rates, gas
fractions, and stellar-to-halo mass ratio are all in the ballpark of the
estimates deduced from observations.

\subsection{The Galaxy Sample and Measurements}
\label{subsec:sample}

The virial and stellar properties of the galaxies
are listed in Table~\ref{tab:sample}.
The virial mass $\Mv$ is the total mass within a sphere of radius
$\Rv$ that encompasses an overdensity of $\Delta(z)=
[18\pi^2-82\oml(z)-39\oml(z)^2]/\omm(z)$,
where $\oml(z)$ and $\omm(z)$ are the cosmological
parameters at $z$ \citep{bryan98,db06}. The stellar mass $\Ms$
is the instantaneous mass in stars within a radius of $0.2\Rv$, 
accounting for past stellar mass loss.

\smallskip
We start the analysis at the cosmological time corresponding to expansion 
factor $a=0.125$ (redshift $z=7$).
As can be seen in Table~\ref{tab:sample}, most galaxies reach $a=0.50$ ($z=1$).
Each galaxy is analyzed at output times separated by a constant interval in
$a$, $\Delta a=0.01$, corresponding at $z=2$ to $\sim100~\Myr$
(roughly half an orbital time at the disc edge).
The sample consists of totally $\sim 1000$ snapshots in the redshift range
$z=7-0.8$ from 35 galaxies that at $z = 2$ span the stellar mass range
$(0.2-6.4)\times10^{11}\Msun$. The half-mass sizes
$\Re$ are determined from the $\Ms$ that are measured within a
radius of $0.2\Rv$ and they range $\Re\simeq0.4-3.2\kpc$ at $z=2$.

\smallskip
The SFR for a simulated galaxy is obtained by
${\rm SFR}=\langle M_{\star}(t_ {\rm age}<t_{\rm max})/t_{\rm max} 
\rangle_{t_{\rm max}}$, where $\Ms(t_{\rm age}<t_{\rm max})$ is the mass
at birth in stars younger than $t_{\rm max}$.
The average $\langle\cdot\rangle_{t_{\rm max}}$ is obtained by averaging over
all $t_{\rm max}$ in the interval $[40,80]\Myr$ in steps of $0.2\Myr$.
The $t_{\rm max}$ in this range are long enough to ensure good statistics.
The SFR ranges from $\sim 1$ to $33 \Msun\yr^{-1}$
at $z\sim2$.

\smallskip
The instantaneous mass of each star particle is derived from its initial mass
at birth and its age using a fitting formula for the mass loss from the stellar
population represented by the star particle,
according to which 10\%, 20\% and 30\% of the mass is lost after
30 Myr, 260 Myr , and 2 Gyr from birth, respectively.
We consistently use here the instantaneous stellar mass, $\Ms$, and
define the specific SFR by
${\rm sSFR}={\rm SFR}/\Ms$.

\smallskip
The determination of the centre of the galaxy is outlined in detail in
appendix B of \citet{mandelker14}.
Briefly, starting form the most bound star, the centre is refined iteratively
  by calculating the centre of mass of stellar particles in spheres of
decreasing radii, updating the centre and decreasing the radius at each
iteration.  We begin with an initial radius of 600 pc, and decrease the radius
by a factor of 1.1 at each iteration. The iteration terminates when the radius
reaches 130 pc or when the number of stellar particles in the sphere drops
below 20.

\smallskip
The disc plane and dimensions are determined iteratively, as detailed in
\citet{mandelker14}. The disc axis is defined by the AM 
of cold gas ($T < 1.5 \times 10^{4}$K), which on average accounts for
$\sim 97\%$ of the total gas mass in the disc.
The radius $\Rd$ is chosen to contain $85\%$ of the cold gas mass in the
galactic mid-plane out to $0.15\Rv$, and the half-height $\Hd$ is defined to
encompass $85\%$ of the cold gas mass in a thick cylinder where both the
radius and half-height equal $\Rd$.
The ratio $\Rd/\Hd$ is used below as one of the measures of gas disciness.

\smallskip 
Another measure of disciness is the kinematic ratio of rotation velocity
to velocity dispersion $\Vrot/\sigma$.
The rotation velocity and the velocity dispersion are computed by
mass-weighted averaging over cells inside a cylinder whose minor axis is along
the AM direction of the cold gas ($T\!<\!4\times 10^4$K) within a
sphere of radius $0.1\Rv$. The cylinder radius is $0.1\Rv$ and its half-height
is $0.25 \Re$, where $\Re$ is the cold-gas half-mass radius
(more details in Kretschmer et al., in prep.).
The radial velocity dispersion is measured with respect to the mean radial
velocity within the cylinder.

\smallskip 
Relevant global properties of the VELA 3 galaxies at $z=2$ are listed
in \tab{sample} and explained in the caption. It includes the global
masses and sizes of the different components, the shape and kinematic
properties.

\smallskip 
We attempt to identify the major event of compaction to a blue nugget 
for each galaxy. This is the one that leads to a significant central gas
depletion and SFR quenching, and marks the transition from dark-matter to
baryon dominance within $\Re$. Following \citet{zolotov15} and
\citet{tacchella16_prof}, the most physical way to identify the compaction
and blue nugget is by the steep rise of gas density (and SFR) within the inner 
$1\kpc$ to the highest peak, as long as it is followed by a significant, 
long-term decline in central gas mass density (and SFR). 
The onset of compaction can also be identified as the start of the steep rise 
of central gas density prior to the blue-nugget peak.
An alternative identification is using the shoulder of the stellar mass density
within $1\kpc$ where its rise due to starburst associated with the compaction
turns into a plateau of maximum long-term compactness slightly after 
the blue-nugget peak of gas density. This is a more practical way to identify
blue nuggets in observations \citep[e.g.][]{barro17}.

\smallskip 
Major mergers in the history of each galaxy, for the limited purpose they are 
used here, are identified in a simplified way by following sudden increases in
the stellar mass. 


\section{Ring detection and properties in the simulations}
\label{sec:app_ring_properties}

\begin{figure*} 
\centering
\includegraphics[width=1.0\textwidth]
{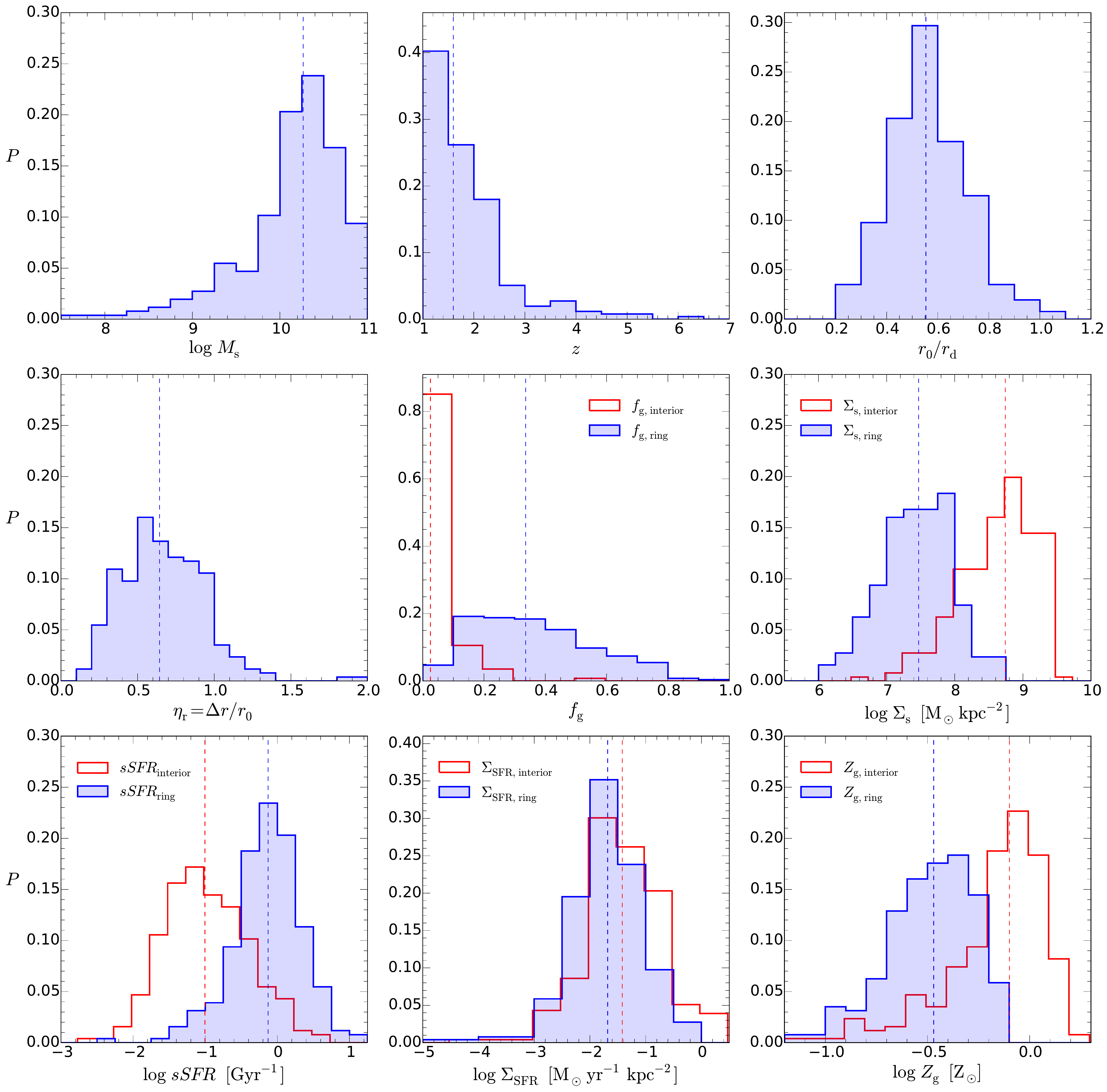}
\caption{
Distribution of ring properties for the significant rings in the simulations,
$\mu_{\rm ring} \sgt 0.3$.
From left to right, top to bottom:
Galaxy stellar mass $\Ms$ and redshift $z$ for these galaxies with rings.
Ring radius $r_0$ with respect to the disc radius,
and the relative ring width $\eta_{\rm r} \seq 2\sigma/r_0$. 
Then, for the ring (within $r_0\!\pm\! 1\sigma$)
and the interior ($r\slt r_0-2\sigma$),
gas fraction $f_{\rm gas}$, 
surface density of stellar mass $\Sigma_{\rm s}$,
specific SFR sSFR, surface density of SFR, 
and gas metallicity $Z_{\rm g}$.
}
\label{fig:app_prop_dist}
\end{figure*}

\begin{figure*} 
\centering
\includegraphics[width=1.0\textwidth]
{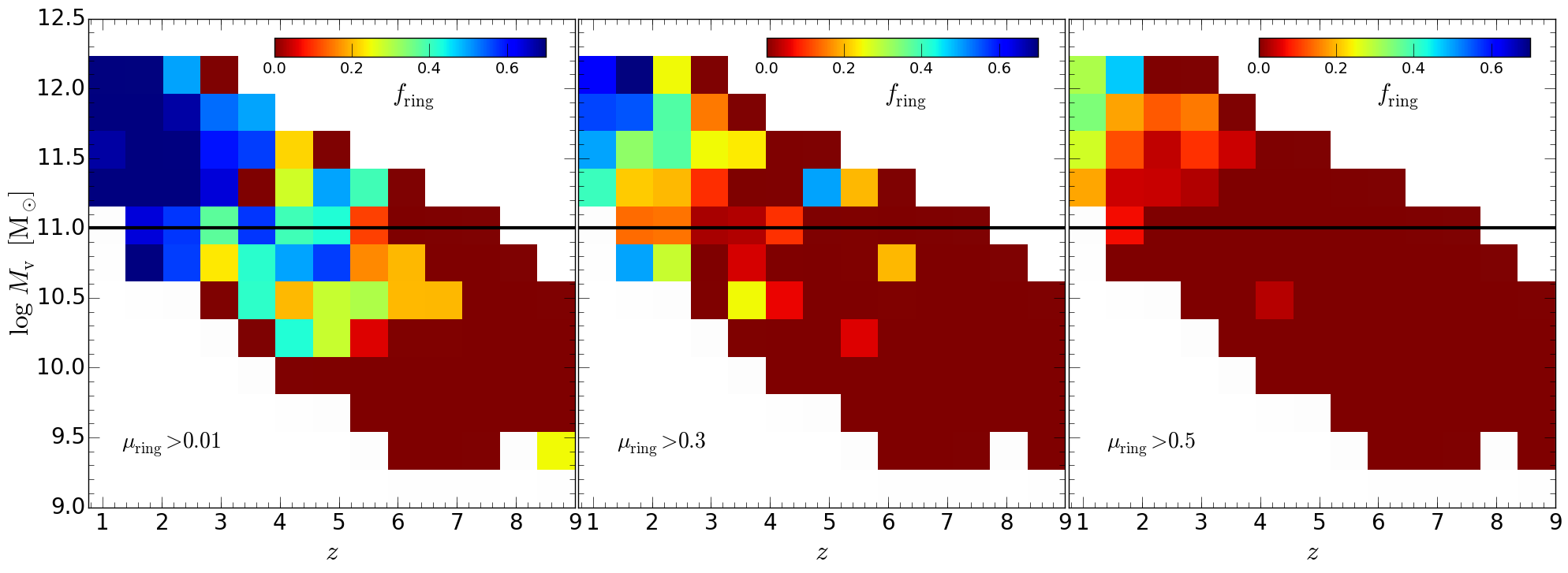}
\caption{
2D distributions of ring fractions
for the simulated galaxies in bins within the $\Mv\sdash z$ plane,
complementing \fig{f_Mz_bins}.
{\bf Left:} all rings with $\mu_{\rm ring}\!>\!0.01$.
{\bf Middle:} significant rings with $\mu_{\rm ring}\!>\!0.3$.
{\bf Right:} pronounced rings with $\mu_{\rm ring}\!>\!0.5$.
This complements the distribution of ring strength in \fig{mu_Mz_bins}.
We see that a high fraction of rings is detected above the threshold mass,
$\Mv \sgt 10^{11}\msun$, where discs survive spin flips (\fig{disc_Mz_bins}),
and at $z\slt 4$.
}
\label{fig:app_f_Mz_bins}
\end{figure*}

\begin{figure} 
\centering
\includegraphics[width=0.46\textwidth]
{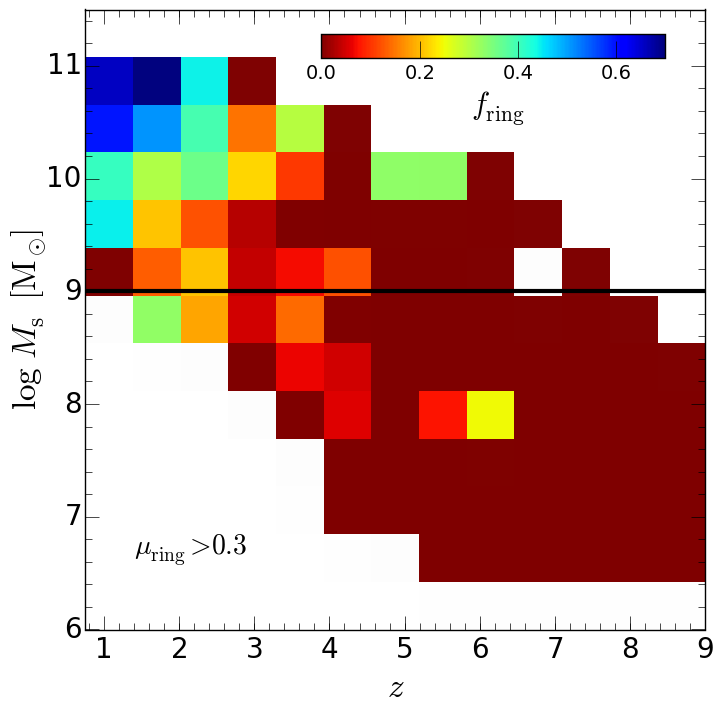}
\caption{
Same as \fig{f_Mz_bins}, the distribution of ring fraction for significant
rings with $\mu_{\rm ring}\sgt 0.3$, but in the plane $\Ms \sdash z$,
for easier comparison to observations.
}
\label{fig:app_f_Msz_bins}
\end{figure}

\subsection{Ring detection}  
\label{sec:app_ring_detection}

In order to measure ring properties for all the simulated galaxies,
we compute for each the gas surface density
profile $\Sigma(r)$ in the plane perpendicular to the angular momentum,
in rings spaced by $100\pc$, and fit to it a function with free parameters
that captures the main
ring with a Gaussian in linear $\Sigma$ versus linear $r$.
Three examples are shown in \fig{fits}, which illustrates the fits.
We consider the radius range between $r_{\rm min}=1.5\kpc$ and the outer
disc radius $\Rd$ (which is typically $\sim\!10\kpc$ or $\sim\!0.1\Rv$),
but limit the outer radius to be $<\!0.15\Rv$.

\smallskip
We first evaluate whether there is no ring, one ring or two rings.
For this we temporarily smooth the profile with a Gaussian window
of standard deviation $\sigma \seq \max\{0.005\Rv, 7.5\Delta x\}$,
where $\Delta x$ is the minimum cell size (varying between $17.5$ and $35\pc$),
giving values of $\sigma\seq (0.2-1)\kpc$.
We then search for maxima that are separated by more than 
$\max\{4\kpc,0.033\Rv\}$.
This also provides initial guesses for the ring radius $r_{\rm in}$,
the peak level $\Sigma_{\rm max}$ and the minimum level at the interior to
the ring $\Sigma_{\rm min}$.

\smallskip
In the case of a single ring, we fit to the raw profile a Gaussian with a
constant background,
\be
\Sigma(r)=\Sigma_0 +\Sigma_{g} \exp\left[-\frac{(r-r_0)^2}{2\sigma^2}\right]\, ,
\ee
with four free parameters.
The fit is performed by minimizing the sum of residuals in radii spaced by
$100\pc$. 
The radius $r_0$ is searched for in the range $r_{\rm in} \pm 1.5\kpc$.
The standard deviation $\sigma$ is allowed to range from
$0.008\Rv$ to $0.016\Rv$ (motivated by the values eventually obtained for the
ring full widths of $4\sigma$.
In the cases of combined double rings (see below)
the value of $\sigma$ can become as large as $0.033\Rv$.
The background level $\Sigma_0$ is not allowed to be smaller than
the minimum value $\Sigma_{\rm min}$ in the interior of the ring, 
such that it will not be biased low
by the background level at the exterior of the ring.

\smallskip
The ring at $r_0$ with a width $\pm \sigma$
can be characterized by its contrast
\be
\delta_{\rm ring} = \frac{\Sigma_{\rm ring}}{\Sigma_{\rm interior}}
= \frac{\Sigma_{\rm g}}{\Sigma_0} \, ,
\ee
ranging from $0$ for no ring,
through $\ll 1$ for a low-contrast ring,
$\gg 1$ for a high contrast ring,
to $\delta_{\rm ring} \rightarrow \infty$ for an ultimate ring with an
empty interior.

\smallskip
The gas mass of the ring $M_{\rm ring}$ is determined by integrating the
density profile in $(r_0-2\sigma,r_0+2\sigma)$ and subtracting the background
mass of surface density $\Sigma_0$.
The corresponding measure of ring strength is the gas mass excess 
\be
\mu_{\rm ring} = \frac{M_{\rm ring}}{M_{\rm gas}(<r_0+2\sigma)} \, ,
\ee
ranging from $\ll 1$ for a negligible ring to
$1$ for an ultimate ring with an empty interior.

\smallskip
In the case of two rings we fit a sum of two Gaussians,
with the same $\Sigma_0$.
If one of the rings is at least three times more massive than the other, 
we choose it as the dominant ring.
Otherwise, for about $10\%$ of the rings, we combine the two rings into one. 
The contrast is set to the average of the two contrasts.
The radius $r_0$ is set to the average of the two radii, $r_{01}<r_{02}$.
When computing $\mu_{\rm ring}$,
for non-overlapping rings the ring mass is the sum of the two ring masses,
and for overlapping rings (within $2\sigma$) 
the ring mass is integrated between $r_{01}-2\sigma_1$
and $r_{02}+2\sigma_2$, with the denominator integrated to
$r_{02}+2\sigma_2$.
The ring width, which is $2\sigma$ for each of the ring, is set to the sum of
the widths in the case of non-overlapping rings, and to half the interval
from $r_{01}-2\sigma_1$ to $r_{02}+2\sigma_2$ in the case of overlapping rings.

\smallskip
Finally, we remove rings with $r_0$ smaller than $400\pc$,
corresponding to four radial bins (four is the number of free parameters 
in the fit). 
We also remove rings when $\mu_{\rm ring} \! \leq \!0$.

\subsection{More ring properties} 
\label{sec:app_ring_prop}

To complement the presentation and discussion of ring properties, especially in
\se{ring_prop},
\fig{app_prop_dist} shows the probability distribution of certain other ring 
properties of interest, as measured by the Gaussian fits in the sample of 
VELA galaxies with significant rings, $\mu_{\rm ring} \sgt 0.3$.

\smallskip
The stellar masses of the galaxies with significant rings are mostly in the 
range $\log \Ms/\msun \!\simeq\! 10.25 \pm 0.75$.
This reflects the preferred occurance of rings above the mass threshold for 
discs by the frequency of merger-driven spin flips, \fig{disc_Mz_bins},
and the characteristic mass
for major compaction into blue nuggets,
\figs{compaction} and (\ref{fig:vela_V}) 
\citep{zolotov15,tomassetti16}.
The main redshift range is $z\ssim 2.6$ with rings also fount out to $z\ssim 4$.
This is largely determined by our sample of galaxies that grow in time and stop
near $z\seq 1$, and reflects the tendency of rings to appear above a 
threshold mass. It also reflects the decrease of $\delta_{\rm d}$ in time due
to the overall decrease in gas fraction, 
\figs{f_Mz_bins} and (\ref{fig:delta_d_Mz_bins}). 

\smallskip
The ring radii are $r_0\!\simeq\! (0.5 \pm 0.3)\,\Rd$. 
They can define the outer edge of the disc, but in some cases they are in inner
radii.
The relative ring widths are 
$\eta_{\rm r} \seq 2\sigma/r_0 \!\simeq\! 0.67 \pm 0.4$.
This means that the rings could be narrow but in some cases they are rather
broad. Some of these broad rings are actually made of two rings.

\smallskip
The gas fraction in the ring ranges from $0.1$ to $0.8$, with the median at
$f_{\rm g} \!\simeq\!0.32$, while in the interior it is typically below $0.1$. 
This reflects the accumulation of gas in the ring while the interior has 
been depleted. 
The typical gas fraction in the ring is somewhat lower than observed for more
massive galaxies at $z\ssim 2$ 
\citep{tacconi10,tacconi13,genzel14_rings,tacconi18}, 
partly because the simulated galaxies are
systematically smaller, as seen in the distribution of $\Ms$, partly because
the simulated rings are at $z\ssim 1$, and partly because the gas fractions are
underestimated in this suite of simulations because of weak feedback that
allows too high SFR at higher redshifts.
The stellar surface density in the rings is 
$\log \Sigma_{\rm s} \!\simeq\! 7.5\pm 0.8$, significantly lower than in the
interior, where it is 
$\log \Sigma_{\rm s} \!\simeq\! 8.9\pm 0.8$ due to the massive bulge.

\smallskip
The sSFR in the ring is 
$\log {\rm sSFR} \!\simeq\! -0.1 \pm 0.5$, typical to the Main Sequence of 
star-forming galaxies at $z\ssim 1\sdash 2$. 
In the interior it is 
$\log {\rm sSFR} \!\simeq\! -1.0 \pm 1.8$, corresponding to both quenched 
red nuggets and star-forming blue nuggets. 
This is also seen in the SFR surface density $\Sigma_{\rm SFR}$, which could be
higher than in the ring for blue nuggets and lower for red nuggets.
In the rings it is
$\log {\rm sSFR} \!\simeq\! -1.6 \pm 1.0 \msun\yr^{-1}\kpc^{-2}$.

\smallskip
Complementing the metallicity profiles shown in \fig{profiles_rings}, 
the gas metallicity in the ring is $\log Z \!\simeq\! -0.5 \pm 0.3$.
This is significantly lower than the interior metallicity of 
$\log Z \!\simeq\! -0.1 \pm 0.2$, consistent with the notion that the ring 
gas is mostly accreted gas.

\smallskip
\Fig{app_f_Mz_bins}, complementing \fig{f_Mz_bins} 
for rings of different
strengths, shows the 2D distributions of ring fractions
for the simulated galaxies in bins within the $\Mv\sdash z$ plane.
{\bf Left:} all rings with $\mu_{\rm ring}\!>\!0.01$.
{\bf Middle:} significant rings with $\mu_{\rm ring}\!>\!0.3$.
{\bf Right:} pronounced rings with $\mu_{\rm ring}\!>\!0.5$.
This also complements the distribution of ring strength in \fig{mu_Mz_bins}.

\smallskip
\Fig{app_f_Msz_bins}, complementing \fig{f_Mz_bins}, 
shows the distribution 
of ring fraction for significant rings with $\mu_{\rm ring}\sgt 0.3$ 
but in the plane $\Ms \sdash z$ instead of $\Mv \sdash z$,
to allow a more straightforward comparison to observations.

\section{Torques by a prolate central body}
\label{sec:app_prolate}

\begin{figure} 
\centering
\includegraphics[width=0.46\textwidth]
{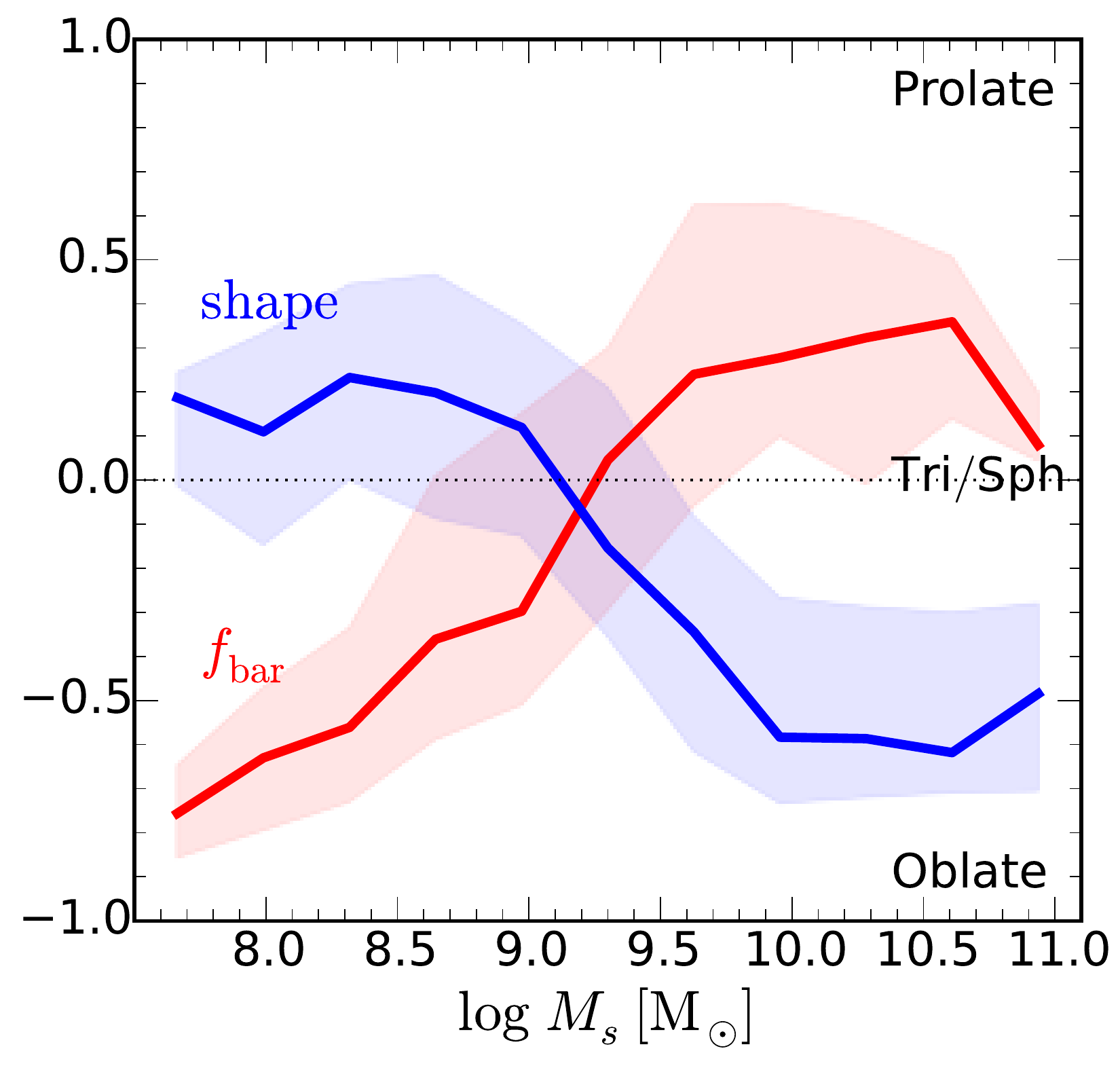}
\caption{
Evolution of shape ($e-f$, see text) of the stellar system (blue) as a
function of stellar mass for stacked VELA simulated galaxies
(median and $1\sigma$ scatter),
It shows a transition from prolate ($>0$) to oblate ($<0$)
at the critical mass for blue nuggets. A similar transition is seen when
plotted against time with respect to the blue-nugget event.
Also shown is the baryon-to-dark matter ratio within the effective radius
(red).  It demonstrates that the transition in shape is associated with a
transition from dark matter to baryon central dominance as a result of the
compaction to a blue nugget \citep{tomassetti16}.
}
\label{fig:app_shape}
\end{figure}

As mentioned in \se{vdi}, torques exerted by a VDI disc in the pre-compaction
stage below the critical mass for major compaction may cause AM loss and thus
shrinkage of the disc.
Another mechanism that could help disrupting a pre-compaction disc in a similar
way involves torques exerted by a central stellar system that is not
cylindrically symmetric, e.g., if it has a prolate shape.
Indeed, as can be seen in \fig{app_shape},
the VELA simulated galaxies tend to be prolate pre compaction
and oblate post compaction, showing a transition about the critical mass
for blue nuggets \citep{ceverino15_shape,tomassetti16}.
The 3D ellipsoidal shape can be measured by the parameters of
``elongation", $e\!=\![1-(b/a)^2]^{1/2}$, and
``flattening", $f\!=\![1-(c/a)^2]^{1/2}$, where $a\geq b \geq c$ are the
principal axes of the mass distribution.
The figure shows the difference $e\!-\!f$, which is useful for
characterizing the shape from extremely oblate at $e\!-\!f=-1$, through pure
triaxial at $e\!-\!f=0$, to extremely prolate at $e\!-\!f=+1$.
A similar transition has been deduced for the shapes of observed CANDELS
galaxies \citep{vanderwel14_shape,zhang19}.
The transition in shape can be explained by the transition from central
dark-matter dominance to baryon dominance as a result of the compaction to
a blue nugget \citep{tomassetti16}.
The prolate pre-compaction bulge may exert non-negligible torques that could
make a significant relative change in AM and thus
help disrupting the gas disc, as we very crudely estimate below.

\smallskip
For a very crude estimate, consider a central body of mass $\gamma M$,
exerting a torque on the mass outside it at a position $(r,\theta,\phi)$.
The factor $\gamma \leq 1$ is the fraction of mass in the body exerting the
torque with respect to the total mass $M$ within the sphere of radius $r$.
The torque per unit mass can be written as
\be
\tau = \mu(r,\theta,\phi)\, \frac{G \gamma M}{r^2} \, r \, ,
\label{eq:tau_mu}
\ee
where $\mu$ depends on the shape of the central body
and on the position where the torque is evaluated.
The change in specific AM caused by this torque
on a circular orbit of velocity $V$ at radius $r$ in a plane of $\phi=const.$,
acting from time $t_1$ to $t_2$,
is $\Delta j = \int_{t_1}^{t_2} \tau \dd t$.
Writing the specific AM in the orbit as $j=Vr$,
approximating $V^2 \simeq GM/r$,
and using ${\dd t} = (r/V)\, \dd \theta$ to relate time and angle,
with the corresponding angle $\theta_1$ and $\theta_2$, we obtain
\be
\frac{\Delta j}{j}=\gamma \int_{\theta_1}^{\theta_2} \mu(r,\theta) \,
\dd \theta \, .
\ee
Note that this expression does not depend on $M$; it depends only on
$\gamma$, the mass fraction in the body that exerts the torque.

\smallskip 
In order to obtain an upper limit for the possible effect, we consider an
extreme prolate system, a dumbbell, made of two equal point masses
separated by a distance $2d$ along the $z$ axis,
and consider a circular orbit of radius $r$ in a plane that includes the
dumbbell where $\phi$ is constant and $\theta$ is varying.
We obtain (for $r$ in units of $d$)
\bea
\mu(r,\theta) = 0.5\,r^2\,\sin\theta \ \
   [& (1+r^2-2r\cos\theta)^{-3/2} \nonumber \\
     &-(1+r^2+2r\cos\theta)^{-3/2} ] \, .
\eea
The torque is periodic, flipping sign in every quadrant of the orbit.
%
%
When evaluated from $0$ to $\pi/2$, the integral gives
\be
\gamma^{-1} \frac{\Delta j}{j} = \int_{0}^{2\pi} \mu(r,\theta) \, \dd \theta
= \frac{r^2}{(r^2-1)} - \frac{r}{(r^2+1)^{1/2}} \, .
\ee
If the dumbbell also dominates the potential, $\gamma=1$,
example values are $\Delta j/j = 5.02, 0.97, 0.44, 0.18$ at
$r/d= 1.1, 1.5, 2, 3$ respectively.
We learn that the relative change of AM during a quadrant of a circular orbit
about a dumbbell, before it flips sign in the following quadrant, can be
significant out to $r \sim 2d$ and beyond.

\smallskip
This crude estimate indicates that the effect of a very prolate central body
could in principle have a non-negligible effect on the survival of a disc.
However, this may be a severe over-estimate for the effect of a more
realistic prolate ellipsoid, which should be computed properly for a
general ellipsoid.

%

\smallskip 
Post compaction, the central body tends toward an oblate shape.
Such a body does not exert torques on masses orbiting in the major plane of the
oblate system, but it does exert torques off this plane.
An extreme oblate system, namely a uniform disc, yields values of
$\Delta j/j \sim  0.1$ per quadrant in a plane perpendicular to the major
plane of the body \citep[][Figure 16]{danovich15}.
This implies that the post-compaction central oblate body, above the critical
mass for blue nuggets, does not affect
significantly the AM of the disc, and is not capable of disrupting it.

\section{Complementary figures}
\label{sec:app_more_figures}

\begin{figure*} 
\centering
\includegraphics[width=0.87\textwidth]
{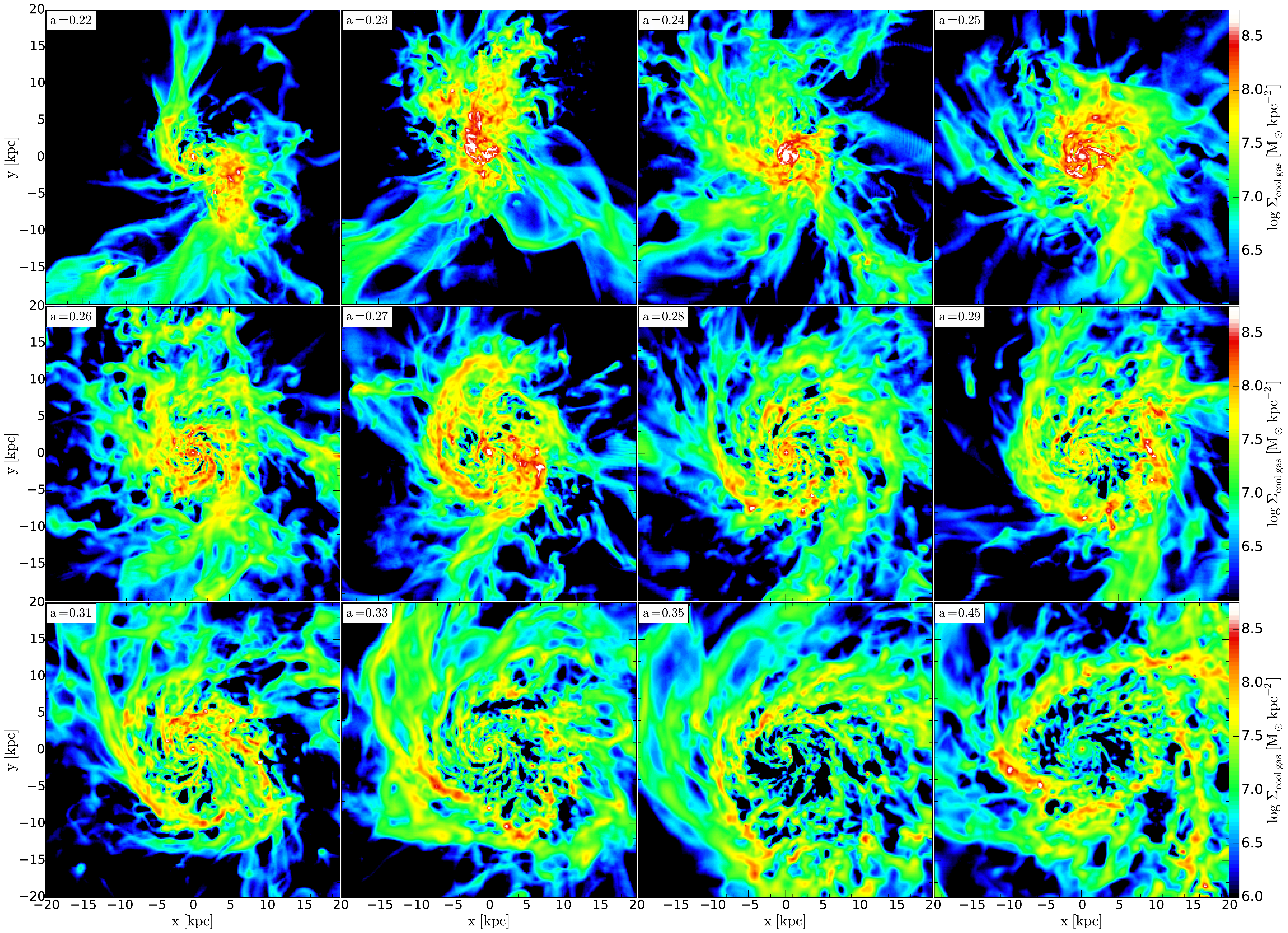}
\vskip -0.3cm
\includegraphics[width=0.87\textwidth]
{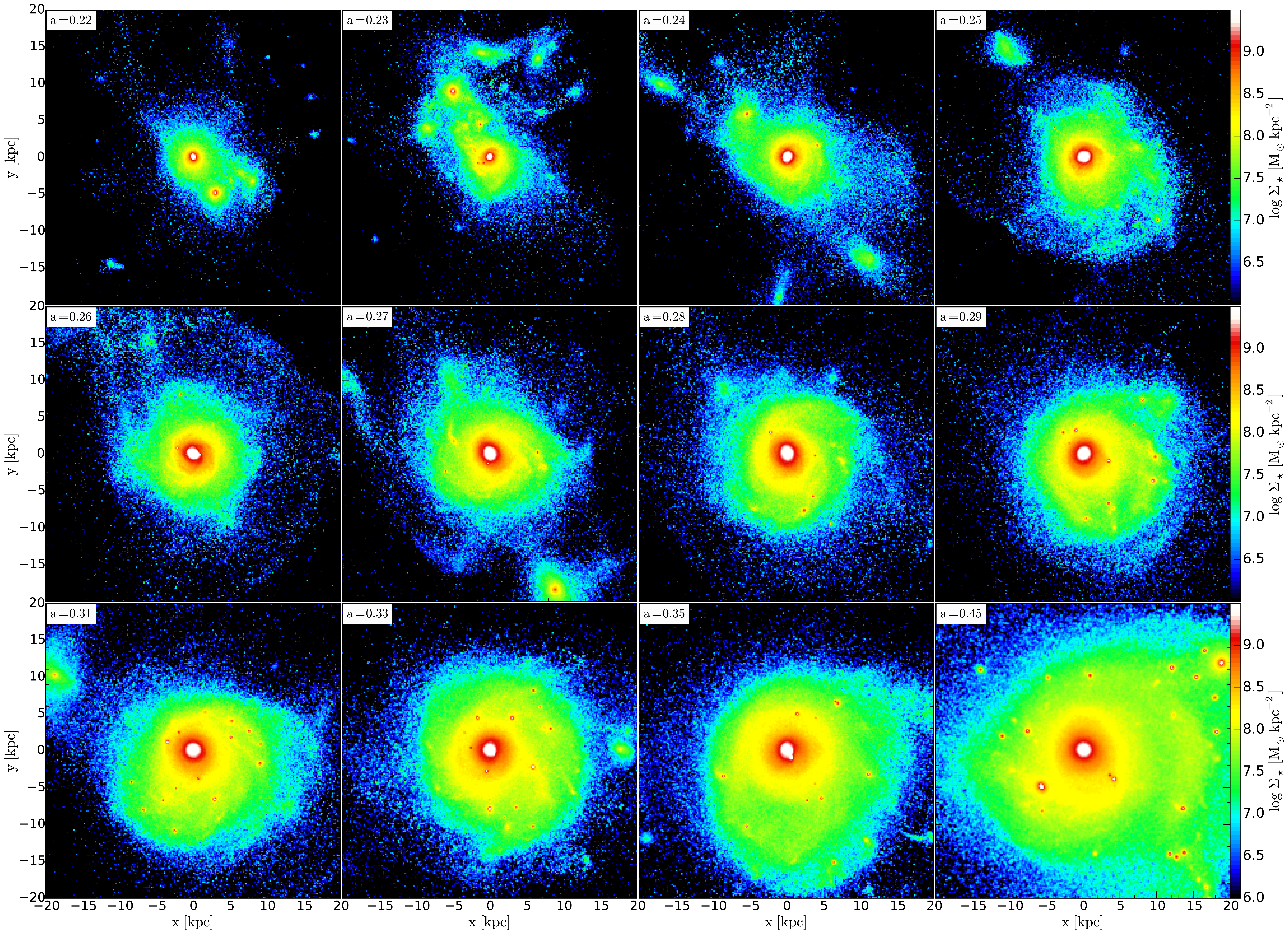}
\vskip -0.25cm
\caption{ 
Evolution from compaction through a blue-nugget to post-blue-nugget
disc and ring, complementing \fig{mosaic_v07}.
Shown are the face-on projected densities of gas (top) and stars (bottom) 
of the simulated galaxy V07.
The first panels at $a\seq 0.22\sdash 0.23$ show the pre-compaction 
phase and the compaction process, leading to a blue nugget at $a\seq 0.24$.
The following few panels show the post-compaction VDI disc.
Already at $a\seq 0.27$ we see the onset of central depletion, and the 
panels from $a\seq 0.29$ and on display a long-lived
extended ring, fed by high-AM cold streams.
A compact stellar bulge forms during and soon after the compaction 
and it remains compact and massive as the stars fade to 
a red nugget surrounded by a stellar envelope.
}
\label{fig:app_mosaic12_v07}
\end{figure*}

Here we show complementary relevant images from the VELA simulations. 

\smallskip
\Fig{app_mosaic12_v07} is a more detailed extension of \fig{mosaic_v07},
showing more stages in the evolution, with emphasis on the post-compaction
discs and rings.

\begin{figure*} 
\centering
\includegraphics[width=1.0\textwidth]
{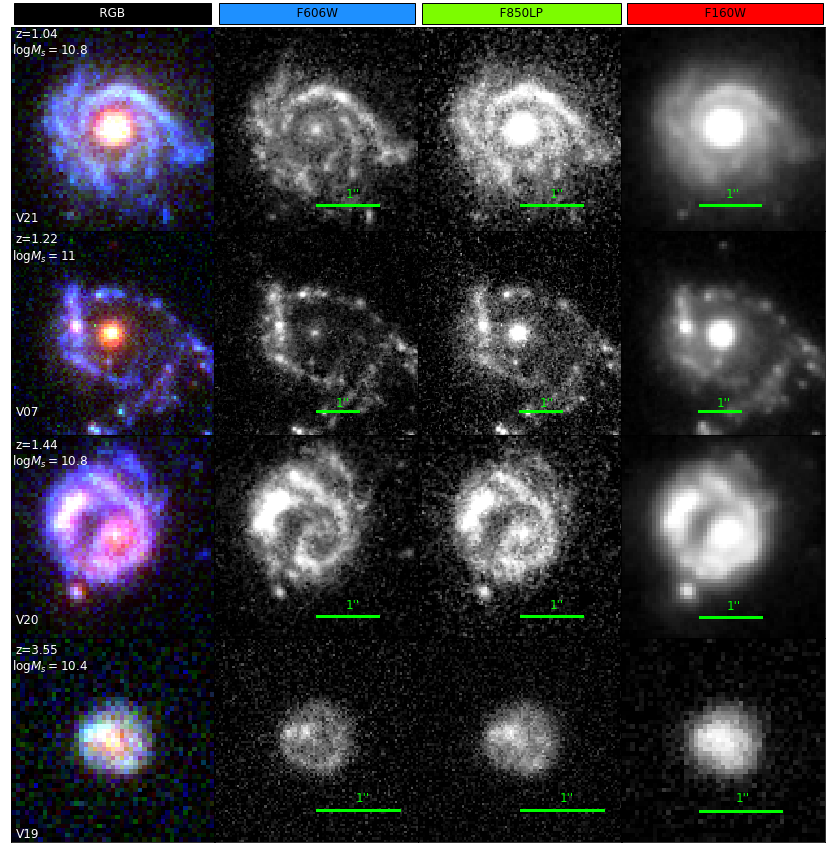}
\caption{
Mock HST three-color images of simulated galaxies.
Complementing \fig{mock_rgb}, this figure presents the 
images in three HST filters separately. 
They show blue rings about red massive bulges
as ``observed" from the four simulated galaxies seen in \fig{rings_gas}.
Dust is incorporated using Sunrise
and the galaxy is observed face-on through the HST filters using the HST
resolution and the noise corresponding to GOODS-S.
}
\label{fig:app_mock_4x4}
\end{figure*}

\begin{figure*} 
\centering
\includegraphics[width=1.0\textwidth]
{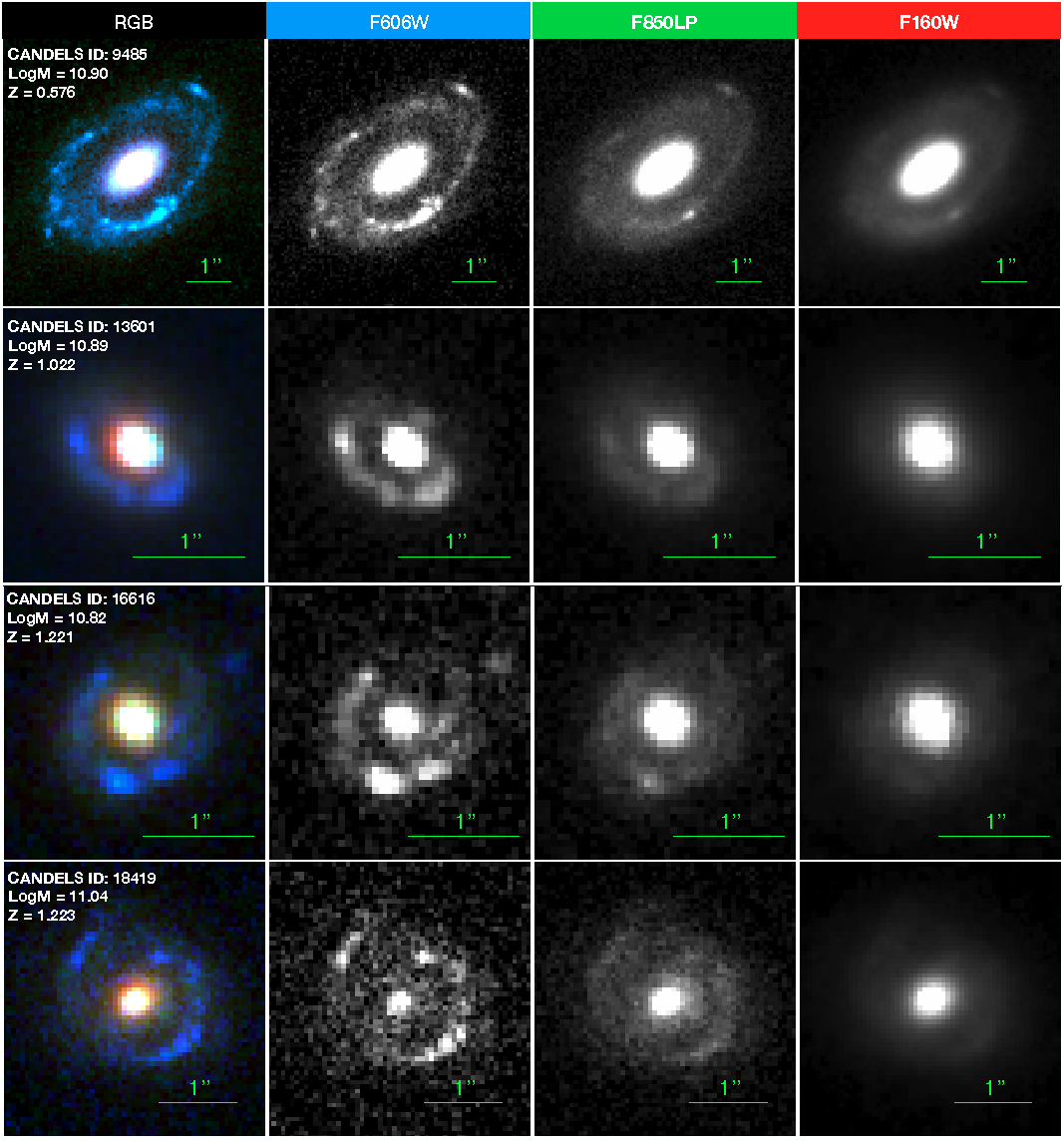}
\caption{
Observed rings in three colors.
Complementing \fig{obs_rgb}, this figure presents the
images in three HST filters separately, 
for galaxies observed in the deepest GOODS-S fields of CANDELS.
The images display extended blue rings about massive bulges, two blue and two
red.
}
\label{fig:app_obs_4x4}
\end{figure*}

\smallskip
\Fig{app_mock_4x4} shows the mock images of the four simulated galaxies
shown in \fig{mock_rgb}, now presenting the images in the three HST filters
F606W, F850LP and F160W.
\Fig{app_obs_4x4} shows the same for the four observed galaxies shown in
\fig{obs_rgb}.

\smallskip


\label{lastpage}
\end{document}